\documentclass[reprint,amsmath,amssymb,aps,prb,superscriptaddress]{revtex4-2}
\usepackage{array}
\usepackage{bm}
\usepackage{color}
\usepackage{float}
\usepackage{graphicx}
\usepackage{hyperref}
\hypersetup{colorlinks=true,allcolors=blue}
\usepackage{natbib}
\usepackage{newtxtext}
\usepackage{newtxmath}
\usepackage{mathrsfs}
\usepackage[caption=false]{subfig}

\begin{document}

\title{Competing quantum spin liquids, gauge fluctuations, and anisotropic interactions \\ in a breathing pyrochlore lattice}

\author{Li Ern Chern}
\affiliation{T.C.M. Group, Cavendish Laboratory, University of Cambridge, Cambridge CB3 0HE, United Kingdom}

\author{Yong Baek Kim}
\affiliation{Department of Physics, University of Toronto, Toronto, Ontario M5S 1A7, Canada}

\author{Claudio Castelnovo}
\affiliation{T.C.M. Group, Cavendish Laboratory, University of Cambridge, Cambridge CB3 0HE, United Kingdom}


\begin{abstract}
We use the projective symmetry group analysis to classify the quantum spin liquids on the $S=1/2$ pyrochlore magnet with a breathing anisotropy. We find 40 $\mathbb{Z}_2$ spin liquids and 16 $U(1)$ spin liquids that respect the $F\bar{4}3m$ space group and the time reversal symmetry. As an application, we consider the antiferromagnetic Heisenberg model, which is proposed to be the dominant interaction in the candidate material Ba$_3$Yb$_2$Zn$_5$O$_{11}$. Focusing on the $U(1)$ spin liquid ansatze, we find that only two of them are physical when restricted to this model. We present an analytical solution to the parton mean field theory for each of these two $U(1)$ spin liquids. It is revealed that one of them has gapless, while the other one has gapped, spinon excitations. The two $U(1)$ spin liquids are equal in energy regardless of the degree of breathing anisotropy, and they can be differentiated by the low-temperature heat capacity contribution from the quadratically-dispersing gapless spinons. We further show that the latter is unaffected by fluctuations of the $U(1)$ gauge field within the random phase approximation. Finally, we demonstrate that a small Dzyaloshinskii-Moriya interaction lifts the degeneracy between the two $U(1)$ spin liquids, and it eventually causes the lattice to decouple into independent tetrahedra at strong coupling. While current model parameters for Ba$_3$Yb$_2$Zn$_5$O$_{11}$ place it indeed in the decoupled regime, other candidate materials may be synthesised in the near future that realize the spin liquid states discussed in our work.
\end{abstract}


\maketitle


\section{\label{section:introduction}Introduction}

Pyrochlore magnets with antiferromagnetically coupled local moments are exemplary frustrated systems that provide fertile grounds for the exploration of spin liquid physics in three dimensions~\cite{nature08917,RevModPhys.89.025003,annurev-conmatphys-031218-013401,science.aay0668}. 
For instance, those with a strong Ising anisotropy realize classical~\cite{nature20619,PhysRevLett.84.3430,science.1064761,nature06433,science.1178868,science.1177582,science.1181510,annurev-conmatphys-020911-125058} or quantum~\cite{PhysRevB.69.064404,PhysRevLett.100.047208,PhysRevLett.108.037202,PhysRevB.86.104412,PhysRevLett.121.067201} spin ice, which displays emergent electromagnetism and magnetic monopoles. 
Further enriching the research into pyrochlore magnets is the abundance of synthetized materials, many of which possess spin interactions that deviate significantly from the spin ice Hamiltonian but nevertheless exhibit unusual behaviors~\cite{RevModPhys.82.53,PhysRevX.1.021002,Gingras_2014,PhysRevB.95.094422,annurev-conmatphys-022317-110520}.

In this work, we study a less symmetric variant of the pyrochlore lattice known as the breathing pyrochlore lattice~\cite{JPSJ.84.104710,PhysRevB.94.075146,ptx023,Essafi_2017,PhysRevLett.120.026801,PhysRevB.99.144406,2206.03707}, which breaks the inversion symmetry but retains all other symmetries of its regular counterpart. Pictorially, the up and down tetrahedra that share vertices with one another have different sizes. Breathing pyrochlore materials, such as spinel oxides~\cite{PhysRevLett.110.097203,PhysRevLett.113.227204,JPSJ.84.043707,PhysRevB.91.174435,PhysRevB.93.174402,PhysRevB.95.134438,JPSJ.87.034709} and Ba$_3$Yb$_2$Zn$_5$O$_{11}$~\cite{PhysRevB.90.060414,JPSJ.85.034721,PhysRevLett.116.257204,PhysRevB.93.220407,Rau_2018,2111.06293}, with $S=3/2$ and $S=1/2$ local moments respectively, were first synthetized and investigated in the context of frustrated magnetism as early as 2012. Recently, there has been a revival of interest in breathing pyrochlore magnets due to proposals that they may stabilize spin liquid phases that are characterized by rank-2 $U(1)$ gauge fields (i.e., the emergent electromagnetic fields are tensors instead of vectors) and fractonic excitations~\cite{PhysRevLett.124.127203,PhysRevB.105.L060408,PhysRevB.105.235120}, and that host an emergent axion field and a $\theta$-term coupled to the emergent QED~\cite{2109.06890}.

Roughly coinciding with these works is the successful classification of symmetric $\mathbb{Z}_2$ and $U(1)$ quantum spin liquids on the regular pyrochlore lattice~\cite{PhysRevB.100.075125,PhysRevB.104.054401,PhysRevB.105.035149,PhysRevB.105.125122}, within the frameworks of bosonic and fermionic parton mean field theories. Such a symmetry-based classification is yet to be extended to the breathing pyrochlore lattice, a task which is taken up by this study. A better understanding of the lower-rank spin liquids is important and interesting in its own right, and, together with the aforementioned rank-2 spin liquids, they will provide a basis for future investigations into exotic phases of matter in breathing pyrochlore magnets.

Here, we classify the possible $\mathbb{Z}_2$ and $U(1)$ quantum spin liquids in the $S=1/2$ breathing pyrochlore magnet, via the projective symmetry group (PSG) analysis~\cite{PhysRevB.65.165113,PhysRevB.83.224413,PhysRevB.84.024420,PhysRevB.95.054404,PhysRevB.95.054410,PhysRevB.97.195141,PhysRevB.104.094413} based on the complex fermion mean field theory~\cite{BASKARAN1987973,PhysRevB.37.580,PhysRevB.38.745,PhysRevB.44.2664}. Both the spatial symmetries of the breathing pyrochlore lattice, i.e., the space group $F4\bar{3}m$, and the time reversal symmetry are enforced. This results in 40 $\mathbb{Z}_2$ spin liquids and 16 $U(1)$ spin liquids. We then explain how these quantum spin liquids are related to those in the regular pyrochlore lattice previously classified by Ref.~\onlinecite{PhysRevB.104.054401}. In particular, we explicitly demonstrate that all of the 16 $U(1)$ spin liquids in the regular pyrochlore lattice are special cases of the 8 $U(1)$ spin liquids in the breathing pyrochlore lattice.

As an application of the PSG classfication results, we consider the antiferromagnetic (AFM) Heisenberg model and look for physical $U(1)$ spin liquid ansatze that have nonzero bond parameters throughout the lattice. There are only two such ansatze out of 16, which we label $U(1)_0$ and $U(1)_\pi$, with trivial and projective realizations of translational symmetries respectively. Interestingly, their corresponding mean field theories admit analytical solutions, owing largely to the existence of flat bands~\cite{Essafi_2017,PhysRevLett.120.026801}. We find that, independently of the ratio between the interactions on the small and large tetrahedra that characterize the breathing anisotropy, the two spin liquids are exactly degenerate. However, the $U(1)_0$ state has gapless spinon excitations with a quadratic dispersion at low energies, while the $U(1)_\pi$ state has gapped spinon excitations, so they can in principle be distinguished by thermodynamic measurements \cite{spinonnote}.

For instance, the exponential (power law) dependence on temperature of the heat capacity may be used to infer the presence (absence) of an excitation gap. We further investigate how the coupling of the $U(1)$ gauge field to the gapless spinons~\cite{PhysRevB.39.8988,PhysRevLett.63.680,PhysRevB.46.5621,POLCHINSKI1994617,PhysRevB.50.17917,PhysRevB.52.17275,PhysRevB.54.2715,PhysRevB.69.035111,PhysRevB.72.045105,PhysRevLett.95.036403,RevModPhys.78.17,PhysRevB.76.165104,PhysRevB.76.235124} may modify the low temperature heat capacity by means of an effective field theory. Within the random phase approximation and the small momentum limit, we find that the heat capacity of the gauge photon has the same scaling $C(T) \sim T^{3/2}$ as that of bare spinons, suggesting that the effect of gauge fluctuations are not important.

The candidate spin model of the $S=1/2$ breathing pyrochlore material Ba$_3$Yb$_2$Zn$_5$O$_{11}$ (abbreviated as BYZO) consists of an AFM Heisenberg interaction and a Dyzaloshinskii-Moriya (DM) interaction roughly five times smaller in magnitude~\cite{PhysRevLett.116.257204,PhysRevB.93.220407}. For a more realistic model, we thus add a subleading DM interaction to the AFM Heisenberg interaction and study its effects on the two $U(1)$ spin liquids. We find that a finite DM interaction lifts the degeneracy between them and favors the $U(1)_0$ state. When the DM interaction becomes sufficiently strong, it drives the system into isolated tetrahedra and neither spin liquids survive. In particular, neither the $U(1)_0$ state nor the $U(1)_\pi$ state can be stabilized in the parameter space relevant for BYZO, where the interactions on the large tetrahedra are very weak (i.e., near the decoupled limit). This result corroborates the fact that single tetrahedron modelling is well suited to understand the physics of BYZO, as previously demonstrated by highly accurate fits to inelastic neutron scattering spectra~\cite{PhysRevLett.116.257204,PhysRevB.93.220407,Rau_2018,2111.06293}. We expect that the $U(1)_0$ and $U(1)_\pi$ spin liquids will be relevant to spin models closer to the Heisenberg limit or with a less severe breathing anisotropy. Materials that statisfy these criteria may be synthetized in the near future, given the growing interests in breathing pyrochlore magnets.

The rest of the paper is organized as follows. In Sec.~\ref{section:symmetry}, we discuss the symmetry of the breathing pyrochlore lattice and set up conventions for the coordinate system. In Sec.~\ref{section:meanfieldtheory}, we briefly introduce the complex fermion mean field theory, before presenting the $\mathbb{Z}_2$ and $U(1)$ quantum spin liquids that result from the PSG analysis. In Sec.~\ref{section:heisenbergmodel}, we consider applications to the AFM Heisenberg model. Focusing on the $U(1)$ spin liquids, we first argue that there are only two physical ansatze, $U(1)_0$ and $U(1)_\pi$, due to constraints from the PSG (Sec.~\ref{section:physicalpsg}). Then we present analytical solutions to the corresponding mean field theories (Sec.~\ref{section:analyticalsolution}). Taking into account the coupling of gapless spinons to the gauge field, we construct an effective field theory for the $U(1)_0$ state, from which we derive the low temperature heat capacity (Sec.~\ref{section:heatcapacity}). After that, we study the effects of adding a Dzyaloshinskii-Moriya interaction (Sec.~\ref{section:dzyaloshinskiimoriya}). In Sec.~\ref{section:discussion}, we summarize our work, discuss its relations to existing theoretical studies, and outline potential future directions.

\section{\label{section:symmetry}Symmetry}

\begin{figure}
\subfloat[]{\label{figure:lattice}
\includegraphics[scale=0.24]{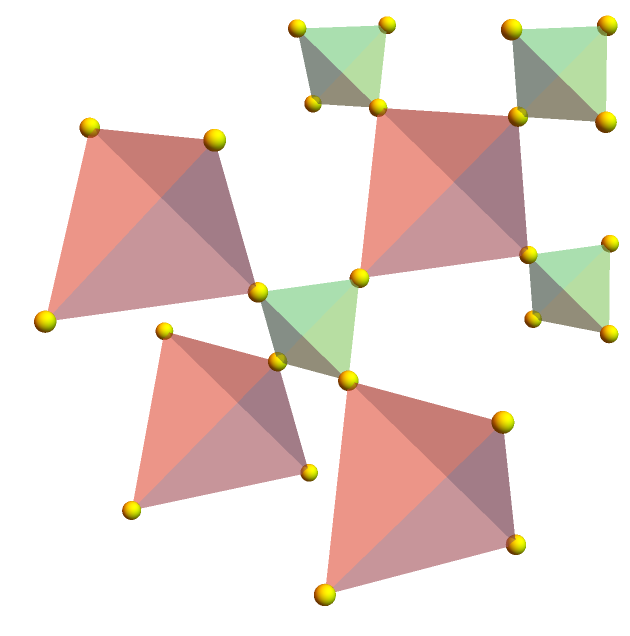}} \quad
\subfloat[]{\label{figure:cube}
\includegraphics[scale=0.28]{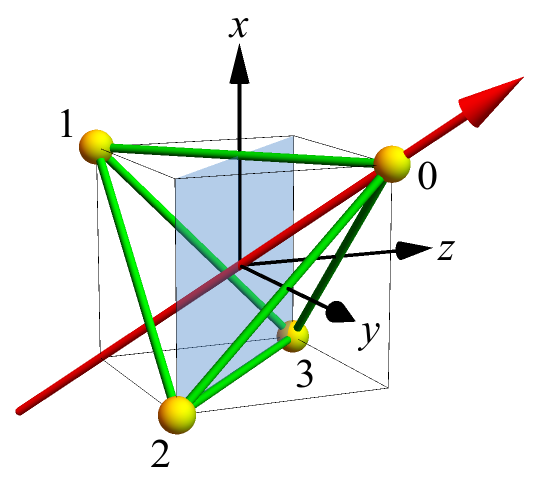}}
\caption{(a) The breathing pyrochlore lattice is characterized by an asymmetry between the up and down tetrahedra, which are drawn in green and red respectively. (b) A tetrahedron can be embedded inside a cube. A threefold rotation $C_3$ about the $[111]$ axis (indicated by the red arrow) and a reflection $\sigma$ across the plane perpendicular to $[011]$ (indicated by the blue plane) generate the entire point group of the breathing pyrochlore lattice. This figure also defines the coordinate system and the sublattice labelings.}
\end{figure}

The pyrochlore lattice is a three dimensional network of corner sharing tetrahedra. By convention, each tetrahedron can be categorized as ``up'' or ``down'' according to its spatial orientation. These two species are related by an inversion symmetry about a site (i.e., a vertex of some tetrahedron). The inversion symmetry can be broken by introducing a breathing anisotropy, so that the up and down tetrahedra now have different sizes, see Fig.~\ref{figure:lattice}. The resulting structure is called the breathing pyrochlore lattice, which belongs to the space group $F 4 \bar{3} m$ (No.~216)~\cite{PhysRevB.90.060414,PhysRevLett.116.257204}. The underlying Bravais lattice is the face centered cubic (FCC) lattice, with the primitive translation vectors
\begin{equation}
\mathbf{a}_1 = \frac{a}{2} (\hat{\mathbf{y}} + \hat{\mathbf{z}}) ,
\mathbf{a}_2 = \frac{a}{2} (\hat{\mathbf{z}} + \hat{\mathbf{x}}) ,
\mathbf{a}_3 = \frac{a}{2} (\hat{\mathbf{x}} + \hat{\mathbf{y}}) ,
\end{equation}
where $a$ is the lattice constant that defines the FCC unit cell. We choose, for concreteness, the up (down) tetrahedra to be the smaller (larger) ones. The coordinates of a site on the breathing pyrochlore lattice can be expressed as
\begin{equation}
\mathbf{r} = r_1 \mathbf{a}_1 + r_2 \mathbf{a}_2 + r_3 \mathbf{a}_3 + \mathbf{d}_s \equiv (r_1,r_2,r_3;s) ,
\end{equation}
where $r_i \in \mathbb{Z}$, $s \in \lbrace 0,1,2,3 \rbrace$ indexes the four sites of a unit cell, and the sublattice coordinates $\mathbf{d}_s$ are
\begingroup
\allowdisplaybreaks
\begin{subequations}
\begin{align}
\mathbf{d}_0 &= \rho a (+ \hat{\mathbf{x}} + \hat{\mathbf{y}} + \hat{\mathbf{z}}) / 8 , \label{sublatticecoordinates0} \\
\mathbf{d}_1 &= \rho a (+ \hat{\mathbf{x}} - \hat{\mathbf{y}} - \hat{\mathbf{z}}) / 8, \label{sublatticecoordinates1} \\
\mathbf{d}_2 &= \rho a (- \hat{\mathbf{x}} + \hat{\mathbf{y}} - \hat{\mathbf{z}}) / 8, \label{sublatticecoordinates2} \\
\mathbf{d}_3 &= \rho a (- \hat{\mathbf{x}} - \hat{\mathbf{y}} + \hat{\mathbf{z}}) / 8.
\label{sublatticecoordinates3}
\end{align}
\end{subequations}
\endgroup
Here $0 < \rho < 1$ parametrizes the breathing anisotropy, with smaller $\rho$ giving a greater difference between the sizes of small and large tetrahedra. Note that with $\rho=1$ in \eqref{sublatticecoordinates0}-\eqref{sublatticecoordinates3}, we restore the inversion symmetry and thus recover the regular pyrochlore lattice.

The point group of $F 4 \bar{3} m$ is the tetrahedral group $T_d$ of 24 elements, which are best visualized by embedding a tetrahedron in a cube \cite{tinkhamtextbook,dresselhaustextbook}, see Fig.~\ref{figure:cube}. These 24 elements can be generated by just two of them, namely $C_3$, a rotation by $2 \pi / 3$ about the $[111]$ axis, and $\sigma$, a reflection across the plane perpendicular to $[011]$ and containing the origin. For example, the  fourfold rotoreflection consisting of a rotation by $\pi / 2$ about the $[100]$ axis and then a reflection across the $yz$ plane can be expressed as $S_4=C_3^2 \sigma C_3^2$. Below we list the action of the space group generators on a generic site with the coordinates $(r_1,r_2,r_3;s)$.
\begingroup
\allowdisplaybreaks
\begin{subequations}
\begin{align}
T_1 &: (r_1,r_2,r_3,s) \longrightarrow (r_1+1,r_2,r_3,s) , \label{T1operation} \\
T_2 &: (r_1,r_2,r_3,s) \longrightarrow (r_1,r_2+1,r_3,s) , \label{T2operation} \\
T_3 &: (r_1,r_2,r_3,s) \longrightarrow (r_1,r_2,r_3+1,s) , \label{T3operation} \\
C_3 &: (r_1,r_2,r_3,s) \longrightarrow (r_3,r_1,r_2,C_3(s)) , \label{C3operation} \\
\sigma &: (r_1,r_2,r_3,s) \longrightarrow (-r_1-r_2-r_3,r_2,r_3,\sigma(s)) . \label{sigmaoperation}
\end{align}
\end{subequations}
\endgroup
When $s=0,1,2,3$, $C_3(s)=0,2,3,1$ and $\sigma(s)=1,0,2,3$. 

Apart from the space group, we also consider time reversal symmetry, which is present in a model with only bilinear spin interactions (or, more generally, involving only products of an even number of dipolar operators). We denote the time reversal operator by $\mathcal{T}$.

The commutation relations between these symmetry generators constitute what are called the algebraic identities, which are listed in App.~\ref{appendix:algebraicidentity}. The algebraic identities will be crucial to the projective symmetry group (PSG) analysis discussed in the next section.

\section{\label{section:meanfieldtheory}Complex fermion mean field theory}

Complex fermion mean field theory~\cite{BASKARAN1987973,PhysRevB.37.580,PhysRevB.38.745,PhysRevB.44.2664} and its gauge structure have been discussed extensively in numerous references~\cite{PhysRevB.65.165113,PhysRevB.83.224413,PhysRevB.84.024420,PhysRevB.95.054404,PhysRevB.95.054410,PhysRevB.97.195141,PhysRevB.104.094413}. We only outline here some key steps that are instrumental to the methodology of this work.

We first represent spins in terms of complex fermions,
\begin{equation} \label{partonrepresentation}
\mathbf{S}_i = \sum_{\alpha \beta} f_{i {\alpha}}^\dagger \frac{[\vec{\sigma}]_{\alpha \beta}}{2} f_{i \beta} 
\, ,
\end{equation}
so that the Hamiltonian, which is assumed to be bilinear in the spins, is quartic in the fermions. We then perform a mean field decoupling to obtain a Hamiltonian quadratic in the fermions. For concreteness, let us consider the antiferromagnetic Heisenberg model,
\begin{equation} \label{heisenbergmodel}
H = \sum_{ij} J_{ij} \: \mathbf{S}_i \cdot \mathbf{S}_j \, , 
\qquad 
J_{ij}>0
\end{equation}
from which we obtain
\begin{equation} \label{meanfieldhamiltonian}
\begin{aligned}[b]
H^\mathrm{MF} = & \sum_{ij} \frac{J_{ij}}{4} \Big[ \chi_{ij}^* ( f_{i \uparrow}^\dagger f_{j \uparrow} + f_{i \downarrow}^\dagger f_{j \downarrow}) + \mathrm{h.c.} - \lvert \chi_{ij} \rvert^2 \\ 
& \quad \quad \quad \: + \Delta_{ij}^* ( f_{i \uparrow} f_{j \downarrow} - f_{i \downarrow} f_{j \uparrow} ) + \mathrm{h.c.} - \lvert \Delta_{ij} \rvert^2 \Big] \\
+ & \sum_i \left[ \lambda_i^{(3)} ( n_i - 1 ) + (\lambda_i^{(1)} + i \lambda_i^{(2)}) f_{i \downarrow} f_{i \uparrow} + \mathrm{h.c.} \right] ,
\end{aligned}
\end{equation}
where $n_i =f_{i \uparrow}^\dagger f_{i \uparrow} + f_{i \downarrow}^\dagger f_{i \downarrow}$ is the number operator, and $\chi_{ij}$ and $\Delta_{ij}$ are variational parameters of the singlet hopping and pairing channels respectively. On-site Lagrange multipliers $\lambda^{(1,2,3)}_i \in \mathbb{R}$ are introduced to enforce the single occupancy constraint (i.e., one fermion per site), as the representation~\eqref{partonrepresentation} has enlarged the original Hilbert space $\mathcal{H} = \otimes_i \lbrace \lvert \uparrow \rangle_i , \lvert \downarrow \rangle_i \rbrace$ by allowing zero or double occupancies, which are unphysical.

Anisotropic spin interactions, such as the Dzyaloshinskii-Moriya interaction that will be discussed in Sec.~\ref{section:dzyaloshinskiimoriya}, usually require triplet hopping and pairing channels~\cite{PhysRevB.80.064410,PhysRevB.85.224428,PhysRevB.86.224417,s41598-019-47517-6} in the parton representation. Importantly, the projective symmetry group classification of quantum spin liquids (discussed in Sec.~\ref{section:psg}) does not depend on the particular spin model, but only on the symmetries of the system.

The parton representation~\eqref{partonrepresentation} introduces an $SU(2)$ gauge redundancy, in which the mean field Hamiltonian~\eqref{meanfieldhamiltonian} is invariant under a symmetry $X$ of the system only up to a gauge transformation $G_X \in SU(2)$. In other words, the symmetries of the system are realized \textit{projectively} at the mean field level. The $2 \times 2$ matrix of variational parameters,
\begin{equation} \label{meanfieldansatz}
u_{ij} = \frac{J_{ij}}{4} \begin{pmatrix} \chi_{ij} & -\Delta_{ij}^* \\ -\Delta_{ij} & - \chi_{ij}^* \end{pmatrix},
\end{equation}
by symmetry and $SU(2)$ gauge redundancy, obeys
\begin{equation} \label{ansatzspacerule}
u_{X(i)X(j)} = G_X (X(i)) u_{ij} G_X^\dagger (X(j)) ,
\end{equation}
for any space group element $X$.

The time reversal symmetry $\mathcal{T}$, being an antiunitary operator, requires some care. We can choose a gauge such that~\cite{PhysRevB.65.165113}
\begin{equation} \label{ansatztimerule}
u_{ij} = - G_\mathcal{T}(i) u_{ij} G_\mathcal{T}^\dagger (j) .
\end{equation}
The $2 \times 2$ matrix $u_{ii}$ for the on-site terms is similarly defined, see~\eqref{onsiteansatz}, and it also obeys~\eqref{ansatzspacerule} and~\eqref{ansatztimerule}. Further details of the complex fermion mean field theory can be found in App.~\ref{appendix:meanfieldtheory}.

\begin{table*}
\caption{\label{table:spinliquid} In conjunction with~\eqref{Z2cellT1gauge}-\eqref{Z2celltimegauge} and~\eqref{U1cellT1angle}-\eqref{U1celltimeangle}, this table lists all the possible $\mathbb{Z}_2$ and $U(1)$ spin liquids. For all $\mathbb{Z}_2$ spin liquids, $g_{C_3} (1,2,3)=1$ and $g_\mathcal{T} (s)=i \tau_2$. For all $U(1)$ spin liquids, $n_\mathcal{T}=1$, $\varphi_{C_3} (s)=0$, $\varphi_\sigma (0,1,2) = 0$, and $\varphi_\mathcal{T} (s)=0$. Below $q_0 \in \lbrace 0,1,2 \rbrace$ and $p_{T_2 T_1}, p_{\sigma T_2}, p_{\sigma C_3} \in \lbrace 0,1 \rbrace$.}
\begin{ruledtabular}
\begin{tabular}{cccc||cccc}
\multicolumn{4}{c||}{$\mathbb{Z}_2$} & \multicolumn{4}{c}{$U(1)$} \\ \hline
$(\eta_{T_2 T_1},\eta_{\sigma T_2},\eta_\mathcal{T},\eta_\sigma,\eta_{\sigma \mathcal{T}}, \eta_{\sigma C_3})$ & $g_{C_3} (0)$ & $g_\sigma (0,1,2)$ & $g_\sigma (3)$ & $n_\sigma$ & $\theta_{T_2 T_1}$ & $\theta_{\sigma T_2}$ & $\varphi_\sigma (3)$ \\ \hline
$(\pm 1, \pm 1, -1, +1, +1, \pm 1)$ & $1$ & $1$ & $\eta_{\sigma C_3}$ & $0$ & $p_{T_2 T_1} \pi$ & $p_{\sigma T_2} \pi$ & $p_{\sigma C_3} \pi$ \\
$(\pm 1, \pm 1, -1, -1, +1, \pm 1)$ & $1$ & $i \tau_2$ & $\eta_{\sigma C_3} (i \tau_2)$ & $1$ & $p_{T_2 T_1} \pi$ & $p_{\sigma T_2} \pi$ & $p_{\sigma C_3} \pi$ \\
$(\pm 1, \pm 1, -1, -1, -1, \pm 1)$ & $e^{-i (2 \pi q_0/3) \tau_2}$ & $i \tau_3$ & $\eta_{\sigma C_3} (i \tau_3) e^{i (2 \pi q_0/3) \tau_2}$ & & & &
\end{tabular}
\end{ruledtabular}
\end{table*}

\subsection{\label{section:psg}Projective symmetry group analysis}

It was first proposed in Ref.~\onlinecite{PhysRevB.65.165113} that, given a particular set of symmetries $\lbrace X \rbrace$, the different possible sets of gauge transformations $\lbrace G_X \rbrace$ provide a means to classify quantum spin liquids. Compound operators of the form $G_X X$ constitute the so called projective symmetry group (PSG). This forms the basis of the PSG analysis~\cite{PhysRevB.83.224413,PhysRevB.84.024420,PhysRevB.95.054404,PhysRevB.95.054410,PhysRevB.97.195141,PhysRevB.104.094413}.

The form of $G_X \in SU(2)$ is not arbitrary but restricted by the symmetries of the system. To understand this, we introduce a special subgroup of PSG known as the invariant gauge group (IGG), which consists of pure gauge transformations that leave the mean field ansatz invariant. The algebraic identities~\eqref{T2T1commute}-\eqref{timesquare}, such as $T_2^{-1} T_1^{-1} T_2 T_1 = e$, constrain $G_X$ via, for example
\begin{equation} \label{psgT2T1commute}
(G_{T_2} T_2)^{-1} (G_{T_1} T_1)^{-1} (G_{T_2} T_2) (G_{T_1} T_1) \in \mathrm{IGG} .
\end{equation}
This is because for expressions like the left hand side of~\eqref{psgT2T1commute}, the net effect of the symmetry operators is an identity, so what remains must amount to a pure gauge transformation that leaves the mean field ansatz invariant (take $X=e$ in~\eqref{ansatzspacerule}, for instance). When both hopping and pairing terms are present in the Hamiltonian, the IGG is $\lbrace +1, -1 \rbrace$, and the resulting spin liquids are called $\mathbb{Z}_2$ spin liquids. When only hopping terms are present in the Hamiltonian, the IGG is $\lbrace e^{i \theta \tau_3} \, \vert \, 0 \leq \theta < 2 \pi \rbrace$, and the resulting spin liquids are called $U(1)$ spin liquids.

We summarize below the results of the PSG classification of the $\mathbb{Z}_2$ and $U(1)$ spin liquids. Details of the calculations can be found in Apps.~\ref{appendix:classifyZ2spinliquid} and \ref{appendix:classifyU1spinliquid}.

\subsubsection{$\mathbb{Z}_2$ spin liquids}

We find $40$ fully symmetric $\mathbb{Z}_2$ spin liquids on the breathing pyrochlore lattice, which are distinguished by the gauge transfomations
\begingroup
\allowdisplaybreaks
\begin{subequations}
\begin{align}
& G_{T_1} (r_1,r_2,r_3;s) = 1, \label{Z2cellT1gauge} \\
& G_{T_2} (r_1,r_2,r_3;s) = \eta_{T_2 T_1}^{r_1}, \label{Z2cellT2gauge} \\
& G_{T_3} (r_1,r_2,r_3;s) = \eta_{T_2 T_1}^{r_1+r_2}, \label{Z2cellT3gauge} \\
& G_{C_3} (r_1,r_2,r_3;s) = \eta_{T_2 T_1}^{r_1(r_2+r_3)} g_{C_3}(s), \label{Z2cellC3gauge} \\
& G_\sigma (r_1,r_2,r_3;s) 
= \eta_{\sigma T_2}^{r_2} \eta_{T_2 T_1}^{r_2 (r_2-1)/2 + r_3 (r_3-1)/2 + r_2 r_3} g_\sigma(s) , \label{Z2cellsigmagauge} \\
& G_\mathcal{T} (r_1,r_2,r_3;s) = g_\mathcal{T} (s), \label{Z2celltimegauge}
\end{align}
\end{subequations}
\endgroup
with $\eta_{\ldots}$ and $g_X (s)$ given in Table~\ref{table:spinliquid}.

\subsubsection{$U(1)$ spin liquids}

For $U(1)$ spin liquids, the gauge transformations have the specific form~\cite{PhysRevB.65.165113,PhysRevB.95.054404,PhysRevB.97.195141,PhysRevB.104.054401}
\begin{equation} \label{U1gaugetransformform}
\begin{aligned}[b]
& G_X (r_1,r_2,r_3;s) = (i \tau_1)^{n_X} e^{i \phi_X (r_1,r_2,r_3;s) \tau_3}, \\
& n_X \in \lbrace 0, 1 \rbrace, \phi_{X} \in [0, 2 \pi) .
\end{aligned}
\end{equation}
We find 16 fully symmetric $U(1)$ spin liquids on the breathing pyrochlore lattice, which are distinguished by the gauge transfomations
\begingroup
\allowdisplaybreaks
\begin{subequations}
\begin{align}
& \phi_{T_1} (r_1,r_2,r_3;s) = 0, \; n_{T_1} = 0, \label{U1cellT1angle} \\
& \phi_{T_2} (r_1,r_2,r_3;s) = r_1 \theta_{T_2 T_1}, \; n_{T_2} = 0, \label{U1cellT2angle} \\
& \phi_{T_3} (r_1,r_2,r_3;s) = (r_2 - r_1) \theta_{T_2 T_1}, \; n_{T_3} = 0, \label{U1cellT3angle} \\
\label{U1cellC3angle}
& \phi_{C_3} (r_1,r_2,r_3;s) = r_1 (r_2 - r_3) \theta_{T_2 T_1} + \varphi_{C_3} (s), \; n_{C_3} = 0,
\\
\begin{split} \label{U1cellsigmaangle}
& \phi_\sigma (r_1,r_2,r_3;s) = \bigg[ \frac{r_2 (r_2+1)}{2} + (-1)^{n_\sigma} \frac{r_3 (r_3+1)}{2} \\
& \quad + (-1)^{n_\sigma} r_2 + r_3 + r_2 r_3 \bigg] \theta_{T_2 T_1} + r_2 \theta_{\sigma T_2} + \varphi_\sigma (s) ,
\end{split} \\
& \phi_\mathcal{T} (r_1,r_2,r_3;s) = \varphi_\mathcal{T} (s) . \label{U1celltimeangle}
\end{align}
\end{subequations}
\endgroup
with $\theta_{\ldots}$ and $\varphi_X (s)$ given in Table \ref{table:spinliquid}.

\subsection{Relation to the isotropic lattice}

We have found 16 fully symmetric $U(1)$ spin liquids in the breathing pyrochlore lattice. Intriguingly, the regular pyrochlore lattice has the same number of $U(1)$ spin liquids~\cite{PhysRevB.104.054401}. In this subsection, we would like to further investigate the relation between the $U(1)$ spin liquids of the regular and breathing pyrochlore lattices. Let A and B be two systems such that the symmetry group of B is a subgroup of the symmetry group of A. Generally, we expect the spin liquids in A to be included, as special cases, among the spin liquids in B~\cite{PhysRevB.95.054410}. In our case, the breathing pyrochlore lattice is obtained from the regular pyrochlore by breaking the inversion symmetry, so we can think of A (B) as the regular (breathing) pyrochlore lattice.

The PSG analysis for the regular pyrochlore lattice is done with the set of space group generators $\lbrace T_1,T_2,T_3,\overline{C}_6,S \rbrace$, where $\overline{C}_6$ is a sixfold rotoinversion and $S$ is a twofold nonsymmorphic screw~\cite{PhysRevB.104.054401}. They can be used to construct the point group generators of the breathing pyrochlore lattice, via $C_3 = \overline{C}_6^4$ and $\sigma = (\overline{C}_6)^4 S \overline{C}_6^{-1}$. Therefore, to see the relation between the spin liquids in the breathing and regular pyrochlore lattices, we can compare the gauge transformation parts of $G_{T_1} T_1$, $G_{T_2} T_2$, $G_{T_3} T_3$, $G_{C_3} C_3$, $G_\sigma \sigma$, and $G_\mathcal{T} \mathcal{T}$ found in Sec.~\ref{section:psg} to those of $G_{T_1} T_1$, $G_{T_2} T_2$, $G_{T_3} T_3$, $(G_{\overline{C}_6} \overline{C}_6)^4$, $(G_{\overline{C}_6} \overline{C}_6)^4 (G_S S) (G_{\overline{C}_6} \overline{C}_6)^{-1}$, and $G_\mathcal{T} \mathcal{T}$ in Ref.~\onlinecite{PhysRevB.104.054401}, respectively~\cite{PhysRevB.95.054410}.

Leaving the detailed calculations to App.~\ref{appendix:connection}, the final result is stated as follows. All 16 $U(1)$ spin liquids of the regular pyrochlore lattice are continuously connected to the 8 $U(1)$ spin liquids of the breathing pyrochlore lattice with $\theta_{\sigma T_2}=0$, in the fashion of a two-to-one mapping. The other 8 $U(1)$ spin liquids in the breathing pyrochlore lattice with $\theta_{\sigma T_2}=\pi$ have no correspondence. A similar exercise can be carried out to clarify the relation between the $\mathbb{Z}_2$ spin liquids in the regular and breathing pyrochlore lattices, but we shall not report it here for simplicity.

\section{\label{section:heisenbergmodel}Antiferromagnetic Heisenberg Model}

Ba$_3$Yb$_2$Zn$_5$O$_{11}$ (abbreviated as BYZO) is so far the only existing $S=1/2$ breathing pyrochlore magnet~\cite{PhysRevB.90.060414,JPSJ.85.034721,PhysRevLett.116.257204,PhysRevB.93.220407,Rau_2018,2111.06293}. There are other breathing pyrochlore materials but with higher spins~\cite{PhysRevLett.110.097203,PhysRevLett.113.227204,JPSJ.84.043707,PhysRevB.91.174435,PhysRevB.93.174402,PhysRevB.95.134438,JPSJ.87.034709}, and they do not fall into the classification scheme presented in this work. The proposed spin model for BYZO consists of a dominant antiferromagnetic (AFM) Heisenberg interaction and a subleading Dzyaloshinskii-Moriya interaction on isolated (small) tetrahedra~\cite{PhysRevLett.116.257204,PhysRevB.93.220407,spinmodelnote}. The inter-tetrahedron coupling, i.e., the interactions on the large tetrahedra, are not well understood at the moment but expected to be insignificant, due to the rather large ratio $\sim 2$ between the bond lengths of the large and small tetrahedra.

As an application of the PSG classification results, we first study the pure AFM Heisenberg model on the breathing pyrochlore lattice, and then consider the effects of adding a Dzyaloshinskii-Moriya interaction. For simplicity, we consider only the fully symmetric $U(1)$ spin liquid candidates.

We denote the nearest neighbor Heisenberg interactions on the up (down) tetrahedra by $J_1$ ($J_2$). The mean field Hamiltonian of the AFM Heisenberg model is given by~\eqref{meanfieldhamiltonian}, which describes a $\mathbb{Z}_2$ spin liquid as both hopping and pairing terms are present (one can think of it as the spinon analog of a BCS superconductor). We set the pairing terms to be zero for a $U(1)$ spin liquid Hamiltonian (essentially a chargeless free fermion model). Notice that both $U(1)$ spin liquids with gapless and gapped spinons are allowed to exist in 3+1 dimensions, while in 2+1 dimensions $U(1)$ spin liquids with gapped spinons are unstable to monopole proliferation~\cite{POLYAKOV1977429}.

\subsection{\label{section:physicalpsg}Physical spin liquid ansatze}

As discussed in Sec.~\ref{section:psg}, the mean field ansatze $u_{ij}$ on different bonds are related by the (projective) symmetries of the system via~\eqref{ansatzspacerule}, which reduces the number of independent $\chi_{ij}$ and/or $\Delta_{ij}$. On the other hand, a symmetry that maps a bond onto itself, e.g., the time reversal symmetry via~\eqref{ansatztimerule}, may impose additional constraints on the bond. If the symmetry, or the conflicting requirements arising from different symmetries, forces $u_{ij}=0$ everywhere, then we get a zero Hamiltonian, which is unphysical and thus discarded~\cite{PhysRevB.83.224413,PhysRevB.95.054404}. We also exclude the case of $u_{ij} \neq 0$ on one species of the tetrahedra (say \textit{up}) and $u_{ij}=0$ on the other (say \textit{down}). This describes a system of decoupled tetrahedra, which reduces to a $4$-spin problem. While the decoupled tetrahedron model is interesting, it does not require complex fermion mean field theory, and we do not consider it explicitly in this work where we focus on spin liquid states with full spatial connectivitiy.

Therefore, having specified the spin model and thus the mean field parameters, we may find that some of the PSGs in Sec.~\ref{section:psg} are unphysical or irrelevant, upon the applications of~\eqref{ansatzspacerule} and~\eqref{ansatztimerule}. For $U(1)$ spin liquids, with the pairing terms $\Delta_{ij}=0$ (as well as the on-site $\lambda^{(1,2)}_i=0$) in~\eqref{meanfieldhamiltonian}, we have
\begin{equation} \label{simplifyhoppingansatz}
u_{ij} = \begin{pmatrix} \chi_{ij} & 0 \\ 0 & - \chi_{ij}^* \end{pmatrix} \, ,
\end{equation}
where, to simplify the notation, we have temporarily dropped the $J_{ij}/4$ factor in~\eqref{simplifyhoppingansatz}. First, we note that $\chi_{ji}=\chi_{ij}^*$ from the definition~\eqref{singlethopping}. Second, all $U(1)$ spin liquids must have real $\chi_{ij}$ by virtue of time reversal symmetry: since $G_\mathcal{T} (i) = i \tau_1$ for all $i$,~\eqref{ansatztimerule} implies $u_{ij}=- (i \tau_1) u_{ij} (- i \tau_1)$, or
\begin{equation}
\begin{pmatrix} \chi_{ij} & 0 \\ 0 & - \chi_{ij}^* \end{pmatrix} = \begin{pmatrix} \chi_{ij}^* & 0 \\ 0 & - \chi_{ij} \end{pmatrix} \, ,
\end{equation}
so $\chi_{ij} = \chi_{ij}^*$. It follows that $u_{ji}=u_{ij}$ for all $i$ and $j$.

Introducing the shorthand notation $u_{st}$ to denote $u_{ij}$ with $i=(0,0,0;s)$ and $j=(0,0,0;t)$, i.e., the mean field ansatz of the bond formed by sublattices $s$ and $t$ in the unit cell at the origin, we demonstrate that the PSGs with $n_\sigma = 1$ is unphysical.~\eqref{ansatzspacerule} with $X=\sigma$ yields
\begin{equation}
u_{10} = (i \tau_1) u_{01} (- i \tau_1) ,
\end{equation}
or $u_{01} = - u_{01} = 0$, and by symmetry all $u_{ij}$ on the up tetrahedra will be zero. We thus rule out all 8 PSGs that carry $n_\sigma = 1$.

On the other hand, for $n_\sigma = 0$, we show that only the PSGs with $p_{\sigma C_3}=0$ and $p_{\sigma T_2}=0$ are relevant. \eqref{ansatzspacerule} with $X=\sigma$ yields
\begin{equation}
u_{23} = u_{23} e^{i p_{\sigma C_3} \pi \tau_3} .
\end{equation}
If $p_{\sigma C_3} = 1$, then $u_{23} = -u_{23} = 0$, and by symmetry all $u_{ij}$ on the up tetrahedra will be zero. We thus discard $p_{\sigma C_3}=1$. Next, we consider the bond formed by sublattices $2$ and $3$ on a down tetrahedron. Let $u_{23}' \equiv u_{(0,1,0;2)(0,0,1;3)}$. By $\sigma$ we have
\begin{equation}
u_{(-1,1,0;2)(-1,0,1;3)} = e^{i p_{\sigma T_2} \pi \tau_3} u_{23}' .
\end{equation}
By $T_1$ we have
\begin{equation}
u_{(-1,1,0;2)(-1,0,1;3)} = u_{23}' .
\end{equation}
If $p_{\sigma T_2} = 1$, then $u_{23}' = -u_{23}' = 0$, and by symmetry all $u_{ij}$ on the up tetrahedra will be zero. We thus discard $p_{\sigma T_2}=1$.

\begin{figure}
\subfloat[]{\label{figure:ansatzU1zero}
\includegraphics[scale=0.36]{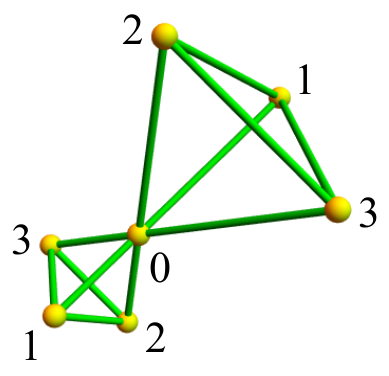}} \quad
\subfloat[]{\label{figure:ansatzU1pi}
\includegraphics[scale=0.36]{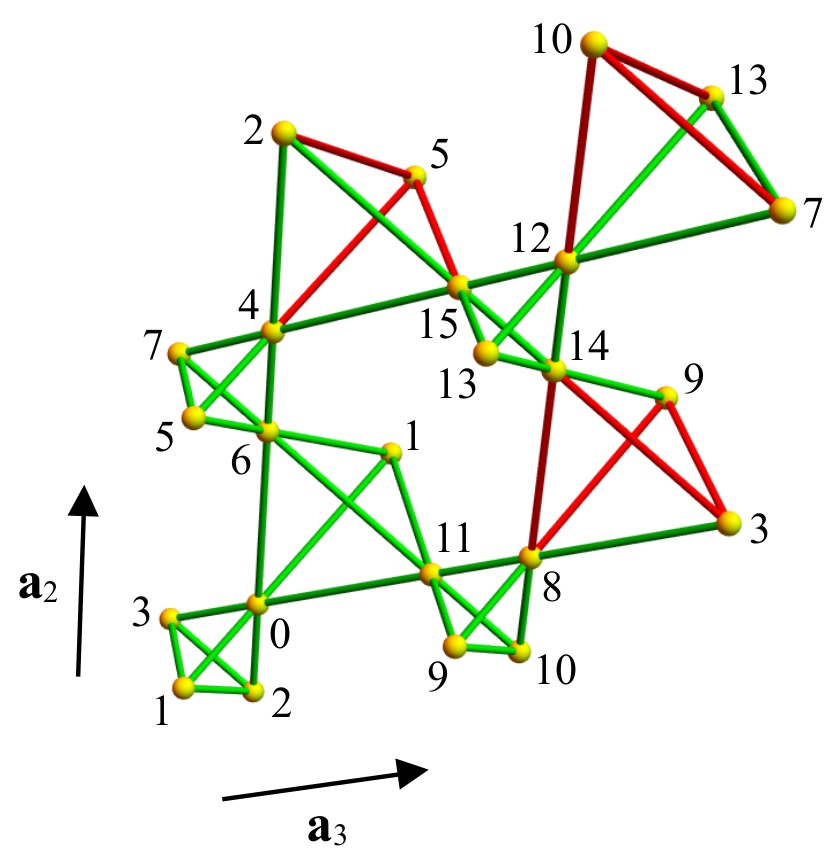}}
\caption{The mean field ansatze of the two possible $U(1)$ spin liquids, the (a) $U(1)_0$ and (b) $U(1)_{\pi}$ states, in the antiferromagnetic Heisenberg model. They contain $\mathcal{N}=4$ and $\mathcal{N}=16$ sites per unit cell respectively, which are labelled by $s=0, \ldots,\mathcal{N}-1$. Green (red) bonds on the down tetrahedra indicate $\chi_{ij}=+\chi_2$ ($\chi_{ij}=-\chi_2$). All bonds on the up tetrahedra have $\chi_{ij}=+\chi_1$ and they are colored in green. In (b), the unit cell is enlarged in the $\mathbf{a}_2$ and $\mathbf{a}_3$ directions, due to the projective realization of the translational symmetry.}
\end{figure}

Out of the $16$ PSGs for the $U(1)$ spin liquids, we are eventually left with only $2$ PSGs with $n_\sigma = 0$, $p_{\sigma T_2}=0$, and $p_{\sigma C_3}=0$. They are distinguished by the value of $\theta_{T_2 T_1}=0,\pi$, so we call them the $U(1)_0$ and $U(1)_\pi$ states respectively. Upon repeated applications of~\eqref{ansatzspacerule}, they lead to the mean field ansatze depicted in Figs.~\ref{figure:ansatzU1zero} and~\ref{figure:ansatzU1pi}.

\subsection{\label{section:analyticalsolution}Analytical solutions}

The mean field theories of the $U(1)_0$ and $U(1)_\pi$ states can be solved analytically, owing to the existence of flat bands. It also follows from the solutions that the degeneracy of these spin liquids persists throughout the range of $J_2/J_1 \in (0,1)$. We will present the solution for the $U(1)_0$ state in this subsection, while relegating that of the $U(1)_\pi$ state to App.~\ref{appendix:analyticalsolution}.

Before dwelving into the analysis, let us first examine the structure of the Fourier transformed Hamiltonian. We use the following convention for Fourier transform
\begin{equation}
f_{\mathbf{k}s\sigma} = \frac{1}{\sqrt{N}} \sum_\mathbf{R} f_{\mathbf{R} s \sigma} e^{i \mathbf{k} \cdot \mathbf{R}} ,
\end{equation}
where $N$ is the total number of unit cells, $\mathbf{R}$ is the unit cell coordinate, $s$ is the sublattice index, and $\sigma \in \lbrace \uparrow , \downarrow \rbrace$. Taking as our basis $\Psi_\mathbf{k} = (f_{\mathbf{k} 0 \uparrow}, \ldots, f_{\mathbf{k} (\mathcal{N}-1) \uparrow},f_{\mathbf{k} 0 \downarrow}, \ldots, f_{\mathbf{k} (\mathcal{N}-1) \downarrow})$, where $\mathcal{N}$ is the number of sublattices per unit cell (4 and 16 for the $U(1)_0$ and $U(1)_\pi$ states respectively), the Hamiltonian has the form
\begin{equation}
H = \sum_\mathbf{k} \left[ \Psi_\mathbf{k}^\dagger D_\mathbf{k} \Psi_\mathbf{k} + \frac{3 \mathcal{N}}{2} \left( \frac{J_1}{4} \chi_1^2 + \frac{J_2}{4} \chi_2^2 \right) \right] .
\end{equation}
Since the Hamiltonian contains only singlet hopping channels, fermions with up and down spins do not mix, and the $2 \mathcal{N} \times 2 \mathcal{N}$ matrix $D_\mathbf{k}$ is block diagonal, consisting of two identical copies of the $\mathcal{N} \times \mathcal{N}$ matrix $d_\mathbf{k}$. The energy eigenvalues of $D_\mathbf{k}$ can thus be obtained by diagonalizing $d_\mathbf{k}$ and imposing a twofold degeneracy. The single occupancy constraint in real space, in which each site is occupied by a fermion, is translated to a half filling constraint in momentum space, in which the lower (upper) half of the energy eigenstates are occupied (empty) in the ground state. Note that if we impose the half filling constraint by hand, then we no longer have to explicitly introduce the Lagrange multiplier $\lambda_i^{(3)}$.

Furthermore, we note that the global gauge transformation $i \tau_1$ flips the signs of both $\chi_1$ and $\chi_2$ (which can be thought of as a particle hole transformation), while leaving the total energy invariant (as it should). We thus assume $\chi_1 < 0$ without loss of generality. While the overall sign is not important, we will see later that the relative sign between $\chi_1$ and $\chi_2$ does make a difference. Finally, we define $\tilde{\chi}_1 = - J_1 \chi_1 / 4 >0$ and $\tilde{\chi}_2 = - J_2 \chi_2 / 4$ for convenience. We also require both $\chi_1$ and $\chi_2$ to be nonzero for the physical solutions that we are interested in (see the discussion in Sec.~\ref{section:physicalpsg}).

For the $U(1)_0$ state, diagonalizing $d_\mathbf{k}$ yields the four eigenvalues $\varepsilon_0, \varepsilon_0, \varepsilon_+, \varepsilon_-$, where
\begin{subequations}
\begin{align}
\varepsilon_0 (\mathbf{k}) &= - \tilde{\chi}_1 - \tilde{\chi}_2 , \label{U1zeroeigenvalue0} \\
\varepsilon_\pm (\mathbf{k}) &= \tilde{\chi}_1 + \tilde{\chi}_2 \pm \sqrt{4 \tilde{\chi}_1^2 + 4 \tilde{\chi}_2^2 - 2 \tilde{\chi}_1 \tilde{\chi}_2 [2 - f (\mathbf{k})]} , \label{U1zeroeigenvaluepm}
\end{align}
\end{subequations}
with
\begin{equation}
\begin{aligned}[b]
f (\mathbf{k}) = & \, \cos (k_1 - k_2) + \cos (k_2 - k_3) + \cos (k_3 - k_1) \\
& + \cos k_1 + \cos k_2 + \cos k_3 \, , \qquad k_i \in [0, 2 \pi) \, .
\end{aligned}
\end{equation}
The maximum and minimum of $f (\mathbf{k})$ are $6$ and $-2$ respectively, an information that will be useful later. The same dispersions~\eqref{U1zeroeigenvalue0} and~\eqref{U1zeroeigenvaluepm} have been obtained in different contexts, namely as eigenvalues of (i) the interaction matrix in the classical analysis of the breathing pyrochlore Heisenberg model~\cite{JPSJ.84.104710} and (ii) the tight binding model on the breathing pyrochlore lattice with real hopping integrals~\cite{PhysRevLett.120.026801}.

\begin{figure*}
\subfloat[]{\label{figure:bandzerofluxJ2080}
\includegraphics[scale=0.3]{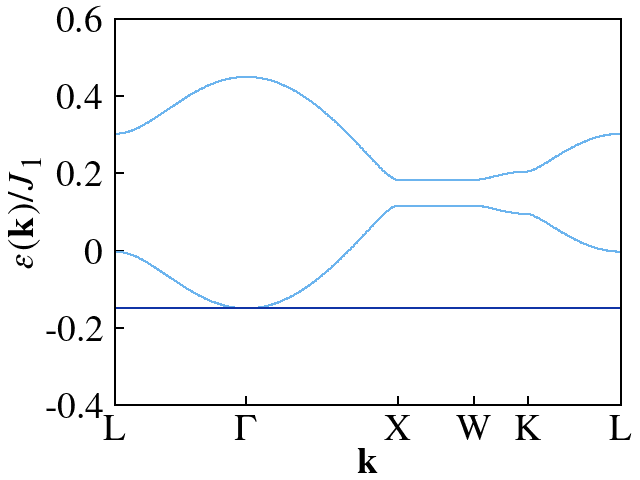}} \quad
\subfloat[]{\label{bandzerofluxJ2050}
\includegraphics[scale=0.3]{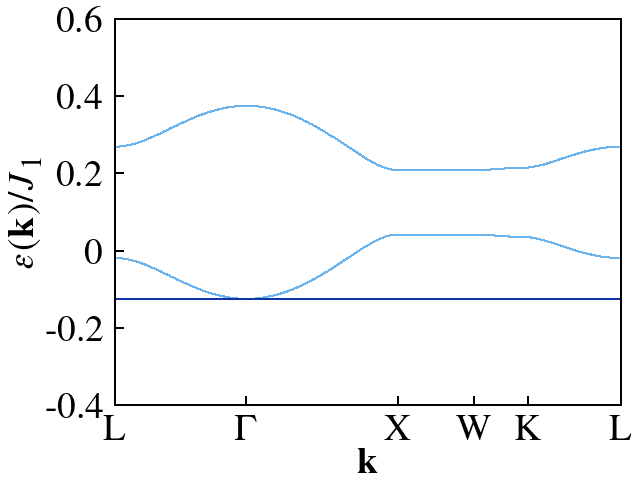}} \quad
\subfloat[]{\label{figure:bandzerofluxJ2020}
\includegraphics[scale=0.3]{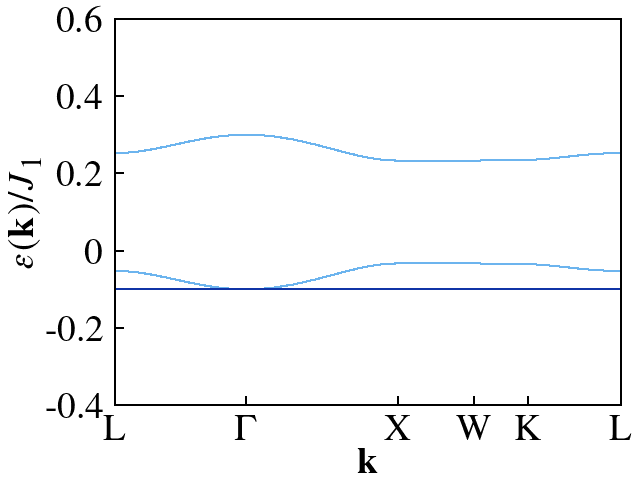}} \\
\subfloat[]{\label{figure:bandpifluxJ2080}
\includegraphics[scale=0.3]{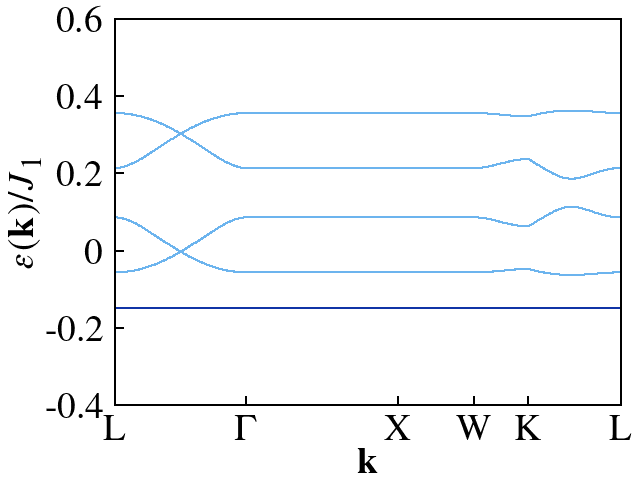}} \quad
\subfloat[]{\label{bandpifluxJ2050}
\includegraphics[scale=0.3]{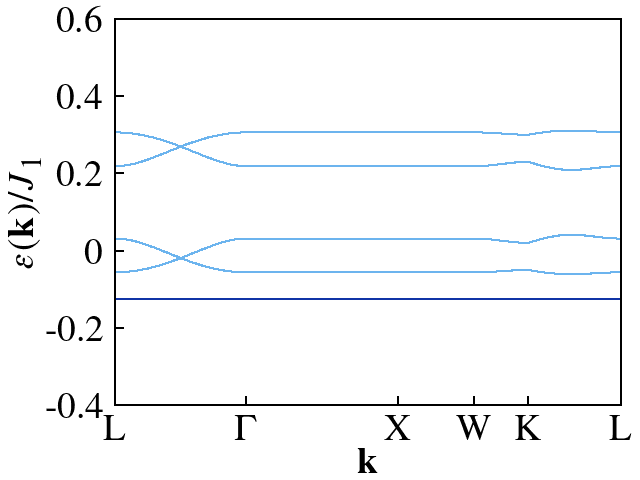}} \quad
\subfloat[]{\label{figure:bandpifluxJ2020}
\includegraphics[scale=0.3]{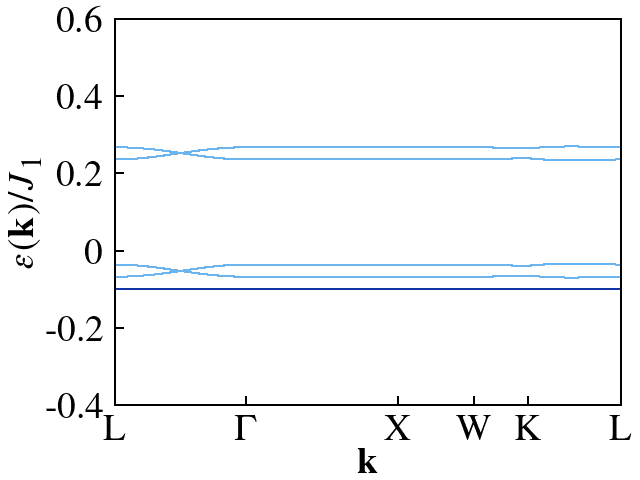}}
\caption{(a-c) The spinon dispersion of the $U(1)_0$ spin liquid, at (a) $J_2/J_1=0.8$, (b) $J_2/J_1=0.5$, and (c) $J_2/J_1=0.2$. The flat band (indicated by darker color) is fourfold degenerate, while each of the two dispersing bands (indicated by lighter color) is doubly degenerate, totalling eight bands (4 sublattices $\times$ 2 spins). At half filling, all flat (dispersing) bands are (un)occupied. (d-f) The spinon dispersion of the $U(1)_\pi$ spin liquid, at (d) $J_2/J_1=0.8$, (e) $J_2/J_1=0.5$, and (f) $J_2/J_1=0.2$. The flat band (indicated by darker color) is 16-fold degenerate, while each of the four dispersing bands (indicated by lighter color) is 4-fold degenerate, totalling 32 bands (16 sublattices $\times$ 2 spins). At half filling, all flat (dispersing) bands are (un)occupied.}
\end{figure*}
 
We first assume that $\tilde{\chi}_2 > 0$, i.e.,~$\chi_1$ and $\chi_2$ have the same sign. We see that $\varepsilon_+ (\mathbf{k}) > \varepsilon_0 (\mathbf{k})$ for all $\mathbf{k}$, so the entire $+$ band is unoccupied. On the other hand,
\begin{equation*}
\begin{aligned}[b]
\min_\mathbf{k} \varepsilon_- (\mathbf{k}) &= \tilde{\chi}_1 + \tilde{\chi}_2 - \sqrt{4 \tilde{\chi}_1^2 + 4 \tilde{\chi}_2^2 + 8 \tilde{\chi}_1 \tilde{\chi}_2} , \\
& = - \tilde{\chi}_1 - \tilde{\chi}_2 ,
\end{aligned}
\end{equation*}
which implies $\varepsilon_0 (\mathbf{k}) \leq \varepsilon_- (\mathbf{k})$ for all $\mathbf{k}$ (equality holds only when $\mathbf{k}=0$). The $\pm$ bands are thus well separated from, and higher in energy than, the two $0$ bands throughout the Brillouin zone except at $\mathbf{k}=0$. Then, filling the lower half of the energy eigenstates, i.e., all the flat bands, the total energy of a system with $N \times \mathcal{N}$ sites is given by
\begin{equation} \label{U1zerofluxtotalenergy}
\begin{aligned}[b]
E_S & = \sum_\mathbf{k} \left[ 4 \varepsilon_0 (\mathbf{k}) + 6 \left( \frac{J_1}{4} \chi_1^2 + \frac{J_2}{4} \chi_2^2 \right) \right] \\
& = \sum_\mathbf{k} \left[ 4 \left( \frac{J_1}{4} \chi_1 + \frac{J_2}{4} \chi_2 \right) + 6 \left( \frac{J_1}{4} \chi_1^2 + \frac{J_2}{4} \chi_2^2 \right) \right] ,
\end{aligned}
\end{equation}
where the factor of $4$ instead of $2$ in front of $\varepsilon_0 (\mathbf{k})$ takes into account the two spin flavors (recall that we have two copies of $d_\mathbf{k}$).

Next, we show that if $\tilde{\chi}_2 < 0$, i.e., $\chi_1$ and $\chi_2$ have opposite signs, then the total energy is always higher than the previously analyzed case of $\tilde{\chi}_2 > 0$. Without loss of generality, we may assume $\lvert \tilde{\chi}_1 \rvert \geq \lvert \tilde{\chi}_2 \rvert$, for if it is otherwise, we can perform the global gauge transformation $i \tau_1$ and interchange $\tilde{\chi}_1$ and $\tilde{\chi}_2$. For clarity, we rewrite the eigenvalues~\eqref{U1zeroeigenvalue0} and~\eqref{U1zeroeigenvaluepm} as
\begingroup
\allowdisplaybreaks
\begin{subequations}
\begin{align*}
\varepsilon_0 (\mathbf{k}) = & - \lvert \tilde{\chi}_1 \rvert + \lvert \tilde{\chi}_2 \rvert , \\
\begin{split}
\varepsilon_\pm (\mathbf{k}) = & \lvert \tilde{\chi}_1 \rvert - \lvert \tilde{\chi}_2 \rvert \pm \sqrt{4 \lvert \tilde{\chi}_1 \rvert^2 + 4 \lvert \tilde{\chi}_2 \rvert^2 + 2 \lvert \tilde{\chi}_1 \rvert \lvert \tilde{\chi}_2 \rvert [2 - f (\mathbf{k})]},
\end{split}
\end{align*}
\end{subequations}
\endgroup
One can again see that $\varepsilon_+ (\mathbf{k}) > \varepsilon_0 (\mathbf{k})$ for all $\mathbf{k}$, so the entire $+$ band is unoccupied. On the other hand,
\begingroup
\allowdisplaybreaks
\begin{subequations}
\begin{align*}
\begin{split}
\max_\mathbf{k} \varepsilon_- (\mathbf{k}) &= \lvert \tilde{\chi}_1 \rvert - \lvert \tilde{\chi}_2 \rvert - \sqrt{4 \lvert \tilde{\chi}_1 \rvert^2 + 4 \lvert \tilde{\chi}_2 \rvert^2 - 8 \lvert \tilde{\chi}_1 \rvert \lvert \tilde{\chi}_2 \rvert} \\
&= - \lvert \tilde{\chi}_1 \rvert + \lvert \tilde{\chi}_2 \rvert ,
\end{split} \\
\begin{split}
\min_\mathbf{k} \varepsilon_- (\mathbf{k}) &= \lvert \tilde{\chi}_1 \rvert - \lvert \tilde{\chi}_2 \rvert - \sqrt{4 \lvert \tilde{\chi}_1 \rvert^2 + 4 \lvert \tilde{\chi}_2 \rvert^2 + 8 \lvert \tilde{\chi}_1 \rvert \lvert \tilde{\chi}_2 \rvert} \\
&= - \lvert \tilde{\chi}_1 \rvert - 3 \lvert \tilde{\chi}_2 \rvert ,
\end{split}
\end{align*}
\end{subequations}
\endgroup
which implies $\varepsilon_0 \geq \varepsilon_- (\mathbf{k})$ for all $\mathbf{k}$ (equality holds only when $\mathbf{k}=0$). The $\pm$ bands are thus well separated from the two $0$ zero bands, with the $+$ ($-$) band lying above (below) the flat bands. Therefore, filling the lower half of the energy eigenstates corresponds to filling the $-$ band and one of the $0$ bands. The total energy for a system of $N \times \mathcal{N}$ sites is given by
\begingroup
\allowdisplaybreaks
\begin{equation}
\begin{aligned}[b]
\begin{split}
E_A &= \sum_\mathbf{k} \bigg[ 2 \varepsilon_- (\mathbf{k}) + 2 \varepsilon_0 (\mathbf{k}) + 6 \left( \frac{J_1}{4} \lvert \chi_1 \rvert^2 + \frac{J_2}{4} \lvert \chi_2 \rvert^2 \right) \bigg] \\
& > \sum_\mathbf{k} \bigg[ 2 \min_\mathbf{k} \varepsilon_- (\mathbf{k}) + 2 \left( - \frac{J_1}{4} \lvert \chi_1 \rvert + \frac{J_2}{4} \lvert \chi_2 \rvert \right) \\
& \qquad \quad + 6 \left( \frac{J_1}{4} \lvert \chi_1 \rvert^2 + \frac{J_2}{4} \lvert \chi_2 \rvert^2 \right) \bigg]
\\
&= \sum_\mathbf{k} \bigg[ - 4 \left( \frac{J_1}{4} \lvert \chi_1 \rvert + \frac{J_2}{4} \lvert \chi_2 \rvert \right) + 6 \left( \frac{J_1}{4} \lvert \chi_1 \rvert^2 + \frac{J_2}{4} \lvert \chi_2 \rvert^2 \right) \bigg] \\
& = E_S \, .
\end{split}
\end{aligned}
\nonumber 
\end{equation}
\endgroup
The inequality in the second line is strict because there is at least one point, $\mathbf{k}=0$, that does not saturate the lower bound of $\varepsilon_- (\mathbf{k})$. We have thus shown that $E_S<E_A$, i.e., the total energy for $\tilde{\chi}_2 < 0$ is always higher than that for $\tilde{\chi}_2>0$, given that $\tilde {\chi}_1>0$. We can therefore exclude the case where $\chi_1$ and $\chi_2$ have opposite signs.

Finally, minimizing~\eqref{U1zerofluxtotalenergy} with respect to $\chi_1$ and $\chi_2$, $\partial E_S / \partial \chi_{1,2} = 0$, yields $\chi_{1,2} = - 1/3$, regardless of the values of $J_1$ and $J_2$ as long as they are positive. The ground state energy per site is thus $-(J_1+J_2)/24$.

The solution for the $U(1)_\pi$ state proceeds along essentially the same line of reasoning. Leaving the details of calculations to App.~\ref{appendix:analyticalsolution}, upon the optimization we also have $\chi_1=\chi_2=\pm 1/3$ and the ground state energy per site is $-(J_1+J_2)/24$, as in the $U(1)_0$ state. We have thus shown the robust ``1/3 quantization'' of the mean field parameters in the ground states of the two $U(1)$ spin liquids, and established their degeneracy, independent of the ratio $J_2 / J_1$ as long as both $J_1$ and $J_2$ are positive.

The spinon spectra of the $U(1)_{0,\pi}$ spin liquids at several values of $J_2/J_1$ along some high symmetry cuts in the Brillouin zone \cite{SETYAWAN2010299} are plotted in Figs.~\ref{figure:bandzerofluxJ2080} to~\ref{figure:bandpifluxJ2020}. In the $U(1)_0$ spin liquid, the spinon gap closes at the $\Gamma$ point, while in the $U(1)_\pi$ spin liquid the spinons are always gapped; these hold throughout the range $J_2/J_1 \in (0,1)$.

We remark that, in the isotropic limit $J_2/J_1=1$, our $U(1)_0$ and $U(1)_\pi$ states respectively reduce to the uniform and $(\pi,\pi)$ states of Ref.~\onlinecite{PhysRevB.79.144432}, where they are also found numerically to be degenerate. Note that our energy is smaller by an overall factor of $4$ than that in Ref.~\onlinecite{PhysRevB.79.144432} due to different schemes of the mean field decoupling.

\subsection{\label{section:heatcapacity}Low Temperature Heat Capacity}

Although the $U(1)_0$ and $U(1)_\pi$ spin liquids are degenerate, there is a qualitative distinction between their spinon spectra, namely the presence or absence of an excitation gap. This allows them to be differentiated by specific heat measurements, for example. Considering the contribution from the spinons alone, i.e., ignoring photons and visons, the heat capacity $C(T)$ is expected to show a power law (exponential) dependence on temperature $T$ for the gapless (gapped) case at low temperatures.

The spinon excitation spectrum of the $U(1)_0$ spin liquid becomes gapless at the $\Gamma$ point ($\mathbf{k}=0$), where the flat bands $\varepsilon_0 (\mathbf{k})$ touch the dispersing bands $\varepsilon_- (\mathbf{k})$, as shown in Sec.~\ref{section:analyticalsolution}. Assuming $\tilde{\chi}_{1,2} \equiv - J_{1,2} \chi_{1,2}/4 > 0$, so that the flat (dispersing) bands are filled (empty), we perform a small $\mathbf{k}$ expansion in the vicinity of the $\Gamma$ point to the lowest nontrivial order, 
\begin{equation} \label{smallkexpansion}
\begin{aligned}[b] \\
\varepsilon_- (\mathbf{k} \approx \mathbf{0}) & \approx \tilde{\chi}_1 + \tilde{\chi}_2 - \sqrt{4 (\tilde{\chi}_1 + \tilde{\chi}_2)^2 - \tilde{\chi}_1 \tilde{\chi}_2 \lvert \mathbf{k} \rvert^2} \\
& \approx - \tilde{\chi}_1 - \tilde{\chi}_2 + \frac{\tilde{\chi}_1 \tilde{\chi}_2}{4 (\tilde{\chi}_1 + \tilde{\chi}_2)} \lvert \mathbf{k} \rvert^2 \, ,
\end{aligned}
\end{equation}
where $\lvert \mathbf{k} \rvert^2=k_x^2+k_y^2+k_z^2$ and we have used
\begin{equation}
\begin{pmatrix} k_1 \\ k_2 \\ k_3 \end{pmatrix} = \frac{1}{2} \begin{pmatrix} 0 & 1 & 1 \\ 1 & 0 & 1 \\ 1 & 1 & 0 \end{pmatrix} \begin{pmatrix} k_x \\ k_y \\ k_z \end{pmatrix} \, .
\end{equation}
Therefore, at low energies the spinon excitations follow a quadratic dispersion, which gives rise to a heat capacity $C(T) \sim T^{3/2}$ at low temperatures if we neglect the effects of the $U(1)$ gauge field.

However, gauge fluctuations may be important in a spin liquid with gapless spinons and lead to a singular correction to the heat capacity coefficient $C(T)/T$ for small $T$. For instance, it has been established that, if spinons in a $U(1)$ spin liquid form a sufficiently large Fermi surface, when their coupling to the gauge field is taken into account, $C(T)$ scales as $T^{2/3}$ and $T \ln (1/T)$ in 2D~\cite{PhysRevB.72.045105,PhysRevLett.95.036403,PhysRevB.76.235124} and 3D~\cite{PhysRevB.69.035111,PhysRevB.8.2649,PhysRevB.40.11571} respectively, in contrast to $T$ for bare spinons. Here, we do not have a Fermi surface, but a flat-quadratic band touching. To understand how gauge fluctuations may modify $C(T)$ in the $U(1)_0$ spin liquid, we construct an effective field theory that includes the interaction between the low energy spinons and the gauge field via minimal coupling. Going to the continuum limit, the Lagrangian and the partition function are~\cite{PhysRevB.76.165104,PhysRevB.76.235124}
\begin{subequations}
\begin{align}
\mathscr{L} &= \bar{\psi}_\sigma (\partial_\tau - i a_0) \psi_\sigma + \frac{1}{2m} \bar{\psi}_\sigma (- i \nabla - \mathbf{a})^2 \psi_\sigma , \label{lagrangian} \\
Z &= \int D a \, D \psi_\sigma \, D \bar{\psi}_\sigma e^{- \int \mathrm{d} \tau \int \mathrm{d}^3 \mathbf{r} \, \mathscr{L}} , \label{action}
\end{align}
\end{subequations}
where $\tau$ is the imaginary time, $\psi_\sigma$ and $\bar{\psi}_\sigma$ are the spinon fields with $\sigma$ being the spin flavor, $a_0$ and $\mathbf{a}$ are the temporal and spatial components of the $U(1)$ gauge field respectively, and $a \equiv (a_0, \mathbf{a})$. The effective mass $m$ of the spinons is determined by the coupling constants $J_{1,2}$ and by the mean field parameters $\chi_{1,2}$ through~\eqref{smallkexpansion}. Note the absence of the Maxwell term
\begin{equation}
\mathscr{L}_g = \frac{1}{g} f_{\mu \nu} f^{\mu \nu} ,
f_{\mu \nu} = \partial_\mu a_\nu - \partial_\nu a_\mu ,
\end{equation}
in \eqref{lagrangian} as $g$ is proportional to the charge gap and we are working in the insulating phase \cite{PhysRevB.76.165104,PhysRevB.76.235124}. Furthermore, since the flat bands are momentum independent, they contribute a constant (taken to be zero here) to the total energy and act as a reservoir of spinons that can be thermally excited to the quadratic spectrum. Note that the fields are functions of spacetime, for example $\psi_\sigma$ in~\eqref{lagrangian} should be understood as $\psi_\sigma (\mathbf{r},\tau)$.

Integrating out the spinons in~\eqref{action} and using the random phase approximation~\cite{RevModPhys.78.17,PhysRevB.76.165104,PhysRevB.76.235124,nagaosatextbook}, we obtain an effective Lagrangian for the $U(1)$ gauge field,
\begin{subequations}
\begin{align}
& \mathscr{L}_\mathrm{eff} = - \sum_{i,j \in \lbrace x,y \rbrace} a_i (-q) \Pi_{ij} (q) a_j (q) , \label{effectivephotonlagrangian} \\
& \Pi_{ij} (q ; l = 0) = - c_1 \sqrt{T} \lvert \mathbf{q} \rvert^2 \delta_{ij} , \label{photonpropagatorstatic} \\
& \Pi_{ij} (q ; l \neq 0) = \delta_{ij} \left( - c_2 T^{3/2} + c_3 T^{5/2} \frac{\lvert \mathbf{q} \rvert^2}{\nu_l^2} \right), \label{photonpropagatordynamic}
\end{align}
\end{subequations}
in the small $\lvert \mathbf{q} \rvert$ limit. Some remarks are in order. The first one is about conventions and notations. We have chosen the Coulomb gauge $\nabla \cdot \mathbf{a} = 0$ and labeled the two transverse components of $\mathbf{a}$ by $x$ and $y$. The inverse photon propagator is identified as $- \Pi_{ij} (q)$, where $q=(\mathbf{q},\nu_l)$ is the four-momentum, $\nu_l=2 \pi l T$ with $l \in \mathbb{Z}$ is the Matsubara frequency, and $\mathbf{q}$ is the momentum, of the photon. The coefficients $c_i$ are positive and depend only on $m$. The second one is that, unlike the case of a spinon Fermi surface~\cite{PhysRevB.69.035111,PhysRevB.76.165104,nagaosatextbook}, we do not have a Fermi wavevector $k_\mathrm{F}$ relative to which the smallness of $\lvert \mathbf{q} \rvert$ can be defined. The natural dimensionless parameter that enables a small $\lvert \mathbf{q} \rvert$ expansion in our case is $\lvert \mathbf{q} \rvert / \sqrt{mT} \ll 1$. Finally, $\Pi_{ij} (q)$ assumes one of the two forms, either~\eqref{photonpropagatorstatic} or~\eqref{photonpropagatordynamic}, according to whether $l$ is zero or finite. Details of the calculations as well as extended discussions can be found in App.~\ref{appendix:effectivefieldtheory}.

In a $U(1)$ spin liquid with a spinon Fermi surface, one finds $\Pi_{ij} (q) = \delta_{ij} (- \Gamma \lvert \nu_l \rvert / \lvert \mathbf{q} \rvert - \chi \lvert \mathbf{q} \rvert^2)$, with positive constants $\Gamma$ and $\chi$ that are determined by the details of the spinon dispersion~\cite{PhysRevB.69.035111}. This is in contrast with our $\Pi_{ij}$, \eqref{photonpropagatorstatic} and~\eqref{photonpropagatordynamic}, where the coefficients $c_i$ are multiplied by powers of $T$ so that they vanish at $T=0$, as expected because there is no thermally excited spinon in the quadratic spectrum at absolute zero.

From~\eqref{effectivephotonlagrangian} we can calculate the partition function and the free energy as~\cite{PhysRevB.69.035111,PhysRevB.76.165104}
\begin{subequations}
\begin{align}
Z &= \int D \mathbf{a} \, e^{ - \sum_{\mathbf{q} l} \mathscr{L}_\mathrm{eff}} , \label{effectivepartitionfunction} \\
F &= - T \ln Z , \label{freeenergy}
\end{align}
\end{subequations}
which yield, for the ``static'' ($l=0$) and the ``dynamic'' ($l \neq 0$) contributions respectively,
\begin{subequations}
\begin{align}
F_\mathrm{sta} &= T \sum_{\mathbf{q}} \ln \bigg( c_1 \sqrt{T} \lvert \mathbf{q} \rvert^2 \bigg) , \label{freeenergystatic} \\
F_\mathrm{dyn} &= T \sum_\mathbf{q} \sum_{l \neq 0} \ln \bigg( c_2 T^{3/2} - c_3 T^{5/2} \frac{\lvert \mathbf{q} \rvert^2}{\nu_l^2} \bigg) . \label{freeenergydynamic}
\end{align}
\end{subequations}
To evaluate~\eqref{freeenergystatic} and~\eqref{freeenergydynamic}, we change the summation over $\mathbf{q}$ to an integral, which has an upper limit of $\sim \sqrt{mT}$ in line with the small $\mathbf{q}$ assumption, and regularize the summation over nonzero $l$ with the Riemann zeta function. Details of the calculation are presented in App.~\ref{appendix:effectivefieldtheory}.

The heat capacity is then given by $C (T) = - T (\partial^2 F / \partial T^2)$. In the low temperature limit, we find $C(T) \sim T^{3/2}$ to leading order. We thus conclude that $U(1)$ gauge fluctuations do not modify the scaling of the bare spinon heat capacity with temperature, unlike the case of a Fermi surface.

\subsection{\label{section:dzyaloshinskiimoriya}Effects of Dzyaloshinskii-Moriya interaction}

For the pure AFM Heisenberg model, we have seen in Sec.~\ref{section:analyticalsolution} that regardless of the strength of breathing anisotropy, we are always able to obtain spatially connected solutions from the mean field theories with the $U(1)_0$ and $U(1)_\pi$ ansatze (even if we take $J_2/J_1 \longrightarrow 0^+$, as long as $J_2$ is not strictly zero). Moreover, these two spin liquids are degenerate throughout the range $J_2/J_1 \in (0,1)$. In this subsection, we investigate how the stability and degeneracy of the $U(1)_0$ and $U(1)_\pi$ spin liquids are affected by a finite Dzyaloshinskii-Moriya interaction.

The spin Hamiltonian is now given by
\begin{equation} \label{JDmodel}
H = \sum_{n=1}^2 \sum_{\langle ij \rangle \in n} \left ( J_n \mathbf{S}_i \cdot \mathbf{S}_j + D_{n} \hat{\mathbf{d}}_{ij} \cdot \mathbf{S}_i \times \mathbf{S}_j \right) ,
\end{equation}
where, as before, the subscripts $1$ or $2$ on the interactions indicate whether they belong to the up or down tetrahedra, and $\hat{\mathbf{d}}_{ij}=\hat{\mathbf{d}}_{st}$ are unit vectors that depend only on the sublattice indices $s$ and $t$ of sites $i$ and $j$ \cite{PhysRevLett.116.257204,PhysRevLett.124.127203},
\begin{equation}
\begin{aligned}[b]
& \hat{\mathbf{d}}_{01} = \frac{-\hat{\mathbf{y}} + \hat{\mathbf{z}}}{\sqrt{2}} \, , \quad 
\hat{\mathbf{d}}_{02} = \frac{-\hat{\mathbf{z}} + \hat{\mathbf{x}}}{\sqrt{2}} \, , \quad 
\hat{\mathbf{d}}_{03} = \frac{-\hat{\mathbf{x}} + \hat{\mathbf{y}}}{\sqrt{2}} \, , \\
& \hat{\mathbf{d}}_{12} = \frac{-\hat{\mathbf{x}} - \hat{\mathbf{y}}}{\sqrt{2}} \, , \quad 
\hat{\mathbf{d}}_{23} = \frac{-\hat{\mathbf{y}} - \hat{\mathbf{z}}}{\sqrt{2}} \, , \quad 
\hat{\mathbf{d}}_{31} = \frac{-\hat{\mathbf{z}} - \hat{\mathbf{x}}}{\sqrt{2}} \, .
\end{aligned}
\end{equation}
In line with the proposed spin models of BYZO in Refs.~\onlinecite{PhysRevLett.116.257204,PhysRevB.93.220407}, we choose $D_1<0$. We further assume $D_2 = (J_2/J_1) D_1$. Using the parton representation of spins~\eqref{partonrepresentation}, the Dzyaloshinskii-Moriya interaction involves both the singlet and triplet channels (see App.~\ref{appendix:dzyaloshinskiimoriya}). If we require the singlet channels associated with the dominant AFM Heisenberg interaction to be nonvanishing, we still have $U(1)_0$ and $U(1)_\pi$ as the only two $U(1)$ spin liquid ansatze to be considered upon adding the Dzyaloshinskii-Moriya interaction.

\begin{figure}
\subfloat[]{\label{figure:phase}
\includegraphics[scale=0.33]{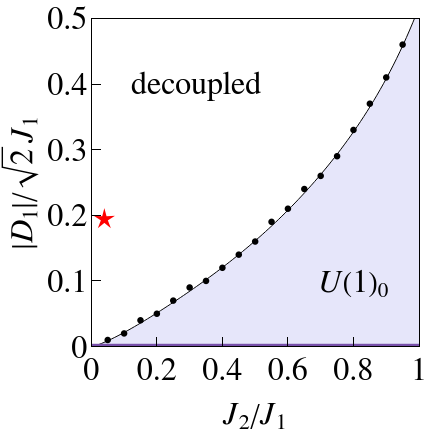}}
\subfloat[]{\label{figure:bandzerofluxJ2080D1020}
\includegraphics[scale=0.29]{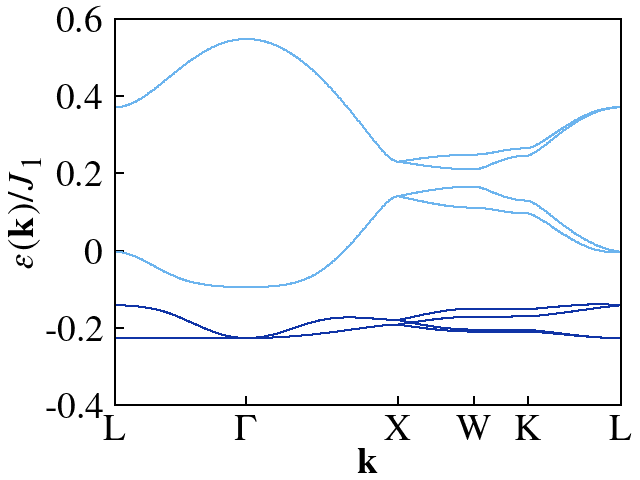}}
\caption{(a) The stability of the $U(1)_0$ and $U(1)_\pi$ spin liquids in the presence of Dzyaloshinskii-Moriya interaction, with $J_1>0$ and $D_1 < 0$. At $D_1=0$, these two spin liquids are degenerate, as indicated by the dark colored thick line. When $D_1$ becomes finite, $U(1)_0$ is preferred over $U(1)_\pi$, as indicated by the light colored area. The blank region labeled by ``decoupled'' is where the mean field theory favors a solution of isolated tetrahedra, instead of a spatially connected $U(1)$ spin liquid. The red star indicates the parameters of the candidate spin model for BYZO~\cite{PhysRevLett.116.257204,PhysRevB.93.220407}, where $\lvert D_1 \rvert / J_1 \approx 0.2 \times \sqrt{2} = 0.28$ and $J_2/J_1 \approx 0$. (b) The spinon dispersion of the $U(1)_0$ spin liquid at $J_2/J_1=0.8$ and $\lvert D_1 \lvert / \sqrt{2} J_1 = 0.2$  (c.f.~Fig.~\ref{figure:bandzerofluxJ2080}). Filled (empty) bands are indicated by darker (lighter) blue lines.}
\end{figure}

For either ansatz, the mean field theory of the model~\eqref{JDmodel} contains only two additional variational parameters, $E_1^y$ and $E_2^y$, on top of $\chi_1$ and $\chi_2$ that are already present in the Heisenberg limit (see App.~\ref{appendix:dzyaloshinskiimoriya}). The theory is no longer analytically tractable, so we solve it computationally by iterating the self consistency equations, e.g., $\chi_{ij} = \sum_\sigma \langle f_{i \sigma}^\dagger f_{j \sigma} \rangle$ (which is done in momentum space, see Ref.~\onlinecite{PhysRevB.86.224417} for example), until the mean field parameters converge. We study the parameter region defined by $J_2/J_1 \in (0,1)$ and $D_1 / \sqrt{2} J_1 \in [-0.5,0]$.

Our result, which is plotted in Fig.~\ref{figure:phase}, has two main features. First, in the presence of a sufficiently small Dzyaloshinskii-Moriya interaction, the degeneracy between the $U(1)_0$ and $U(1)_\pi$ spin liquids is lifted, such that the $U(1)_0$ spin liquid is lower in energy. The Dzyaloshinskii-Moriya interaction also gaps out the spinon spectrum of the $U(1)_0$ spin liquid. In particular, the highest occupied band and the lowest unoccupied band no longer touch at the $\Gamma$ point, see Fig.~\ref{figure:bandzerofluxJ2080D1020}. Second, as the Dzyaloshinskii-Moriya interaction increases in magnitude, there exists a critical value $\lvert D_1 \rvert_\mathrm{c}$ above which the mean field theory favors a solution of decoupled tetrahedra, such that the mean field parameters $\chi_2$ and $E_2^y$ ($\chi_1$ and $E_1^y$) associated with the down (up) tetrahedra converge to zero (finite values). Since the spatial connectivity is lost, we do not call the resulting state a $U(1)$ spin liquid. Moreover, $\lvert D_1 \rvert_\mathrm{c}$ increases as $J_2/J_1$ increases. It seems reasonable to speculate $\lvert D_1 \rvert_\mathrm{c} / J_1 \longrightarrow \infty$ as $J_2 / J_1 \longrightarrow 1$, i.e., we expect that a spatially connected solution is always favorable in the isotropic limit.

\section{\label{section:discussion}Discussion}

In summary, we have investigated the possible quantum spin liquids in the $S=1/2$ breathing pyrochlore magnet using the complex fermion mean field theory and the projective symmetry group (PSG) analysis. We identify 40 $\mathbb{Z}_2$ spin liquids and 16 $U(1)$ spin liquids that are subjected to the $F\bar{4}3m$ space group of the breathing pyrochlore lattice as well as the time reversal symmetry. As an application, we consider the antiferromagnetic (AFM) Heisenberg model, and identify the physical $U(1)$ spin liquid ansatze. Most of the 16 $U(1)$ states are constrained by PSG to have vanishing bond parameters, leaving only two cases, which we label $U(1)_0$ and $U(1)_\pi$. Their corresponding mean field theories admit analytical solutions, as shown in the text. We find that the $U(1)_0$ state has gapless spinon excitations, while the $U(1)_\pi$ state exhibits a spinon gap; these two states are however degenerate regardless of the ratio $J_2/J_1$ between the interactions on the large and small tetrahedra. While the spinon contribution to the heat capacity of the $U(1)_\pi$ state is exponentially suppressed at low temperatures in the presence of an excitation gap, the quadratically dispersing spinons in the $U(1)_0$ state give rise to a heat capacity contribution $C(T) \sim T^{3/2}$ at low temperatures. Using an effective field theory, we demonstrate that such power law dependence is unchanged by small momentum gauge fluctuations within the random phase approximation.

The degeneracy between the $U(1)_0$ and $U(1)_\pi$ states is lifted by a finite Dzyaloshinskii-Moriya interaction, which favors the $U(1)_0$ state and gaps out its spectrum. It is also found that when the DM interaction becomes sufficiently large, neither $U(1)$ spin liquids survive as the bond parameters on the large tetrahedra tend to zero upon solving the self consistent mean field equations iteratively. The parameter region that favors such a decoupled tetrahedron solution contains the candidate spin model~\cite{PhysRevLett.116.257204,PhysRevB.93.220407} for the material Ba$_3$Yb$_2$Zn$_5$O$_{11}$ (BYZO), whose interactions on the large tetrahedra appear to be orders of magnitude smaller than those on the small tetrahedra. The $U(1)_0$ and $U(1)_\pi$ spin liquids are thus more relevant for spin models closer to the Heisenberg limit or with a weaker breathing anisotropy. While BYZO is the only $S=1/2$ breathing pyrochlore material discovered so far, similar materials with parameters that span the phase diagram in our work may be synthetized in near future.

The AFM Heisenberg model on the breathing pyrochlore lattice had been investigated in several prior works, first as a route to understand the same model but on the regular pyrochlore lattice, and later for its own sake when breathing pyrochlore materials made their appearance in experimental laboratories. 
It is therefore worth briefly discussing our work in relation to the existing literature. 

Starting from decoupled tetrahedra with one $S=1/2$ moment per vertex, Refs.~\onlinecite{JPSJ.70.640,PhysRevB.65.024415} treat the intertetrahedron interaction perturbatively up to third order, and arrive at an effective Hamiltonian with one pseudospin-$1/2$ degree of freedom per tetrahedron and interactions that involve three nearby pseudospins. A mean field approximation, which essentially treats the pseudospins as classical vectors, yields a partially dimerized ground state, leaving one disordered pseudospin for every four pseudospins. This is different from our approach, which is nonperturbative, and our spin liquid ansatze preserves the lattice symmetry, while the dimerized state breaks it. Such a perturbative reconnection method first appeared in Ref.~\onlinecite{PhysRevLett.80.2933} (see also Ref.~\onlinecite{PhysRevB.61.1149}) for $S=1/2$, and it was also extended to $S=1$ and $S=3/2$ in Ref.~\onlinecite{ptx023}. 

Ref.~\onlinecite{PhysRevB.94.075146} uses gauge mean field theory~\cite{PhysRevLett.108.037202,PhysRevB.86.104412} to study the $S=1/2$ XXZ model on the breathing pyrochlore lattice. While gauge mean field theory provides an excellent description of the quantum spin ice state near the Ising limit, it is unclear how accurately the theory captures other types of quantum spin liquid in a more generic parameter region. For example, it predicts an antiferromagnetically ordered state in the AFM Heisenberg limit, where the strong frustration warrants the consideration of possible quantum spin liquid candidates, which can be systematically classified by PSG.

Ref.~\onlinecite{JPSJ.84.104710} studies the Heisenberg model on the breathing pyrochlore lattice in the classical limit, where the signs of $J_1$ and $J_2$ are allowed to be different. Interestingly, the classical ground state depends only on the signs of the interactions but not on their relative magnitude. In particular, for $J_1>0$ and $J_2>0$, the ground state is a Coulombic spin liquid~\cite{annurev-conmatphys-070909-104138} subject to the divergenceless condition that the vector summation of the four spins on every tetrahedron is zero, which is the continuous version of the 2-in-2-out ice rule. This gives us reason to look for quantum spin liquids in the corresponding $S=1/2$ model.

More interestingly, a recent $SU(2)$ density matrix renormalization group study \cite{PhysRevLett.126.117204} of the $S=1/2$ AFM Heisenberg model on the regular pyrochlore lattice suggests a disordered ground state with spontaneously broken inversion symmetry, as evidenced by a difference in energy between the two species of tetrahedra. Assuming that none of the remaining symmetries is broken, the quantum spin liquids classified in this work may even be considered as candidate ground states of the regular pyrochlore lattice.

We close by mentioning several potentially interesting future directions. While we have focused on the $U(1)$ spin liquids in this work, it will also be useful to conduct a thorough examination of the $\mathbb{Z}_2$ spin liquids, though they are greater in number. Besides, we have only considered here time reversal symmetric spin liquid ansatze. It may be worthwhile to study ansatze that break time reversal symmetry, like the monopole flux state in Ref.~\onlinecite{PhysRevB.79.144432}, as they may give rise to stable chiral spin liquids. Moreover, it is known that mean field theory does not give highly accurate energetics, and advanced numerical techniques, specifically variational Monte Carlo with Gutzwiller projection, may be employed to obtain more reliable estimates of the energies of the various spin liquids, magnetic orders, and other candidate ground states. Our PSG analysis lay the foundation on which the variational wave functions of the quantum spin liquids can be constructed for such comparisons. 

\begin{acknowledgements}
This work was supported in part by the Engineering and Physical Sciences Research Council grants No.~EP/P034616/1, No.~EP/T028580/1, and No.~EP/V062654/1 (CC and LEC). YBK was supported by the NSERC of Canada, the Center for Quantum Materials at the University of Toronto, the Simons Fellowship from the Simons Foundation, and the Guggenheim Fellowship from the John Simon Guggenheim Memorial Foundation.
\end{acknowledgements}

\appendix

\section{\label{appendix:algebraicidentity}Algebraic identities}

In this appendix, we list the algebraic identities of the breathing pyrochlore lattice: 
\begingroup
\allowdisplaybreaks
\begin{subequations}
\begin{align}
T_2^{-1} T_1^{-1} T_2 T_1 &= e , \label{T2T1commute} \\
T_3^{-1} T_2^{-1} T_3 T_2 &= e , \label{T3T2commute} \\
T_1^{-1} T_3^{-1} T_1 T_3 &= e , \label{T1T3commute} \\
C_3^{-1} T_2^{-1} C_3 T_1 &= e , \label{C3T1commute} \\
C_3^{-1} T_3^{-1} C_3 T_2 &= e , \label{C3T2commute} \\
C_3^{-1} T_1^{-1} C_3 T_3 &= e , \label{C3T3commute} \\
\sigma^{-1} T_1 \sigma T_1 &= e , \label{sigmaT1commute} \\
\sigma^{-1} T_2^{-1} T_1 \sigma T_2 &= e , \label{sigmaT2commute} \\
\sigma^{-1} T_3^{-1} T_1 \sigma T_3 &= e , \label{sigmaT3commute} \\
C_3^3 &= e , \label{C3cube} \\
\sigma^2 &= e , \label{sigmasquare} \\
\left( \sigma C_3 \right)^4 &= e , \label{sigmaC3commute} \\
\mathcal{O}^{-1} \mathcal{T}^{-1} \mathcal{O} \mathcal{T} &= e , \label{spacetimecommute} \\
\mathcal{T}^2 &= e , \label{timesquare}
\end{align}
\end{subequations}
\endgroup
where $\mathcal{O}$ in~\eqref{spacetimecommute} denotes any space group generator, and $e$ is the identity element.

\section{\label{appendix:meanfieldtheory}Complex fermion mean field theory}

Using the parton representation~\eqref{partonrepresentation}, the antiferromagnetic Heisenberg model~\eqref{heisenbergmodel} can be written as, up to some constant,
\begin{equation}
H = - \sum_{ij} \frac{J_{ij}}{4} \left( \hat{\chi}_{ij}^\dagger \hat{\chi}_{ij} + \hat{\Delta}_{ij}^\dagger \hat{\Delta}_{ij} \right) \, ,
\end{equation}
where
\begin{subequations}
\begin{align}
\hat{\chi}_{ij} &= f_{i \uparrow}^\dagger f_{j \uparrow} + f_{i \downarrow}^\dagger f_{j \downarrow} 
\, , 
\label{singlethopping} \\
\hat{\Delta}_{ij} &= f_{i \uparrow} f_{j \downarrow} - f_{i \downarrow} f_{j \uparrow} 
\, , 
\label{singletpairing}
\end{align}
\end{subequations}
are the singlet hopping and pairing operators respectively. A mean field decoupling then leads to~\eqref{meanfieldhamiltonian}. 

Ref.~\onlinecite{PhysRevB.38.745} first noted an alternative form of \eqref{partonrepresentation},
\begin{equation} \label{spinor}
\mathbf{S}_i = \frac{1}{4} \mathrm{Tr} \left[ \Psi_i^\dagger \vec{\sigma} \Psi_i \right], \Psi_i = \begin{pmatrix} f_{i \uparrow} & f_{i \downarrow}^\dagger \\ f_{i \downarrow} & - f_{i \uparrow}^\dagger \end{pmatrix} ,
\end{equation}
from which it is easy to see that an $SU(2)$ gauge transformation $G_i$, such that $\Psi_i \longrightarrow \Psi_i G_i$, leaves the spin operator invariant. Note that $G_i$ preserves the anticommutation relation of fermions. On the other hand, the mean field Hamiltonian~\eqref{meanfieldhamiltonian} can be written as
\begin{equation} \label{singlettrace}
H^\mathrm{MF} = \sum_{ij} \mathrm{Tr} \left[ \Psi_i u_{ij} \Psi_j^\dagger \right] ,
\end{equation}
where $u_{ij}$ is defined in~\eqref{meanfieldansatz} and the on-site $u_{ii}$ is given by
\begin{equation} \label{onsiteansatz}
u_{ii} = -\frac{1}{2} \begin{pmatrix} \lambda_i^{(3)} & \lambda_i^{(1)}+i \lambda_i^{(2)} \\ \lambda_i^{(1)} - i \lambda_i^{(2)} & - \lambda_i^{(3)} \end{pmatrix} 
\, .
\end{equation}
As stated in the main text, due to the $SU(2)$ gauge redundancy, the mean field Hamiltonian is invariant under a symmetry $X$ of the system only up to a gauge transformation $G_X \in SU(2)$. From Ref.~\onlinecite{PhysRevB.95.054404}, 
\begin{equation} \label{compoundoperator}
\begin{aligned}[b]
H^\mathrm{MF} & \overset{X}{\longrightarrow} \sum_{ij} \Psi_{X(i)} u_{ij} \Psi_{X(j)}^\dagger \\
& \overset{G_X}{\longrightarrow} \sum_{ij} \Psi_{X(i)}G_X(X(i)) u_{ij} G_X^\dagger (X(j))\Psi_{X(j)}^\dagger
\, , 
\end{aligned}
\end{equation}
we see that the mean field ansatz should obey~\eqref{ansatzspacerule}.

\section{\label{appendix:classifyZ2spinliquid}Classification of $\mathbb{Z}_2$ spin liquids}

We solve
\begingroup
\allowdisplaybreaks
\begin{subequations}
\begin{align}
& G_{T_2}^\dagger (T_1^{-1} (i)) G_{T_1}^\dagger (i) G_{T_2}(i) G_{T_1} (T_2^{-1} (i)) = \eta_{T_2 T_1} , \label{Z2T2T1commute} \\
& G_{T_3}^\dagger (T_2^{-1} (i)) G_{T_2}^\dagger (i) G_{T_3}(i) G_{T_2} (T_3^{-1} (i)) = \eta_{T_3 T_2} , \label{Z2T3T2commute} \\
& G_{T_1}^\dagger (T_3^{-1} (i)) G_{T_3}^\dagger (i) G_{T_1}(i) G_{T_3} (T_1^{-1} (i)) = \eta_{T_1 T_3} , \label{Z2T1T3commute} \\
& G_{C_3}^\dagger (T_2^{-1} (i)) G_{T_2}^\dagger (i) G_{C_3}(i) G_{T_1} (C_3^{-1} (i)) = \eta_{C_3 T_1} , \label{Z2C3T1commute} \\
& G_{C_3}^\dagger (T_3^{-1} (i)) G_{T_3}^\dagger (i) G_{C_3}(i) G_{T_2} (C_3^{-1} (i)) = \eta_{C_3 T_2} , \label{Z2C3T2commute} \\
& G_{C_3}^\dagger (T_1^{-1} (i)) G_{T_1}^\dagger (i) G_{C_3}(i) G_{T_3} (C_3^{-1} (i)) = \eta_{C_3 T_3} , \label{Z2C3T3commute} \\
& G_\sigma^\dagger (i) G_{T_1} (i) G_\sigma (T_1^{-1}(i)) G_{T_1} (\sigma^{-1} T_1^{-1} (i)) = \eta_{\sigma T_1} , \label{Z2sigmaT1commute} \\
\begin{split} \label{Z2sigmaT2commute}
& G_\sigma^\dagger (T_2^{-1}(i)) G_{T_2}^\dagger (i) G_{T_1} (i) \\ & \quad G_\sigma (T_1^{-1}(i)) G_{T_2} (\sigma^{-1} T_1^{-1} (i)) = \eta_{\sigma T_2} ,
\end{split} \\
\begin{split} \label{Z2sigmaT3commute}
& G_\sigma^\dagger (T_3^{-1}(i)) G_{T_3}^\dagger (i) G_{T_1} (i) \\ & \quad G_\sigma (T_1^{-1}(i)) G_{T_3} (\sigma^{-1} T_1^{-1} (i)) = \eta_{\sigma T_3} ,
\end{split} \\
& G_{C_3} (C_3 (i)) G_{C_3} (i) G_{C_3} (C_3^{-1} (i)) = \eta_{C_3} , \label{Z2C3cube} \\
& G_\sigma (\sigma(i)) G_\sigma (i) = \eta_\sigma , \label{Z2sigmasquare} \\
\begin{split} \label{Z2sigmaC3commute}
& G_\sigma ((\sigma C_3)^3 (i)) G_{C_3} (C_3 (\sigma C_3)^2 (i)) \\
& \quad G_\sigma ((\sigma C_3)^2 (i)) G_{C_3} (C_3 \sigma C_3 (i)) G_\sigma (\sigma C_3 (i)) \\
& \qquad G_{C_3} (C_3 (i)) G_\sigma (i) G_{C_3} (\sigma^{-1} (i)) = \eta_{\sigma C_3} ,
\end{split} \\
& G_{\mathcal{O}}^\dagger (i) G_\mathcal{T}^\dagger (i) G_\mathcal{O} (i) G_\mathcal{T} (\mathcal{O}^{-1} (i)) = \eta_\mathcal{OT} , \label{Z2spacetimecommute} \\
& [G_\mathcal{T} (i) ]^2 = \eta_\mathcal{T} , \label{Z2timesquare}
\end{align}
\end{subequations}
\endgroup
for $G_X (r_1,r_2,r_3;s)$, where each $\eta_{\ldots}$ on the right hand side is either $+1$ or $-1$. 

First, notice that for a given $X$ we are free to multiply $G_X$ by an element of IGG. We exploit this IGG freedom of $G_{T_1}$ to fix $\eta_{\sigma T_3}=1$, of $G_{T_2}$ to fix $\eta_{C_3 T_2}=1$, of $G_{T_3}$ to fix $\eta_{C_3 T_3}=1$, and of $G_{C_3}$ to fix $\eta_{C_3}=1$. Then, using
\begin{equation} \label{gaugetransformgauge}
G_X (i) \longrightarrow W(i) G_X(i) W^\dagger(X^{-1}(i))
\, , \;\;\;\; W(i) \in SU(2),
\end{equation}
we fix 
\begin{subequations}
\begin{align*}
& G_{T_1} (r_1,r_2,r_3;s) = 1, \\
& G_{T_2} (0,r_2,r_3;s) = 1, \\
& G_{T_3} (0,0,r_3;s) = 1 \, .
\end{align*}
\end{subequations}

Then, \eqref{Z2T2T1commute} yields $G_{T_2}^\dagger (r_1-1,r_2,r_3;s) G_{T_2} (r_1,r_2,r_3;s) = \eta_{T_2 T_1}$, or
\begin{equation}
G_{T_2} (r_1,r_2,r_3;s) = \eta_{T_2 T_1}^{r_1} \, .
\end{equation}
Eqs.~\eqref{Z2T3T2commute} and~\eqref{Z2T1T3commute} yield $G_{T_3} (r_1,r_2,r_3;s) = \eta_{T_3 T_2}^{r_2} G_{T_3} (r_1,0,r_3;s)$ and $G_{T_3} (r_1,r_2,r_3;s) = \eta_{T_1 T_3}^{r_1} G_{T_3} (0,r_2,r_3;s)$ respectively, so
\begin{equation}
G_{T_3} (r_1,r_2,r_3;s) = \eta_{T_1 T_3}^{r_1} \eta_{T_3 T_2}^{r_2} \, .
\end{equation}
Eqs.~\eqref{Z2C3T1commute},~\eqref{Z2C3T2commute}, and~\eqref{Z2C3T3commute} yield
\begin{subequations}
\begin{align}
& G_{C_3} (r_1,r_2,r_3;s) = \eta_{C_3 T_1}^{r_2} \eta_{T_2 T_1}^{r_1 r_2} G_{C_3} (r_1,0,r_3;s) , \\
& G_{C_3} (r_1,r_2,r_3;s) = \eta_{T_1 T_3}^{r_1 r_3} \eta_{T_3 T_2}^{r_2 r_3} \eta_{T_2 T_1}^{r_2 r_3} G_{C_3} (r_1,r_2,0;s) , \\
& G_{C_3} (r_1,r_2,r_3;s) = \eta_{T_1 T_3}^{r_2 r_1} \eta_{T_3 T_2}^{r_3 r_1} G_{C_3} (0,r_2,r_3;s) ,
\end{align}
\end{subequations}
which further lead to
\begingroup
\allowdisplaybreaks
\begin{subequations}
\begin{align}
& G_{C_3} (r_1,r_2,r_3;s) = \eta_{C_3 T_1}^{r_2} \eta_{T_2 T_1}^{r_1 r_2} \eta_{T_1 T_3}^{r_1 r_3} g_{C_3} (s) , \\
& G_{C_3} (r_1,r_2,r_3;s) = \eta_{T_1 T_3}^{r_2 r_1} \eta_{T_3 T_2}^{r_3 (r_1+r_2)} \eta_{T_2 T_1}^{r_2 r_3} \eta_{C_3 T_2}^{r_2} g_{C_3} (s) , \\
& G_{C_3} (r_1,r_2,r_3;s) = \eta_{T_1 T_3}^{r_2 r_1} \eta_{T_3 T_2}^{r_3 r_1} \eta_{C_3 T_1}^{r_2} g_{C_3} (s) , \\
& G_{C_3} (r_1,r_2,r_3;s) = \eta_{C_3 T_1}^{r_2} \eta_{T_2 T_1}^{r_1 r_2} \eta_{T_3 T_2}^{r_3 r_1} g_{C_3} (s) , \\
& G_{C_3} (r_1,r_2,r_3;s) = \eta_{T_1 T_3}^{r_1 (r_2+r_3)} \eta_{T_3 T_2}^{r_2 r_3} \eta_{T_2 T_1}^{r_2 r_3} \eta_{C_3 T_1}^{r_2} g_{C_3} (s) , \\
& G_{C_3} (r_1,r_2,r_3;s) = \eta_{T_1 T_3}^{r_1 r_3} \eta_{T_3 T_2}^{r_2 r_3} \eta_{T_2 T_1}^{r_2 (r_3+r_1)} \eta_{C_3 T_1}^{r_2} g_{C_3} (s) ,
\end{align}
\end{subequations}
\endgroup
where $g_X (s) \equiv G_X (0,0,0;s)$. The right hand sides of these six equations must be equal to each other, which forces $\eta_{T_1 T_3} = \eta_{T_3 T_2} = \eta_{T_2 T_1}$. It follows that
\begin{equation}
G_{C_3} (r_1,r_2,r_3;s) = \eta_{C_3 T_1}^{r_2} \eta_{T_2 T_1}^{r_1 (r_2+r_3)} g_{C_3} (s) \, .
\end{equation}
Eq.~\eqref{Z2C3cube}, with $i=(r_1,r_2,r_3;0)$, yields
\begin{equation}
\eta_{C_3 T_1}^{r_3+r_1+r_2} [g_{C_3} (0)]^3 = 1 \, .
\end{equation}
Since the right hand side has no coordinate dependence, we must have $\eta_{C_3 T_1}=1$, which is anticipated since~\eqref{C3T1commute} is implied by~\eqref{C3T2commute} and~\eqref{C3T3commute}.
On the other hand, with $i=(r_1,r_2,r_3;s=1,2,3)$, we get
\begin{equation}
g_{C_3} (3) g_{C_3} (2) g_{C_3} (1) = 1 \, .
\end{equation}
Eq.~\eqref{Z2sigmaT1commute} yields
\begin{equation}
G_\sigma (r_1,r_2,r_3;s) = \eta_{\sigma T_1}^{r_1} G_\sigma (0,r_2,r_3;s) .
\end{equation}
Subsequently,~\eqref{Z2sigmaT2commute} and~\eqref{Z2sigmaT3commute} yield
\begin{subequations}
\begin{align*}
& G_\sigma (0,r_2,r_3;s) = \eta_{\sigma T_2}^{r_2} \eta_{\sigma T_1}^{r_2} \eta_{T_2 T_1}^{r_2(r_2+1)/2+r_2(r_3+1)} G_\sigma (0,0,r_3;s) , \\
& G_\sigma (0,r_2,r_3;s) = \eta_{\sigma T_1}^{r_3} \eta_{T_2 T_1}^{r_3(r_3+1)/2+r_3(r_2+1)} G_\sigma (0,r_2,0;s) ,
\end{align*}
\end{subequations}
which together imply
\begin{equation*}
G_\sigma (r_1,r_2,r_3;s) = \eta_{\sigma T_1}^{r_1+r_2+r_3} \eta_{\sigma T_2}^{r_2} \eta_{T_2 T_2}^{r_2(r_2-1)/2 + r_3(r_3-1)/2 + r_2 r_3} g_\sigma(s) \, .
\end{equation*}
Eq.~\eqref{Z2sigmasquare}, with $i=(r_1,r_2,r_3;0)$ yields
\begin{equation}
\eta_{\sigma T_1}^{r_2+r_3} g_\sigma (1) g_\sigma (0) = \eta_\sigma ,
\end{equation}
which implies $\eta_{\sigma T_1}=0$. We also have
\begin{equation}
[g_\sigma (2)]^2=\eta_\sigma, [g_\sigma (3)]^2=\eta_\sigma,
\end{equation}
with $i=(r_1,r_2,r_3;s=2,3)$. Finally,~\eqref{Z2sigmaC3commute} yields
\begin{equation*}
g_\sigma(0) g_{C_3}(1) g_\sigma(3) g_{C_3}(3) g_\sigma(2) g_{C_3}(2) g_\sigma(1) g_{C_3}(0) = \eta_{\sigma C_3} .
\end{equation*}

We now proceed to the parts that involve the time reversal symmetry.  Eq.~\eqref{Z2spacetimecommute} with $\mathcal{O}=T_1,T_2,T_3$ yields
\begin{subequations}
\begin{align}
& G_\mathcal{T} (r_1,r_2,r_3;s) = \eta_{T_1 \mathcal{T}}^{r_1} G_\mathcal{T} (0,r_2,r_3;s) , \\
& G_\mathcal{T} (r_1,r_2,r_3;s) = \eta_{T_2 \mathcal{T}}^{r_2} G_\mathcal{T} (r_1,0,r_3;s) , \\
& G_\mathcal{T} (r_1,r_2,r_3;s) = \eta_{T_3 \mathcal{T}}^{r_3} G_\mathcal{T} (r_1,r_2,0;s) ,
\end{align}
\end{subequations}
which together imply
\begin{equation}
G_\mathcal{T} (r_1,r_2,r_3;s) = \eta_{T_1 \mathcal{T}}^{r_1} \eta_{T_2 \mathcal{T}}^{r_2} \eta_{T_3 \mathcal{T}}^{r_3} g_\mathcal{T} (s) .
\end{equation}
Eq.~\eqref{Z2spacetimecommute} with $\mathcal{O}=C_3$ and $i=(r_1,r_2,r_3;0)$ yields
\begin{equation}
\eta_{T_1 \mathcal{T}}^{r_1+r_2} \eta_{T_2 \mathcal{T}}^{r_2+r_3} \eta_{T_3 \mathcal{T}}^{r_3+r_1} g_{C_3}^\dagger (0) g_\mathcal{T}^\dagger (0) g_{C_3} (0) g_\mathcal{T} (0) = \eta_{C_3 \mathcal{T}} ,
\end{equation}
which implies $\eta_{T_1 \mathcal{T}}=\eta_{T_2 \mathcal{T}}=\eta_{T_3 \mathcal{T}}$. We also have
\begin{subequations}
\begin{align}
g_{C_3}^\dagger (2) g_\mathcal{T}^\dagger (2) g_{C_3} (2) g_\mathcal{T} (1) = \eta_{C_3 \mathcal{T}}, \\
g_{C_3}^\dagger (3) g_\mathcal{T}^\dagger (3) g_{C_3} (3) g_\mathcal{T} (2) = \eta_{C_3 \mathcal{T}}, \\
g_{C_3}^\dagger (1) g_\mathcal{T}^\dagger (1) g_{C_3} (1) g_\mathcal{T} (3) = \eta_{C_3 \mathcal{T}},
\end{align}
\end{subequations}
with $i=(r_1,r_2,r_3;s=1,2,3)$.
Eq.~\eqref{Z2spacetimecommute} with $\mathcal{O}=\sigma$ and $i=(r_1,r_2,r_3;0)$ yields
\begin{equation}
\eta_{T_1 \mathcal{T}}^{r_2+r_3} g_\sigma^\dagger (0) g_\mathcal{T}^\dagger (0) g_\sigma (0) g_\mathcal{T} (1) = \eta_{\sigma \mathcal{T}},
\end{equation}
which implies $\eta_{T_1 \mathcal{T}}=1$. We also have
\begin{subequations}
\begin{align}
& g_\sigma^\dagger (1) g_\mathcal{T}^\dagger (1) g_\sigma (1) g_\mathcal{T} (0) = \eta_{\sigma \mathcal{T}}, \\
& g_\sigma^\dagger (2) g_\mathcal{T}^\dagger (2) g_\sigma (2) g_\mathcal{T} (2) = \eta_{\sigma \mathcal{T}}, \\
& g_\sigma^\dagger (3) g_\mathcal{T}^\dagger (3) g_\sigma (3) g_\mathcal{T} (3) = \eta_{\sigma \mathcal{T}},
\end{align}
\end{subequations}
with $i=(r_1,r_2,r_3;s=1,2,3)$.
Eq.~\eqref{Z2timesquare} yields
\begin{equation}
[g_\mathcal{T} (s)]^2 = \eta_\mathcal{T} \, .
\end{equation}

\subsubsection{Grand summary}

Before we proceed to gauge fixing, it is worthwhile to recollect the results obtained thus far. They are~\eqref{Z2cellT1gauge}-\eqref{Z2celltimegauge} shown in the main text, together with the ``sublattice constraints'': 
\begingroup
\allowdisplaybreaks
\begin{subequations}
\begin{align}
& [g_{C_3}(0)]^3 = 1, \label{Z2sublatticeC30cube} \\
& g_{C_3} (3) g_{C_3} (2) g_{C_3} (1) = 1, \label{Z2sublatticeC3123} \\
& g_\sigma (1) g_\sigma (0) = \eta_\sigma, \label{Z2sublatticesigma01} \\
& [g_\sigma (2)]^2 = \eta_\sigma, \label{Z2sublatticesigma2square} \\
& [g_\sigma (3)]^2 = \eta_\sigma, \label{Z2sublatticesigma3square} \\
\begin{split} \label{Z2sublatticesigmaC3}
& g_\sigma(0) g_{C_3}(1) g_\sigma(3) g_{C_3}(3) \\ 
& \quad g_\sigma(2) g_{C_3}(2) g_\sigma(1) g_{C_3}(0) = \eta_{\sigma C_3} ,
\end{split} \\
& g_{C_3}^\dagger (0) g_\mathcal{T}^\dagger (0) g_{C_3} (0) g_\mathcal{T} (0) = \eta_{C_3 \mathcal{T}}, \label{Z2sublatticeC3time0} \\
& g_{C_3}^\dagger (2) g_\mathcal{T}^\dagger (2) g_{C_3} (2) g_\mathcal{T} (1) = \eta_{C_3 \mathcal{T}}, \label{Z2sublatticeC3time1} \\
& g_{C_3}^\dagger (3) g_\mathcal{T}^\dagger (3) g_{C_3} (3) g_\mathcal{T} (2) = \eta_{C_3 \mathcal{T}}, \label{Z2sublatticeC3time2} \\
& g_{C_3}^\dagger (1) g_\mathcal{T}^\dagger (1) g_{C_3} (1) g_\mathcal{T} (3) = \eta_{C_3 \mathcal{T}}, \label{Z2sublatticeC3time3} \\
& g_\sigma^\dagger (0) g_\mathcal{T}^\dagger (0) g_\sigma (0) g_\mathcal{T} (1) = \eta_{\sigma \mathcal{T}} , \label{Z2sublatticesigmatime1} \\
& g_\sigma^\dagger (1) g_\mathcal{T}^\dagger (1) g_\sigma (1) g_\mathcal{T} (0) = \eta_{\sigma \mathcal{T}}, \label{Z2sublatticesigmatime0} \\
& g_\sigma^\dagger (2) g_\mathcal{T}^\dagger (2) g_\sigma (2) g_\mathcal{T} (2) = \eta_{\sigma \mathcal{T}}, \label{Z2sublatticesigmatime2} \\
& g_\sigma^\dagger (3) g_\mathcal{T}^\dagger (3) g_\sigma (3) g_\mathcal{T} (3) = \eta_{\sigma \mathcal{T}}, \label{Z2sublatticesigmatime3} \\
& [g_\mathcal{T} (s)]^2 = \eta_\mathcal{T} 
\, . 
\label{Z2sublatticetimesquare}
\end{align}
\end{subequations}
\endgroup

\subsubsection{Gauge fixing}

We now gauge fix $g_X (s) \equiv G_X (0,0,0;s), s=0,1,2,3$, subject to the constraints~\eqref{Z2sublatticeC30cube}-\eqref{Z2sublatticetimesquare}. By virtue of~\eqref{gaugetransformgauge} and~\eqref{Z2sublatticeC3123}, we perform a sublattice dependent gauge transformation, $W_s$, such that
\begin{equation}
W_{0,3}=1 \, , \;\; W_1=g_{C_3}^\dagger (1) \, , \;\; W_2=g_{C_3}^\dagger (1) g_{C_3}^\dagger (2) \, ,
\end{equation}
to fix $g_{C_3}(1,2,3)=1$. Then,~\eqref{Z2sublatticeC3time1}-\eqref{Z2sublatticeC3time3} yield
\begin{equation}
\eta_{C_3 \mathcal{T}}^3 = [g_\mathcal{T}^\dagger (1) g_\mathcal{T} (3)] [g_\mathcal{T}^\dagger (3) g_\mathcal{T} (2)] [g_\mathcal{T}^\dagger (2) g_\mathcal{T} (1)]
\end{equation}
or $\eta_{C_3 \mathcal{T}}=1$. It also follows that $g_\mathcal{T} (1) = g_\mathcal{T} (2) = g_\mathcal{T} (3)$. 


\smallskip
\noindent
\textbf{Case 1.}~$\eta_\mathcal{T}=+1$.
Eq.~\eqref{Z2sublatticetimesquare} implies $g_\mathcal{T} (0)=\xi_0$, $g_\mathcal{T} (1,2,3) = \xi_1$, where $\xi_{0,1} \in \lbrace +1, -1 \rbrace$. Eq.~\eqref{Z2sublatticesigmatime2} or~\eqref{Z2sublatticesigmatime3} implies $\eta_{\sigma \mathcal{T}}=1$. Eq.~\eqref{Z2sublatticesigmatime1} or~\eqref{Z2sublatticesigmatime0} then implies $\xi_0=\xi_1$. We thus have $g_\mathcal{T} (s)=\xi_0$ for all $s$, which is further fixed to $1$ by the IGG freedom of $G_\mathcal{T}$. 

\smallskip
\noindent
\textbf{Case 1.1.}~$\eta_\sigma=+1$.
Eqs.~\eqref{Z2sublatticesigma2square} and \eqref{Z2sublatticesigma3square} imply $g_\sigma (2) = \xi_2$ and $g_\sigma (3) = \xi_3$ respectively, with $\xi_{2,3} \in \lbrace +1, -1 \rbrace$. Perform a sublattice gauge transformation $W_0 = \xi_2 g^\dagger_\sigma (0)$, $W_{1,2,3}=1$ to fix $g_\sigma (0)=\xi_2$, without affecting the previously fixed gauges. It follows from~\eqref{Z2sublatticesigma01} that $g_\sigma (1)=\xi_2$. We use the IGG freedom of $G_\sigma$ to fix $g_\sigma (0,1,2)=1$ and $g_\sigma (3)=\xi_2 \xi_3$. Since the product $\xi_2 \xi_3 = \pm 1$, let us just call it $\xi_3$ for simplicity. Eq.~\eqref{Z2sublatticesigmaC3} yields $g_{C_3} (0)=\xi_3 \eta_{\sigma C_3}$, which together with~\eqref{Z2sublatticeC30cube} imply $g_{C_3} (0)=1$ and $\xi_3=\eta_{\sigma C_3}$. 

\smallskip
\noindent
\textbf{Case 1.2.}~$\eta_\sigma=-1$. 
First, we perform a sublattice dependent gauge transformation $W_0=g_\sigma^\dagger (0)$, $W_{1,2,3}=1$ to fix $g_\sigma (0)=1$. It follows from~\eqref{Z2sublatticesigma01} that $g_\sigma (1)=-1$. Eqs.~\eqref{Z2sublatticesigma2square} and~\eqref{Z2sublatticesigma3square} imply $g_\sigma (2) = i \hat{\mathbf{n}}_2 \cdot \vec{\tau}$ and $g_\sigma (3) = i \hat{\mathbf{n}}_3 \cdot \vec{\tau}$, where $\hat{\mathbf{n}}_{2,3}$ are unit vectors and $\vec{\tau}=(\tau_1,\tau_2,\tau_3)$ is the vector of Pauli matrices. We can rotate $\hat{\mathbf{n}}_2$ to $(0,1,0)$, such that $g_\sigma (2) = i \tau_2$, by a global gauge transformation (think about the relation between $SU(2)$ transformations and $SO(3)$ rotations), without changing any previously fixed gauge which is proportional to identity. On the other hand, the most generic solution to \eqref{Z2sublatticeC30cube} is $g_{C_3} (0) = \exp [i (2 \pi q_0 / 3) \hat{\mathbf{n}}_0 \cdot \vec{\tau}]$, where $q_0 \in \lbrace 0,1,2 \rbrace$ and $\hat{\mathbf{n}}_0$ is a unit vector. Eq.~\eqref{Z2sublatticesigmaC3} leads to $g_{\sigma} (3) = - \eta_{\sigma C_3} g_{C_3}^\dagger (0) g_\sigma^\dagger (2)$. \eqref{Z2sublatticesigma3square} then gives $g_{C_3}^\dagger (0) \tau_2 = \tau_2 g_{C_3} (0)$, which implies that $g_{C_3} (0)$ cannot have a finite $\tau_2$ component. We further rotate $\hat{\mathbf{n}}_0$ to $(0,0,1)$ via a global gauge transformation of the form $e^{i \theta \tau_2}$, i.e.,~a uniform gauge rotation about the $\tau_2$ axis, which does not affect any of the previously fixed gauges. Thus $g_{C_3}(0) = \exp [i (2 \pi q_0/3) \tau_3]$ and $g_\sigma (3) = \eta_{\sigma C_3} (i \tau_2) \exp [i (2 \pi q_0/3) \tau_3]$. 

\smallskip
\noindent
\textbf{Case 2.}~$\eta_\mathcal{T}=-1$.
Eq.~\eqref{Z2sublatticetimesquare} implies $g_\mathcal{T} (0) = i \hat{\mathbf{n}}_0 \cdot \vec{\tau}$ and $g_\mathcal{T} (1,2,3) = i \hat{\mathbf{n}}_1 \cdot \vec{\tau}$, where $\hat{\mathbf{n}}_{0,1}$ are unit vectors and $\vec{\tau}=(\tau_1,\tau_2,\tau_3)$. We rotate $\hat{\mathbf{n}}_1$ to $(0,1,0)$, such that $g_\mathcal{T} (1,2,3) = i \tau_2$, by a global gauge transformation, without changing any previously fixed gauge which is proportional to identity. (Note: $\hat{\mathbf{n}}_0$ here is unrelated to that defined in 1.2. We merely recycle the notation. As this is quite clear from the context, we will not give such warnings should similar situations arise later.) 

\smallskip
\noindent
\textbf{Case 2.1.}~$\eta_\sigma=+1$.
Eqs.~\eqref{Z2sublatticesigma2square} and~\eqref{Z2sublatticesigma3square} imply $g_\sigma (2) = \xi_2$ and $g_\sigma (3) = \xi_3$ respectively, with $\xi_{2,3} \in \lbrace +1, -1 \rbrace$. We perform a sublattice gauge transformation $W_0 = \xi_2 g^\dagger_\sigma (0)$, $W_{1,2,3}=1$ to fix $g_\sigma (0)=\xi_2$, without affecting the previously fixed gauges. It follows from~\eqref{Z2sublatticesigma01} that $g_\sigma (1)=\xi_2$. We use the IGG freedom of $G_\sigma$ to fix $g_\sigma (0,1,2)=1$ and $g_\sigma (3)=\xi_2 \xi_3$. Since the product $\xi_2 \xi_3 = \pm 1$, let us just call it $\xi_3$ for simplicity. Eq.~\eqref{Z2sublatticesigmaC3} yields $g_{C_3} (0)=\xi_3 \eta_{\sigma C_3}$, which together with~\eqref{Z2sublatticeC30cube} implies $g_{C_3} (0)=1$ and $\xi_3=\eta_{\sigma C_3}$. (Note that the procedures outlined above are exactly same as those in Case 1.1.) Finally,~\eqref{Z2sublatticesigmatime2} or~\eqref{Z2sublatticesigmatime3} imply $\eta_{\sigma \mathcal{T}}=1$, which together with~\eqref{Z2sublatticesigmatime1} or~\eqref{Z2sublatticesigmatime0} imply $g_\mathcal{T}(0)=i \tau_2$. 

\smallskip
\noindent
\textbf{Case 2.2.}~$\eta_\sigma=-1$.
Eqs.~\eqref{Z2sublatticesigma2square} and~\eqref{Z2sublatticesigma3square} imply $g_\sigma (2) = i \hat{\mathbf{n}}_2 \cdot \vec{\tau}$ and $g_\sigma (3) = i \hat{\mathbf{n}}_3 \cdot \vec{\tau}$ respectively, where $\hat{\mathbf{n}}_{2,3}$ are unit vectors. 

\smallskip
\noindent
\textbf{Case 2.2.1.}~$\eta_{\sigma \mathcal{T}}=+1$.
From~\eqref{Z2sublatticesigmatime2}, we see that $g_\sigma (2)$ commutes with $\tau_2$, so $g_\sigma (2)=\xi_2 (i \tau_2)$, $\xi_2=\pm 1$. Similarly, from~\eqref{Z2sublatticesigmatime3} we have $g_\sigma (3) = \xi_3 (i \tau_2)$, $\xi_3 = \pm 1$. We then perform a sublattice dependent gauge transformation $W_0 = \xi_2 g_\sigma^\dagger (0)$, $W_{1,2,3}=1$ to fix $g_\sigma (0)=\xi_2$. It follows from~\eqref{Z2sublatticesigma01} that $g_\sigma (1)=-\xi_2$. Using the IGG freedom of $G_\sigma$, we can eliminate $\xi_2$ and redefine $\xi_2 \xi_3 \longrightarrow \xi_3$ as in Case 2.1. Eq.~\eqref{Z2sublatticesigmaC3} yields $g_{C_3} (0)=\xi_3 \eta_{\sigma C_3}$, which together with \eqref{Z2sublatticeC30cube} imply $g_{C_3} (0)=1$ and $\xi_3=\eta_{\sigma C_3}$. Finally,~\eqref{Z2sublatticesigmatime1} implies $g_\mathcal{T} (0) = i \tau_2$. To render the solutions in a neater form, we further perform a sublattice dependent gauge transformation $W_0 = i \tau_2$, $W_{1,2,3}=1$ such that $g_{\sigma} (0,1) \longrightarrow i \tau_2$, while others are unaffected. 

\smallskip
\noindent
\textbf{Case 2.2.2.}~$\eta_{\sigma \mathcal{T}}=-1$.
We first perform a sublattice dependent gauge transformation, $W_0 = g_\sigma^\dagger (0)$, $W_{1,2,3}=1$ to fix $g_\sigma (0)=1$. It follows from~\eqref{Z2sublatticesigma01} that $g_\sigma (1)=-1$. Eq.~\eqref{Z2sublatticesigmatime1} or~\eqref{Z2sublatticesigmatime0} then implies $g_\mathcal{T} (0)=-i \tau_2$. From~\eqref{Z2sublatticesigmatime2}, we see that $g_\sigma (2)$ anticommutes with $\tau_2$, which implies that $\hat{\mathbf{n}}_2$ cannot have a finite $\tau_2$ component, i.e., it has the form $(\sin \theta, 0, \cos \theta)$. We further rotate $\hat{\mathbf{n}}_2$ to (0,0,1) via a global gauge transformation $\exp (-i \theta \tau_2/2)$, which does not affect any of the previously fixed gauges. Thus $g_\sigma (2) = i \tau_3$. The most generic solution to~\eqref{Z2sublatticeC30cube} is $g_{C_3} (0) = \exp [i (2 \pi q_0 / 3) \hat{\mathbf{n}}_0 \cdot \vec{\tau}]$, where $q_0=0,1,2$ and $\hat{\mathbf{n}}_0$ is a unit vector. But~\eqref{Z2sublatticeC3time0} requires $g_{C_3} (0)$ to commute with $i \tau_2$, so $\hat{\mathbf{n}}_0=(0, \pm 1,0)$, and we can further specialize to the plus sign without loss of generality. Finally,~\eqref{Z2sublatticesigmaC3} yields $g_\sigma (3) = \eta_{\sigma C_3} (i \tau_3) \exp [i (2 \pi q_0 / 3) \tau_2]$, which anticommutes wth $\tau_2$ as required by~\eqref{Z2sublatticesigmatime3}. To render the solutions in a neater form, we further perform a sublattice dependent gauge transformation $W_0 = i \tau_3$, $W_{1,2,3}=1$, such that $g_{C_3}(0) \longrightarrow \exp [- i (2 \pi q_0 / 3) \tau_2]$, $g_\sigma (0,1) \longrightarrow i \tau_3$, and $g_\mathcal{T} (0) \longrightarrow i \tau_2$, while others are unaffected. 

\smallskip
We will exclude the solutions with $\eta_\mathcal{T}=+1$ (Case 1) because they have $G_\mathcal{T} (i)=1$ for all sites $i$, which forces $u_{ij}=0$ for any pair of sites $i$ and $j$ by~\eqref{ansatztimerule}. These solutions, which lead to vanishing mean field ansatze and thus a zero Hamiltonian, are unphysical. Let us count the remaining solutions (Case 2). Each of 2.1, 2.2.1, 2.2.2 has three $\mathbb{Z}_2$ variables $\eta_{T_2 T_1}$, $\eta_{\sigma T_2}$, and $\eta_{\sigma C_3}$. For 2.2.2, there is an additional $\mathbb{Z}_3$ variable $q_0$. There are in total $2^3 \times (1+1+3) =40$ gauge inequivalent solutions, i.e.,~40 possible $\mathbb{Z}_2$ spin liquids. They are listed in Table~\ref{table:spinliquid}. 

\section{\label{appendix:classifyU1spinliquid}Classification of $U(1)$ spin liquids}

We solve
\begingroup
\allowdisplaybreaks
\begin{subequations}
\begin{align}
& G_{T_2}^\dagger (T_1^{-1} (i)) G_{T_1}^\dagger (i) G_{T_2}(i) G_{T_1} (T_2^{-1} (i)) = e^{i \theta_{T_2 T_1} \tau_3} , \label{U1T2T1commute} \\
& G_{T_3}^\dagger (T_2^{-1} (i)) G_{T_2}^\dagger (i) G_{T_3}(i) G_{T_2} (T_3^{-1} (i)) = e^{i \theta_{T_3 T_2} \tau_3} , \label{U1T3T2commute} \\
& G_{T_1}^\dagger (T_3^{-1} (i)) G_{T_3}^\dagger (i) G_{T_1}(i) G_{T_3} (T_1^{-1} (i)) = e^{i \theta_{T_1 T_3} \tau_3} , \label{U1T1T3commute} \\
& G_{C_3}^\dagger (T_2^{-1} (i)) G_{T_2}^\dagger (i) G_{C_3}(i) G_{T_1} (C_3^{-1} (i)) = e^{i \theta_{C_3 T_1} \tau_3} , \label{U1C3T1commute} \\
& G_{C_3}^\dagger (T_3^{-1} (i)) G_{T_3}^\dagger (i) G_{C_3}(i) G_{T_2} (C_3^{-1} (i)) = e^{i \theta_{C_3 T_2} \tau_3} , \label{U1C3T2commute} \\
& G_{C_3}^\dagger (T_1^{-1} (i)) G_{T_1}^\dagger (i) G_{C_3}(i) G_{T_3} (C_3^{-1} (i)) = e^{i \theta_{C_3 T_3} \tau_3} , \label{U1C3T3commute} \\
\begin{split}
& G_\sigma^\dagger (i) G_{T_1} (i) \\
& \quad G_\sigma (T_1^{-1}(i)) G_{T_1} (\sigma^{-1} T_1^{-1} (i)) = e^{i \theta_{\sigma T_1} \tau_3} , \label{U1sigmaT1commute}
\end{split} \\
\begin{split} \label{U1sigmaT2commute}
& G_\sigma^\dagger (T_2^{-1}(i)) G_{T_2}^\dagger (i) G_{T_1} (i) \\ & \quad G_\sigma (T_1^{-1}(i)) G_{T_2} (\sigma^{-1} T_1^{-1} (i)) = e^{i \theta_{\sigma T_2} \tau_3} ,
\end{split} \\
\begin{split} \label{U1sigmaT3commute}
& G_\sigma^\dagger (T_3^{-1}(i)) G_{T_3}^\dagger (i) G_{T_1} (i) \\ & \quad G_\sigma (T_1^{-1}(i)) G_{T_3} (\sigma^{-1} T_1^{-1} (i)) = e^{i \theta_{\sigma T_3} \tau_3} ,
\end{split} \\
& G_{C_3} (C_3 (i)) G_{C_3} (i) G_{C_3} (C_3^{-1} (i)) = e^{i \theta_{C_3} \tau_3} , \label{U1C3cube} \\
& G_\sigma (\sigma(i)) G_\sigma (i) = e^{i \theta_\sigma \tau_3} , \label{U1sigmasquare} \\
\begin{split} \label{U1sigmaC3commute}
& G_\sigma ((\sigma C_3)^3 (i)) G_{C_3} (C_3 (\sigma C_3)^2 (i)) \\
& \quad G_\sigma ((\sigma C_3)^2 (i)) G_{C_3} (C_3 \sigma C_3 (i)) G_\sigma (\sigma C_3 (i)) \\
& \qquad G_{C_3} (C_3 (i)) G_\sigma (i) G_{C_3} (\sigma^{-1} (i)) = e^{i \theta_{\sigma C_3} \tau_3} ,
\end{split} \\
& G_{\mathcal{O}}^\dagger (i) G_\mathcal{T}^\dagger (i) G_\mathcal{O} (i) G_\mathcal{T} (\mathcal{O}^{-1} (i)) = e^{i \theta_\mathcal{OT} \tau_3} , \label{U1spacetimecommute} \\
& [G_\mathcal{T} (i) ]^2 = e^{i \theta_\mathcal{T} \tau_3} 
\, , 
\label{U1timesquare}
\end{align}
\end{subequations}
\endgroup
for $G_X (r_1,r_2,r_3;s)$, where each $\theta_{\ldots}$ on the right hand side is a continuous variable in the interval $[0, 2 \pi)$. For $U(1)$ spin liquids, the gauge transformations have the specific form~\eqref{U1gaugetransformform}. For a symmetry operator $X$ that appear an odd number of times in an algebraic identity, if any other symmetry operators appear an even number of times in the same algebraic identity, we must have $n_X=0$, otherwise the equality of the corresponding equation in~\eqref{U1T2T1commute}-\eqref{U1timesquare} will not hold. Therefore,~\eqref{U1sigmaT2commute} or~\eqref{U1sigmaT3commute} force $n_{T_1}=0$, and~\eqref{U1C3cube} forces $n_{C_3}=0$. Since $G_{T_1}$ and $G_{C_3}$ do not carry $i \tau_1$ with them,~\eqref{U1C3T1commute} and~\eqref{U1C3T3commute} then force $n_{T_2}=0$ and $n_{T_3}=0$ respectively. Next, we use the IGG freedoms (see App.~\ref{appendix:classifyZ2spinliquid}) of $G_{T_1}$, $G_{T_2}$, and $G_{T_3}$ to fix $\theta_{C_3 T_2}=0$, $\theta_{C_3 T_3}=0$, and $\theta_{\sigma T_3}=0$. Notice that equalities involving $U(1)$ variables, which are abundant in this appendix, are defined modulo $2 \pi$. 

Through~\eqref{gaugetransformgauge} we fix
\begin{subequations}
\begin{align}
& \phi_{T_1} (r_1,r_2,r_3;s)=0, \\
& \phi_{T_2} (0,r_2,r_3;s)=0, \\
& \phi_{T_3} (0,0,r_3;s)=0 
\, . 
\end{align}
\end{subequations}
Eq.~\eqref{U1T2T1commute} yields $-\phi_{T_2} (r_1-1,r_2,r_3;s)+\phi_{T_2}(r_1,r_2,r_3;s)=\theta_{T_2 T_1}$, or
\begin{equation}
\phi_{T_2 T_1} (r_1,r_2,r_3;s) = r_1 \theta_{T_2 T_1} \, .
\end{equation}
Eqs.~\eqref{U1T3T2commute} and~\eqref{U1T1T3commute} respectively yield
\begin{subequations}
\begin{align*}
& \phi_{T_3} (r_1,r_2,r_3;s) = - r_1 \theta_{T_1 T_3} + \phi_{T_3} (0,r_2,r_3;s) , \\
& \phi_{T_3} (r_1,r_2,r_3;s) = r_2 \theta_{T_3 T_2} + \phi_{T_3} (r_1,0,r_3;s) ,
\end{align*}
\end{subequations}
so
\begin{equation}
\phi_{T_3} (r_1,r_2,r_3;s) = r_2 \theta_{T_3 T_2} - r_1 \theta_{T_1 T_3} \, .
\end{equation}

Eqs.~\eqref{U1C3T1commute},~\eqref{U1C3T2commute}, and~\eqref{U1C3T3commute} yield
\begingroup
\allowdisplaybreaks
\begin{subequations}
\begin{align*}
\phi_{C_3} (r_1,r_2,r_3;s) = \, & r_2 \theta_{C_3 T_1} + r_1 r_2 \theta_{T_2 T_1} + \phi_{C_3} (r_1,0,r_3;s) , \\
\begin{split}
\phi_{C_3} (r_1,r_2,r_3;s) = \, & - r_3 r_1 \theta_{T_1 T_3} + r_2 r_3 ( \theta_{T_3 T_2} - \theta_{T_2 T_1} ) \\
& + \phi_{C_3} (r_1,r_2,0;s) ,
\end{split} \\
\phi_{C_3} (r_1,r_2,r_3;s) = \, & r_1 r_2 \theta_{T_1 T_3} - r_3 r_1 \theta_{T_3 T_2} + \phi_{C_3} (0,r_2,r_3;s) ,
\end{align*}
\end{subequations}
\endgroup
which further lead to
\begingroup
\allowdisplaybreaks
\begin{subequations}
\begin{align*}
\begin{split}
 \phi_{C_3} (r_1,r_2,r_3;s) = & \: r_2 \theta_{C_3 T_1} + r_1 r_2 \theta_{T_2 T_1} - r_3 r_1 \theta_{T_1 T_3} + \varphi_{C_3} (s) ,
\end{split} \\
\begin{split}
 \phi_{C_3} (r_1,r_2,r_3;s) = & \: r_2 \theta_{C_3 T_1} + r_1 r_2 \theta_{T_2 T_1} - r_3 r_1 \theta_{T_3 T_2} + \varphi_{C_3} (s) ,
\end{split} \\
\begin{split}
 \phi_{C_3} (r_1,r_2,r_3;s) = & - r_3 r_1 \theta_{T_1 T_3} + r_2 r_3 ( \theta_{T_3 T_2} - \theta_{T_2 T_1} ) \\ & + r_2 \theta_{C_3 T_1} + r_1 r_2 \theta_{T_2 T_1} + \varphi_{C_3} (s) ,
\end{split} \\
\begin{split}
 \phi_{C_3} (r_1,r_2,r_3;s) = & - r_3 r_1 \theta_{T_1 T_3} + r_2 r_3 ( \theta_{T_3 T_2} - \theta_{T_2 T_1} ) \\ & + r_1 r_2 \theta_{T_1 T_3} + r_2 \theta_{C_3 T_1} + \varphi_{C_3} (s),
\end{split} \\
\begin{split}
 \phi_{C_3} (r_1,r_2,r_3;s) = & \: r_1 r_2 \theta_{T_1 T_3} - r_3 r_1 \theta_{T_3 T_2} + r_2 \theta_{C_3 T_1} + \varphi_{C_3} (s) ,
\end{split} \\
\begin{split}
 \phi_{C_3} (r_1,r_2,r_3;s) = & \: r_1 r_2 \theta_{T_1 T_3} - r_3 r_1 \theta_{T_3 T_2} + r_2 r_3 ( \theta_{T_3 T_2} - \theta_{T_2 T_1} ) \\ & + r_2 \theta_{C_3 T_1} + \varphi_{C_3} (s) .
\end{split}
\end{align*}
\end{subequations}
\endgroup
where $\varphi_X (s) \equiv \phi_X (0,0,0;s)$. The right hand sides of these six equations must be equal to each other, which forces $\theta_{T_1 T_3}=\theta_{T_3 T_2}=\theta_{T_2 T_1}$. 

Eq.~\eqref{U1C3cube} with $i=(r_1,r_2,r_3;0)$ yields
\begin{equation}
(r_1+r_2+r_3) \theta_{C_3 T_1} + 3 \varphi_{C_3} (0) = \theta_{C_3} \, .
\end{equation}
Since the right hand side has no coordinate dependence, we must have $\theta_{C_3 T_1}=0$, which is anticipated since~\eqref{C3T1commute} is implied by~\eqref{C3T2commute} and by~\eqref{C3T3commute}. We also have
\begin{equation}
\varphi_{C_3} (3) + \varphi_{C_3} (2) + \varphi_{C_3} (1) = \theta_{C_3} \, ,
\end{equation}
with $i=(r_1,r_2,r_3;s=1,2,3)$. 

\smallskip
We now proceed to the parts that involve $\sigma$. 

\smallskip
\noindent
\textbf{Case 1}.~$n_\sigma=0$. 
Eq.~\eqref{U1sigmaT1commute} yields
\begin{equation}
\phi_\sigma (r_1,r_2,r_3;s) = -r_1 \theta_{\sigma T_1} + \phi_\sigma (0,r_2,r_3;s) .
\end{equation}
Subsequently,~\eqref{U1sigmaT2commute} and~\eqref{U1sigmaT3commute} yield
\begingroup
\allowdisplaybreaks
\begin{subequations}
\begin{align*}
\begin{split}
\phi_\sigma (0,r_2,r_3;s) = & \left[ \frac{r_2 (r_2+1)}{2} + r_2 (2 r_1 + r_3 + 1) \right] \theta_{T_2 T_1} \\ 
& + r_2 (\theta_{\sigma T_2} - \theta_{\sigma T_1}) + \phi_\sigma (0,0,r_3;s) ,
\end{split} \\
\begin{split}
\phi_\sigma (0,r_2,r_3;s) = & - \left[ \frac{r_3 (r_3+1)}{2} + r_3 (2 r_1 + r_2 + 1) \right] \theta_{T_2 T_1} \\ 
& - r_3 \theta_{\sigma T_1} + \phi_\sigma (0,r_2,0;s) ,
\end{split}
\end{align*}
\end{subequations}
\endgroup
which further lead to
\begingroup
\allowdisplaybreaks
\begin{subequations}
\begin{align*}
\begin{split}
\phi_\sigma (r_1,r_2,r_3;s) = & \: \bigg[ \frac{r_2 (r_2+1)}{2} - \frac{r_3 (r_3+1)}{2} \\
& + (2 r_1 + 1) (r_2 - r_3) + r_2 r_3 \bigg] \theta_{T_2 T_1} \\ 
& - (r_1+r_2+r_3) \theta_{\sigma T_1} + r_2 \theta_{\sigma T_2} + \varphi_\sigma (s) ,
\end{split} \\
\begin{split}
\phi_\sigma (r_1,r_2,r_3;s) = & \: \bigg[ \frac{r_2 (r_2+1)}{2} - \frac{r_3 (r_3+1)}{2} \\
& + (2 r_1 + 1) (r_2 - r_3) - r_2 r_3 \bigg] \theta_{T_2 T_1} \\
& - (r_1+r_2+r_3) \theta_{\sigma T_1} + r_2 \theta_{\sigma T_2} + \varphi_\sigma (s) ,
\end{split}
\end{align*}
\end{subequations}
\endgroup
The right hand sides of these two equations must be equal to each other, which forces $\theta_{T_2 T_1}=p_{T_2 T_1} \pi$, $p_{T_2 T_1} \in \lbrace 0,1 \rbrace$. 

Eq.~\eqref{U1sigmasquare} with $i=(r_1,r_2,r_3;0)$ yields
\begin{equation}
- (r_2 + r_3) \theta_{\sigma T_1} + 2 r_2 \theta_{\sigma T_2} + \varphi_\sigma (1) + \varphi_\sigma (0) = \theta_\sigma .
\end{equation}
which implies $\theta_{\sigma T_1} = 0$, $\theta_{\sigma T_2} = p_{\sigma T_2} \pi$, $p_{\sigma T_2} \in {0,1}$. We also have
\begin{equation}
2 \varphi_\sigma (2) = \theta_\sigma , 2 \varphi_\sigma (3) = \theta_\sigma ,
\end{equation}
with $i=(r_1,r_2,r_3;s=2,3)$. 

Eq.~\eqref{U1sigmaC3commute} yields
\begin{equation}
\begin{aligned}[b]
& \varphi_\sigma (0) + \varphi_{C_3} (1) + \varphi_\sigma (3) + \varphi_{C_3} (3) \\
& \quad + \varphi_\sigma (2) + \varphi_{C_3} (2) + \varphi_\sigma (1) + \varphi_{C_3} (0) = \theta_{\sigma C_3} \, .
\end{aligned}
\end{equation}

\smallskip
\noindent
\textbf{Case 2.}~$n_\sigma = 1$.
Eq.~\eqref{U1sigmaT1commute} yields
\begin{equation}
\phi_\sigma (r_1,r_2,r_3;s) = - r_1 \theta_{\sigma T_1} + \phi_\sigma (0,r_2,r_3;s) .
\end{equation}
Subsequently, \eqref{U1sigmaT2commute} and \eqref{U1sigmaT3commute} yield
\begingroup
\allowdisplaybreaks
\begin{subequations}
\begin{align*}
\begin{split}
\phi_\sigma (0,r_2,r_3;s) = & \left[ \frac{r_2 (r_2+1)}{2} + r_2 (r_3 - 1) \right] \theta_{T_2 T_1} \\ 
& + r_2 (\theta_{\sigma T_2} - \theta_{\sigma T_1}) + \phi_\sigma (0,0,r_3;s) ,
\end{split} \\
\begin{split}
\phi_\sigma (0,r_2,r_3;s) = & - \left[ \frac{r_3 (r_3+1)}{2} + r_3 (3 r_2 - 1) \right] \theta_{T_2 T_1} \\ 
& - r_3 \theta_{\sigma T_1} + \phi_\sigma (0,r_2,0;s) ,
\end{split}
\end{align*}
\end{subequations}
\endgroup
which further lead to
\begingroup
\allowdisplaybreaks
\begin{subequations}
\begin{align*}
\begin{split}
\phi_\sigma (r_1,r_2,r_3;s) = & \: \bigg[ \frac{r_2 (r_2+1)}{2} - \frac{r_3 (r_3+1)}{2} \\ & - r_2 + r_3 + r_2 r_3 \bigg] \theta_{T_2 T_1} \\ 
& - (r_1+r_2+r_3) \theta_{\sigma T_1} + r_2 \theta_{\sigma T_2} + \varphi_\sigma (s) ,
\end{split} \\
\begin{split}
\phi_\sigma (r_1,r_2,r_3;s) = & \: \bigg[ \frac{r_2 (r_2+1)}{2} - \frac{r_3 (r_3+1)}{2} \\
& - r_2 + r_3 - 3 r_2 r_3 \bigg] \theta_{T_2 T_1} \\
& - (r_1+r_2+r_3) \theta_{\sigma T_1} + r_2 \theta_{\sigma T_2} + \varphi_\sigma (s) .
\end{split}
\end{align*}
\end{subequations}
\endgroup
The right hand sides of these two equations must be equal to each other, which forces $\theta_{T_2 T_1} = 2 p_{T_2 T_1} \pi / 4$, $p_{T_2 T_1} \in \lbrace 0,1,2,3 \rbrace$. 

Eq.~\eqref{U1sigmasquare} with $i=(r_1,r_2,r_3;0)$ yields
\begin{equation}
- (2 r_1+r_2+r_3) \theta_{\sigma T_1} - \varphi_\sigma (1) + \varphi_\sigma (0) = \theta_\sigma + \pi ,
\end{equation}
which implies $\theta_{\sigma T_1}=0$. The additive factor of $\pi$ on the right hand side comes from $(i \tau_1)^2=-1$. We also have
\begin{equation}
- \varphi_\sigma (2) + \varphi_\sigma (2) = \theta_\sigma + \pi
\end{equation}
with $i=(r_1,r_2,r_3;2)$, which implies $\theta_\sigma = \pi$. 

Eq.~\eqref{U1sigmaC3commute} with $\sigma^{-1} (i) = (r_1,r_2,r_3;0)$ yields
\begin{equation}
\begin{aligned}[b]
& 2 (r_1 + r_2) \theta_{\sigma T_2} - \varphi_\sigma (0) - \varphi_{C_3} (1) + \varphi_\sigma (3) + \varphi_{C_3} (3) \\
& \quad - \varphi_\sigma (2) - \varphi_{C_3} (2) + \varphi_\sigma (1) + \varphi_{C_3} (0) = \theta_{\sigma C_3} ,
\end{aligned}
\end{equation}
which implies $\theta_{\sigma T_2} = p_{\sigma T_2} \pi$, $p_{\sigma T_2} \in \lbrace 0,1 \rbrace$. We also have
\begin{equation}
\begin{aligned}[b]
& - \varphi_\sigma (3) - \varphi_{C_3} (3) + \varphi_\sigma (2) + \varphi_{C_3} (2) \\
& \quad - \varphi_\sigma (1) - \varphi_{C_3} (0) + \varphi_\sigma (0) + \varphi_{C_3} (1) = \theta_{\sigma C_3}
\end{aligned}
\end{equation}
with $\sigma^{-1} (i) = (r_1,r_2,r_3;1)$, which implies $\theta_{\sigma C_3}=p_{\sigma C_3} \pi$, $p_{\sigma C_3} \in \lbrace 0,1 \rbrace$.

\smallskip
We now proceed to consider the parts that involve $\mathcal{T}$. 

\smallskip
\noindent 
\textbf{Case \textit{x}.1.}~$n_\mathcal{T}=0$.
Eq.~\eqref{U1spacetimecommute} with $\mathcal{O}=T_1,T_2,T_3$ yields
\begingroup
\allowdisplaybreaks
\begin{subequations}
\begin{align}
& \phi_\mathcal{T} (r_1,r_2,r_3;s) = - r_1 \theta_{T_1 \mathcal{T}} + \phi_\mathcal{T} (0,r_2,r_3;s) , \\
& \phi_\mathcal{T} (r_1,r_2,r_3;s) = - r_2 \theta_{T_2 \mathcal{T}} + \phi_\mathcal{T} (r_1,0,r_3;s) , \\
& \phi_\mathcal{T} (r_1,r_2,r_3;s) = - r_3 \theta_{T_3 \mathcal{T}} + \phi_\mathcal{T} (r_1,r_2,0;s) ,
\end{align}
\end{subequations}
\endgroup
which together imply
\begin{equation}
\phi_\mathcal{T} (r_1,r_2,r_3;s) = -r_1 \theta_{T_1 \mathcal{T}} - r_2 \theta_{T_2 \mathcal{T}} - r_3 \theta_{T_3 \mathcal{T}} + \varphi_\mathcal{T} (s) .
\end{equation}

Eq.~\eqref{U1spacetimecommute} with $\mathcal{O}=C_3$ and $C_3^{-1}(i)=(r_1,r_2,r_3;0)$ yields
\begin{equation}
(r_3-r_1) \theta_{T_1 \mathcal{T}} + (r_1-r_2) \theta_{T_2 \mathcal{T}} + (r_2-r_3) \theta_{T_3 \mathcal{T}} = \theta_{C_3 \mathcal{T}} ,
\end{equation}
which implies $\theta_{T_1 \mathcal{T}}=\theta_{T_2 \mathcal{T}}=\theta_{T_3 \mathcal{T}}$, and subsequently $\theta_{C_3 \mathcal{T}}=0$. We also have
\begin{subequations}
\begin{align}
& - \varphi_\mathcal{T} (2) + \varphi_\mathcal{T} (1) = 0 , \\
& - \varphi_\mathcal{T} (3) + \varphi_\mathcal{T} (2) = 0 , \\
& - \varphi_\mathcal{T} (1) + \varphi_\mathcal{T} (3) = 0 ,
\end{align}
\end{subequations}
with $C_3^{-1} (i) = (r_1,r_2,r_3;s=1,2,3)$. 

\smallskip
\noindent
\textbf{Case 1.1.}~$n_\sigma=0$.
Eq.~\eqref{U1spacetimecommute} with $\mathcal{O}=\sigma$ and $\sigma^{-1} (i) = (r_1,r_2,r_3;0)$ yields
\begin{equation}
- (2 r_1 + r_2 + r_3) \theta_{T_1 \mathcal{T}} - \varphi_\mathcal{T} (1)  + \varphi_\mathcal{T} (0) = \theta_{\sigma \mathcal{T}} ,
\end{equation}
which implies $\theta_{T_1 \mathcal{T}}=0$. We also have
\begin{equation}
- \varphi_\mathcal{T} (2) + \varphi_\mathcal{T} (2) = \theta_{\sigma \mathcal{T}} ,
\end{equation}
with $\sigma^{-1} (i)=(r_1,r_2,r_3;2)$, which implies $\theta_{\sigma \mathcal{T}}=0$. 

Eq.~\eqref{U1timesquare} then yields
\begin{equation}
2 \varphi_\mathcal{T} (s) = \theta_\mathcal{T} .
\end{equation}

\smallskip
\noindent
\textbf{Case 2.1.}~$n_\sigma=1$.
Eq.~\eqref{U1spacetimecommute} with $\mathcal{O}=\sigma$ and $\sigma^{-1} (i)=(r_1,r_2,r_3;0)$ yields
\begin{equation}
- (r_2 + r_3) \theta_{T_1 \mathcal{T}} + \varphi_\mathcal{T} (1) + \varphi_\mathcal{T} (0) = \theta_{\sigma \mathcal{T}} ,
\end{equation}
which forces $\theta_{T_1 \mathcal{T}}=0$. We also have
\begin{subequations}
\begin{align}
2 \varphi_\mathcal{T} (2) = \theta_{\sigma \mathcal{T}} , \\
2 \varphi_\mathcal{T} (3) = \theta_{\sigma \mathcal{T}} ,
\end{align}
\end{subequations}
with $\sigma^{-1} (i) = (r_1,r_2,r_3;s=2,3)$. \\

Eq.~\eqref{U1timesquare} then yields
\begin{equation}
2 \varphi_\mathcal{T} (s) = \theta_\mathcal{T} .
\end{equation}

\smallskip
\noindent
\textbf{Case \textit{x}.2.}~$n_\mathcal{T}=1$.

\smallskip
\noindent
\textbf{Case 1.2.}~$n_\sigma=0$.
Eq.~\eqref{U1spacetimecommute} with $\mathcal{O}=T_1,T_2,T_3$ yields
\begingroup
\allowdisplaybreaks
\begin{subequations}
\begin{align}
& \phi_\mathcal{T} (r_1,r_2,r_3;s) = - r_1 \theta_{T_1 \mathcal{T}} + \phi_\mathcal{T} (0,r_2,r_3;s) , \\
& \phi_\mathcal{T} (r_1,r_2,r_3;s) = - r_2 \theta_{T_2 \mathcal{T}} + \phi_\mathcal{T} (r_1,0,r_3;s) , \\
& \phi_\mathcal{T} (r_1,r_2,r_3;s) = - r_3 \theta_{T_3 \mathcal{T}} + \phi_\mathcal{T} (r_1,r_2,0;s) ,
\end{align}
\end{subequations}
\endgroup
which together imply
\begin{equation}
\phi_\mathcal{T} (r_1,r_2,r_3;s) = -r_1 \theta_{T_1 \mathcal{T}} - r_2 \theta_{T_2 \mathcal{T}} - r_3 \theta_{T_3 \mathcal{T}} + \varphi_\mathcal{T} (s) \, .
\end{equation}
Eq.~\eqref{U1spacetimecommute} with $\mathcal{O}=C_3$ and $i=(r_1,r_2,r_3;0)$ yields
\begin{equation}
\begin{aligned}
& (r_1-r_2) \theta_{T_1 \mathcal{T}} + (r_2-r_3) \theta_{T_2 \mathcal{T}} + (r_3-r_1) \theta_{T_3 \mathcal{T}} \\
& \quad - 2 \varphi_{C_3} (0) = \theta_{C_3 \mathcal{T}} ,
\end{aligned}
\end{equation}
which implies $\theta_{T_1 \mathcal{T}}=\theta_{T_2 \mathcal{T}}=\theta_{T_3 \mathcal{T}}$. We also have
\begingroup
\allowdisplaybreaks
\begin{subequations}
\begin{align}
& - 2 \varphi_{C_3} (1) - \varphi_\mathcal{T} (1) + \varphi_\mathcal{T} (3) = \theta_{C_3 \mathcal{T}} , \\
& - 2 \varphi_{C_3} (2) - \varphi_\mathcal{T} (2) + \varphi_\mathcal{T} (1) = \theta_{C_3 \mathcal{T}} , \\
& - 2 \varphi_{C_3} (3) - \varphi_\mathcal{T} (3) + \varphi_\mathcal{T} (2) = \theta_{C_3 \mathcal{T}} ,
\end{align}
\end{subequations}
\endgroup
with $i=(r_1,r_2,r_3;s=1,2,3)$. 

Eq.~\eqref{U1spacetimecommute} with $\mathcal{O}=\sigma$ and $i=(r_1,r_2,r_3;0)$ yields
\begin{equation}
(2 r_1 + r_2 + r_3) \theta_{T_1 \mathcal{T}} - 2 \varphi_\sigma (0) - \varphi_\mathcal{T} (0) + \varphi_\mathcal{T} (1) = \theta_{\sigma \mathcal{T}} ,
\end{equation}
which implies $\theta_{T_1 \mathcal{T}}=0$. We also have
\begingroup
\allowdisplaybreaks
\begin{subequations}
\begin{align}
& - 2 \varphi_\sigma (1) - \varphi_\mathcal{T} (1) + \varphi_\mathcal{T} (0) = \theta_{\sigma \mathcal{T}} , \\
& - 2 \varphi_\sigma (2) - \varphi_\mathcal{T} (2) + \varphi_\mathcal{T} (2) = \theta_{\sigma \mathcal{T}} , \\
& - 2 \varphi_\sigma (3) - \varphi_\mathcal{T} (3) + \varphi_\mathcal{T} (3) = \theta_{\sigma \mathcal{T}} ,
\end{align}
\end{subequations}
\endgroup
with $i=(r_1,r_2,r_3;s=1,2,3)$. 

Eq.~\eqref{U1timesquare} then yields
\begin{equation}
\pi = \theta_\mathcal{T} ,
\end{equation}
where $\pi$ comes from $(i \tau_1)^2=-1$. 

\smallskip
\noindent
\textbf{Case 2.2.}~$n_\sigma=1$.
Eq.~\eqref{U1spacetimecommute} with $\mathcal{O}=T_1,T_2,T_3$ yields
\begingroup
\allowdisplaybreaks
\begin{subequations}
\begin{align*}
& \phi_\mathcal{T} (r_1,r_2,r_3;s) = - r_1 \theta_{T_1 \mathcal{T}} + \phi_\mathcal{T} (0,r_2,r_3;s) , \\
& \phi_\mathcal{T} (r_1,r_2,r_3;s) = - 2 r_1 r_2 \theta_{T_2 T_1} - r_2 \theta_{T_2 \mathcal{T}} + \phi_\mathcal{T} (r_1,0,r_3;s) , \\
& \phi_\mathcal{T} (r_1,r_2,r_3;s) = 2 r_3 (r_1 - r_2) \theta_{T_2 T_1} - r_3 \theta_{T_3 \mathcal{T}} + \phi_\mathcal{T} (r_1,r_2,0;s) ,
\end{align*}
\end{subequations}
\endgroup
which further lead to
\begingroup
\allowdisplaybreaks
\begin{subequations}
\begin{align*}
\phi_\mathcal{T} (r_1,r_2,r_3;s) = &- r_1 \theta_{T_1 \mathcal{T}} - r_2 \theta_{T_2 \mathcal{T}} - r_3 \theta_{T_3 \mathcal{T}} + \varphi_\mathcal{T} (s) , \\
\begin{split}
\phi_\mathcal{T} (r_1,r_2,r_3;s) = & - r_1 \theta_{T_1 \mathcal{T}} - r_2 \theta_{T_2 \mathcal{T}} - r_3 \theta_{T_3 \mathcal{T}} + \varphi_\mathcal{T} (s) \\
& - 2 r_2 r_3 \theta_{T_2 T_1} ,
\end{split} \\
\begin{split}
\phi_\mathcal{T} (r_1,r_2,r_3;s) = & - r_1 \theta_{T_1 \mathcal{T}} - r_2 \theta_{T_2 \mathcal{T}} - r_3 \theta_{T_3 \mathcal{T}} + \varphi_\mathcal{T} (s) \\
& - 2 r_1 r_2 \theta_{T_2 T_1} ,
\end{split} \\
\begin{split}
\phi_\mathcal{T} (r_1,r_2,r_3;s) = & - r_1 \theta_{T_1 \mathcal{T}} - r_2 \theta_{T_2 \mathcal{T}} - r_3 \theta_{T_3 \mathcal{T}}  + \varphi_\mathcal{T} (s) \\
& - 2 r_1 (r_2 - r_3) \theta_{T_2 T_1} ,
\end{split} \\
\begin{split}
\phi_\mathcal{T} (r_1,r_2,r_3;s) = & - r_1 \theta_{T_1 \mathcal{T}} - r_2 \theta_{T_2 \mathcal{T}} - r_3 \theta_{T_3 \mathcal{T}} + \varphi_\mathcal{T} (s) \\
& + 2 (r_3 r_1 - r_1 r_2 - r_2 r_3) \theta_{T_2 T_1} ,
\end{split} \\
\begin{split}
\phi_\mathcal{T} (r_1,r_2,r_3;s) = & - r_1 \theta_{T_1 \mathcal{T}} - r_2 \theta_{T_2 \mathcal{T}} - r_3 \theta_{T_3 \mathcal{T}} + \varphi_\mathcal{T} (s) \\
& + 2 r_3 (r_1 - r_2) \theta_{T_2 T_1} .
\end{split}
\end{align*}
\end{subequations}
\endgroup
The right hand sides of these six equations must be equal to each other, which only allows $p_{T_2 T_1}=0,2$ in $\theta_{T_2 T_1} = 2 p_{T_2 T_1} \pi /4$. We redefine $ p_{T_2 T_1}$ such that $\theta_{T_2 T_1} = p_{T_2 T_1} \pi$, $p_{T_2 T_1} \in \lbrace 0,1 \rbrace$. 

Eq.~\eqref{U1spacetimecommute} with $\mathcal{O}=C_3$ and $i=(r_1,r_2,r_3;0)$ yields
\begin{equation}
\begin{aligned}
& (r_1-r_2) \theta_{T_1 \mathcal{T}} + (r_2-r_3) \theta_{T_2 \mathcal{T}} + (r_3-r_1) \theta_{T_3 \mathcal{T}} \\
& \quad - 2 \varphi_{C_3} (0) = \theta_{C_3 \mathcal{T}} ,
\end{aligned}
\end{equation}
which implies $\theta_{T_1 \mathcal{T}}=\theta_{T_2 \mathcal{T}}=\theta_{T_3 \mathcal{T}}$. We also have
\begingroup
\allowdisplaybreaks
\begin{subequations}
\begin{align}
& - 2 \varphi_{C_3} (1) - \varphi_\mathcal{T} (1) + \varphi_\mathcal{T} (3) = \theta_{C_3 \mathcal{T}} , \\
& - 2 \varphi_{C_3} (2) - \varphi_\mathcal{T} (2) + \varphi_\mathcal{T} (1) = \theta_{C_3 \mathcal{T}} , \\
& - 2 \varphi_{C_3} (3) - \varphi_\mathcal{T} (3) + \varphi_\mathcal{T} (2) = \theta_{C_3 \mathcal{T}} ,
\end{align}
\end{subequations}
\endgroup
with $i=(r_1,r_2,r_3;s=1,2,3)$. 

Eq.~\eqref{U1spacetimecommute} with $\mathcal{O}=\sigma$ and $i=(r_1,r_2,r_3;0)$ yields
\begin{equation}
- (r_2 + r_3) \theta_{T_1 \mathcal{T}} - 2 \varphi_\sigma (0) + \varphi_\mathcal{T} (0) + \varphi_\mathcal{T} (1) = \theta_{\sigma \mathcal{T}} ,
\end{equation}
which forces $\theta_{T_1 \mathcal{T}}=0$. We also have
\begingroup
\allowdisplaybreaks
\begin{subequations}
\begin{align}
& - 2 \varphi_\sigma (1) + \varphi_\mathcal{T} (1) + \varphi_\mathcal{T} (0) = \theta_{\sigma \mathcal{T}} , \\
& - 2 \varphi_\sigma (2) + 2 \varphi_\mathcal{T} (2) = \theta_{\sigma \mathcal{T}} , \\
& - 2 \varphi_\sigma (3) + 2 \varphi_\mathcal{T} (3) = \theta_{\sigma \mathcal{T}} ,
\end{align}
\end{subequations}
\endgroup
with $i=(r_1,r_2,r_3;s=1,2,3)$. 

Eq.~\eqref{U1timesquare} then yields
\begin{equation}
\pi = \theta_\mathcal{T} ,
\end{equation}
where $\pi$ comes from $(i \tau_1)^2=-1$.

\subsubsection{Grand Summary}

It is worthwhile to recollect the results obtained thus far before we proceed to gauge fixings. They are~\eqref{U1cellT1angle}-\eqref{U1celltimeangle} shown in the main text and the ``sublattice constraints'' listed below in a case by case manner.
\begin{subequations}
\begin{align}
3 \varphi_{C_3} (0) = \theta_{C_3}, \label{U1sublatticeC30cube} \\
\sum_{s=1}^3 \varphi_{C_3} (s) = \theta_{C_3} . \label{U1sublatticeC3123}
\end{align}
\end{subequations}

\smallskip
\noindent
\textbf{Case 1.}~$n_\sigma=0$.
\begingroup
\allowdisplaybreaks
\begin{subequations}
\begin{align}
& \theta_{T_2 T_1} = p_{T_2 T_1} \pi, p_{T_2 T_1} \in \lbrace 0,1 \rbrace , \\
& \theta_{\sigma T_2} = p_{\sigma T_2} \pi , p_{\sigma T_2} \in \lbrace 0,1 \rbrace , \\
& \varphi_\sigma (1) + \varphi_\sigma (0) = \theta_\sigma, \label{U1sublattice0sigma01} \\
& 2 \varphi_\sigma (2) = \theta_\sigma, \label{U1sublattice0sigma2square} \\
& 2 \varphi_\sigma (3) = \theta_\sigma, \label{U1sublattice0sigma3square} \\
& \sum_{s=0}^3 \left[ \varphi_\sigma (s) + \varphi_{C_3} (s) \right] = \theta_{\sigma C_3} . \label{U1sublattice0sigmaC3}
\end{align}
\end{subequations}
\endgroup

\smallskip
\noindent
\textbf{Case 1.1.}~$n_\mathcal{T}=0$.
\begin{subequations}
\begin{align}
& \varphi_\mathcal{T} (0) = \varphi_\mathcal{T} (1) = \varphi_\mathcal{T} (2) = \varphi_\mathcal{T} (3) , \label{U1sublattice00time0123} \\
& 2 \varphi_\mathcal{T} (s) = \theta_\mathcal{T} \label{U1sublattice00timesquare} .
\end{align}
\end{subequations}

\smallskip
\noindent
\textbf{Case 1.2.}~$n_\mathcal{T}=1$.
\begingroup
\allowdisplaybreaks
\begin{subequations}
\begin{align}
& - 2 \varphi_{C_3} (0) = \theta_{C_3 \mathcal{T}} , \label{U1sublattice01C3time0} \\
& - 2 \varphi_{C_3} (1) - \varphi_\mathcal{T} (1) + \varphi_\mathcal{T} (3) = \theta_{C_3 \mathcal{T}} , \label{U1sublattice01C3time1} \\
& - 2 \varphi_{C_3} (2) - \varphi_\mathcal{T} (2) + \varphi_\mathcal{T} (1) = \theta_{C_3 \mathcal{T}} , \label{U1sublattice01C3time2} \\
& - 2 \varphi_{C_3} (3) - \varphi_\mathcal{T} (3) + \varphi_\mathcal{T} (2) = \theta_{C_3 \mathcal{T}} , \label{U1sublattice01C3time3} \\
& - 2 \varphi_\sigma (0) - \varphi_\mathcal{T} (0) + \varphi_\mathcal{T} (1) = \theta_{\sigma \mathcal{T}} , \label{U1sublattice01sigmatime0} \\
& - 2 \varphi_\sigma (1) - \varphi_\mathcal{T} (1) + \varphi_\mathcal{T} (0) = \theta_{\sigma \mathcal{T}} , \label{U1sublattice01sigmatime1} \\
& - 2 \varphi_\sigma (2) = \theta_{\sigma \mathcal{T}} , \label{U1sublattice01sigmatime2} \\
& - 2 \varphi_\sigma (3) = \theta_{\sigma \mathcal{T}} . \label{U1sublattice01sigmatime3}
\end{align}
\end{subequations}
\endgroup

\smallskip
\noindent
\textbf{Case 2.}~$n_\sigma=1$. 
\begingroup
\allowdisplaybreaks
\begin{subequations}
\begin{align}
& \theta_{\sigma T_2} = p_{\sigma T_2} \pi , p_{\sigma T_2} \in \lbrace 0,1 \rbrace , \\
& \varphi_\sigma (0) = \varphi_\sigma (1) , \label{U1sublattice1sigma01} \\
\begin{split} \label{U1sublattice1sigmaC3}
& - \varphi_\sigma (0) - \varphi_{C_3} (1) + \varphi_\sigma (3) + \varphi_{C_3} (3) \\ & - \varphi_\sigma (2) - \varphi_{C_3} (2) + \varphi_\sigma (1) + \varphi_{C_3} (0) \\ & = p_{\sigma C_3} \pi, p_{\sigma C_3} \in \lbrace 0,1 \rbrace .
\end{split}
\end{align}
\end{subequations}
\endgroup

\smallskip
\noindent
\textbf{Case 2.1.}~$n_\mathcal{T}=0$.
\begingroup
\allowdisplaybreaks
\begin{subequations}
\begin{align}
& \theta_{T_2 T_1} = \frac{2 p_{T_2 T_1} \pi}{4}, p_{T_2 T_1} \in \lbrace 0,1,2,3 \rbrace , \\
& \varphi_\mathcal{T} (1) + \varphi_\mathcal{T} (0) = \theta_{\sigma \mathcal{T}} , \label{U1sublattice10sigmatime0} \\
& 2 \varphi_\mathcal{T} (2) = \theta_{\sigma \mathcal{T}} , \label{U1sublattice10sigmatime2} \\
& 2 \varphi_\mathcal{T} (3) = \theta_{\sigma \mathcal{T}} , \label{U1sublattice10sigmatime3} \\
& \varphi_\mathcal{T} (1) = \varphi_\mathcal{T} (2) = \varphi_\mathcal{T} (3) , \label{U1sublattice10time123} \\
& 2 \varphi_\mathcal{T} (s) = \theta_{\mathcal{T}} . \label{U1sublattice10timesquare}
\end{align}
\end{subequations}
\endgroup

\smallskip
\noindent
\textbf{Case 2.2.}~$n_\mathcal{T}=1$.
\begingroup
\allowdisplaybreaks
\begin{subequations}
\begin{align}
& \theta_{T_2 T_1} = p_{T_2 T_1} \pi, p_{T_2 T_1} \in \lbrace 0,1 \rbrace , \\
& - 2 \varphi_{C_3} (0) = \theta_{C_3 \mathcal{T}} , \label{U1sublattice11C3time0} \\
& - 2 \varphi_{C_3} (1) - \varphi_\mathcal{T} (1) + \varphi_\mathcal{T} (3) = \theta_{C_3 \mathcal{T}} , \label{U1sublattice11C3time1} \\
& - 2 \varphi_{C_3} (2) - \varphi_\mathcal{T} (2) + \varphi_\mathcal{T} (1) = \theta_{C_3 \mathcal{T}} , \label{U1sublattice11C3time2} \\
& - 2 \varphi_{C_3} (3) - \varphi_\mathcal{T} (3) + \varphi_\mathcal{T} (2) = \theta_{C_3 \mathcal{T}} , \label{U1sublattice11C3time3} \\
& - 2 \varphi_\sigma (0) + \varphi_\mathcal{T} (0) + \varphi_\mathcal{T} (1) = \theta_{\sigma \mathcal{T}} , \label{U1sublattice11sigmatime0} \\
& - 2 \varphi_\sigma (1) + \varphi_\mathcal{T} (1) + \varphi_\mathcal{T} (0) = \theta_{\sigma \mathcal{T}} , \label{U1sublattice11sigmatime1} \\
& - 2 \varphi_\sigma (2) + 2 \varphi_\mathcal{T} (2) = \theta_{\sigma \mathcal{T}} , \label{U1sublattice11sigmatime2} \\
& - 2 \varphi_\sigma (3) + 2 \varphi_\mathcal{T} (3) = \theta_{\sigma \mathcal{T}} . \label{U1sublattice11sigmatime3}
\end{align}
\end{subequations}
\endgroup

\subsubsection{Gauge fixing}

We now gauge fix $\varphi_X (s) \equiv \phi_X (0,0,0;s)$, $s=0,1,2,3$, subject to the constraints listed above. First and foremost, we use the IGG freedom of $G_{C_3}$ to fix $\varphi_{C_3} (0) = 0$. It follows from~\eqref{U1sublatticeC30cube} that $\theta_{C_3}=0$. Using~\eqref{gaugetransformgauge} and~\eqref{U1sublatticeC3123}, we perform a sublattice dependent gauge transformation $W_s$
\begin{equation*}
W_{0,3} = 1, W_1 = e^{- i \varphi_{C_3} (1) \tau_3}, W_2 = e^{-i [\varphi_{C_3} (1) + \varphi_{C_3} (2)] \tau_3} ,
\end{equation*}
to fix $\varphi_{C_3} (1,2,3)=0$. 

\smallskip
\noindent
\textbf{Case 1.}~$n_\sigma=0$.
We use the IGG freedom of $G_\sigma$ to fix $\varphi_\sigma (2) = 0$. \eqref{U1sublattice0sigma2square} then implies $\theta_\sigma=0$, and \eqref{U1sublattice0sigma3square} subsequently implies $\varphi_\sigma (3) = q_3 \pi$, $q_3 \in \lbrace 0,1 \rbrace$. We perform a sublattice dependent gauge transformation $W_0=\exp [- i \varphi_\sigma (0) \tau_3]$, $W_{1,2,3}=1$ to fix $\varphi_\sigma (0,1)=0$, without affecting any previously fixed gauge. \eqref{U1sublattice0sigmaC3} then yields $\theta_{\sigma C_3} = q_3 \pi$, so we rename $q_3$ as $p_{\sigma C_3}$. \\

\smallskip
\noindent
\textbf{Case 1.1.}~$n_\mathcal{T}=0$. 
We use the IGG freedom of $G_\mathcal{T}$ to fix $\varphi_\mathcal{T} (0) = 0$. By \eqref{U1sublattice00time0123}, we also have $\varphi_\mathcal{T} (1,2,3)=0$. \\

\smallskip
\noindent
\textbf{Case 1.2.}~$n_\mathcal{T}=1$. 
Eq.~\eqref{U1sublattice01C3time0}-\eqref{U1sublattice01C3time3} yield $\theta_{C_3 \mathcal{T}} = 0$ and
\begin{subequations}
\begin{align}
& - \varphi_\mathcal{T} (1) + \varphi_\mathcal{T} (3) = 0 , \\
& - \varphi_\mathcal{T} (2) + \varphi_\mathcal{T} (1) = 0 , \\
& - \varphi_\mathcal{T} (3) + \varphi_\mathcal{T} (2) = 0 ,
\end{align}
\end{subequations}
or $\varphi_\mathcal{T} (1) = \varphi_\mathcal{T} (2) = \varphi_\mathcal{T} (3)$. Eq.~\eqref{U1sublattice01sigmatime2} or~\eqref{U1sublattice01sigmatime3} imply $\theta_{\sigma \mathcal{T}}=0$, which together with~\eqref{U1sublattice01sigmatime0} or ~\eqref{U1sublattice01sigmatime1} imply $\varphi_\mathcal{T} (0) = \varphi_\mathcal{T} (1)$. We use the IGG freedom of $G_\mathcal{T}$ to fix $\varphi_\mathcal{T} (0)=0$, and it follows that $\varphi_\mathcal{T} (1,2,3) = 0$. 

\smallskip
\noindent
\textbf{Case 2.}~$n_\sigma=1$.
We perform a sublattice dependent gauge transformation $W_0 = \exp [ - i (\varphi_\sigma (2) - \varphi_\sigma (0)) \tau_3 ]$, $W_{1,2,3}=1$, such that $\varphi_\sigma (0,1) = \varphi_\sigma (2)$, without affecting any previously fixed gauge. Then, we use the IGG freedom of $G_\sigma$ to fix $\varphi_\sigma (2)=0$, and rename $\varphi_\sigma (3) - \varphi_\sigma (2)$ as $\varphi_\sigma (3)$. Eq.~\eqref{U1sublattice1sigmaC3} implies $\varphi_{\sigma} (3) = p_{\sigma C_3} \pi$, $p_{\sigma C_3} \in \lbrace 0,1 \rbrace$. 

\smallskip
\noindent
\textbf{Case 2.1.}~$n_\mathcal{T}=0$.
We use the IGG freedom of $G_\mathcal{T}$ to fix $\varphi_\mathcal{T} (1) = 0$. It follows from~\eqref{U1sublattice10time123} that $\varphi_\mathcal{T} (2,3)=0$. We also have $\theta_{\sigma \mathcal{T}}=0$ and $\theta_\mathcal{T}=0$ from \eqref{U1sublattice10sigmatime2} and \eqref{U1sublattice10timesquare} respectively. \eqref{U1sublattice10sigmatime0} then implies $\varphi_\mathcal{T} (0)=0$. 

\smallskip
\noindent
\textbf{Case 2.2.}~$n_\mathcal{T}=1$.
Eqs.~\eqref{U1sublattice11C3time0}-\eqref{U1sublattice11C3time3} yield $\theta_{C_3 \mathcal{T}} = 0$ and
\begin{subequations}
\begin{align}
& - \varphi_\mathcal{T} (1) + \varphi_\mathcal{T} (3) = 0 , \\
& - \varphi_\mathcal{T} (2) + \varphi_\mathcal{T} (1) = 0 , \\
& - \varphi_\mathcal{T} (3) + \varphi_\mathcal{T} (2) = 0 ,
\end{align}
\end{subequations}
or $\varphi_\mathcal{T} (1) = \varphi_\mathcal{T} (2) = \varphi_\mathcal{T} (3)$. We use the IGG freedom of $G_\mathcal{T}$ to fix $\varphi_\mathcal{T} (1)=0$, and it follows that $\varphi_\mathcal{T} (2,3)=0$. Eq.~\eqref{U1sublattice11sigmatime2} or~\eqref{U1sublattice11sigmatime3} imply $\theta_{\sigma \mathcal{T}}=0$, and~\eqref{U1sublattice11sigmatime0} or~\eqref{U1sublattice11sigmatime1} subsequently imply $\varphi_\mathcal{T} (0)=0$. 

\smallskip
We will exclude the solutions with $n_\mathcal{T}=0$ (Cases 1.1 and 2.1) because they have $G_\mathcal{T} (i) = 1$ for all sites $i$, which force $u_{ij}=0$ for any pair of sites $i$ and $j$ by~\eqref{ansatztimerule}. These solutions, which lead to vanishing mean field ansatze and thus a zero Hamiltonian, are unphysical. Let us count the remaining solutions. Each of 1.2 and 2.2 has three $\mathbb{Z}_2$ variables $p_{T_2 T_1}$, $p_{\sigma T_2}$, and $p_{\sigma C_3}$. Therefore, we have in total $2 \times 2^3 = 16$ gauge inequivalent solutions, i.e.,~16 possible $U(1)$ spin liquids. They are listed in Table~\ref{table:spinliquid}.

\subsection{\label{appendix:connection}Relation to isotropic lattice}

We first note that Ref.~\onlinecite{PhysRevB.104.054401} (see also Ref.~\onlinecite{PhysRevB.100.075125}) have chosen a coordinate system in which the unit cell at the origin, $(r_1,r_2,r_3)=(0,0,0)$, is associated with a \textit{down} tetrahedron, whereas in our coordinate system it is associated with an \textit{up} tetrahedron. However, it can be straightforwardly shown that if we instead choose a down tetrahedron at the origin and adopt the primitive translation vectors and sublattice labelings of Ref.~\onlinecite{PhysRevB.104.054401}, the expressions~\eqref{T1operation}-\eqref{sigmaoperation} of the action of the five space group generators remain invariant, so exactly the same PSG solutions (see Table \ref{table:spinliquid}) for the breathing pyrochlore lattice will follow. This observation allows us to assume the coordinate system of Ref.~\onlinecite{PhysRevB.104.054401} in this appendix and directly compare our PSG solutions to those of Ref.~\onlinecite{PhysRevB.104.054401}. 

The sixfold rotoinversion $\overline{C}_6 = \mathcal{I} C_3$ and the twofold nonsymmorphic screw $S$ act on a generic site with coordinates $(r_1,r_2,r_3;s)$ as~\cite{PhysRevB.100.075125,PhysRevB.104.054401}
\begingroup
\allowdisplaybreaks
\begin{subequations}
\begin{align}
\begin{split}
\overline{C}_6 : (r_1,r_2,r_3;s) \longrightarrow ( & \! - r_3 - \delta_{s,3},-r_1 - \delta_{s,1}, \\
& \! -r_2 - \delta_{s,2};\bar{C}_6 (s)), \label{C6operation}
\end{split} \\
\begin{split}
S : (r_1,r_2,r_3;s) \longrightarrow ( & \! -r_1 - \delta_{s,1}, -r_2 - \delta_{s,2}, \\
& r_1+r_2+r_3 + 1 - \delta_{s,0};S(s)) . \label{Soperation}
\end{split}
\end{align}
\end{subequations}
\endgroup
When $s=0,1,2,3$, $\overline{C}_6 (s) = 0,2,3,1$ and $S(s) = 3,1,2,0$. Recall that, for $U(1)$ spin liquids, the gauge transformation $G_X$ associated with a symmetry operator $X$ has the specific form~\eqref{U1gaugetransformform}. The distinct $U(1)$ spin liquids that respect both the $Fd\bar{3}m$ space group of the regular pyrochlore lattice and the time reversal symmetry are characterized by the PSGs in (20a)-(20c), (21a)-(21e), and Table I of Ref.~\onlinecite{PhysRevB.104.054401}, which we quote below,
\begingroup
\allowdisplaybreaks
\begin{subequations}
\begin{align}
\phi_{T_1} (r_1,r_2,r_3;s) = & \:  0 , n_{T_1} = 0, \label{regularU1cellT1angle} \\
\phi_{T_2} (r_1,r_2,r_3;s) = & - \chi_1 r_1 , n_{T_2} = 0, \label{regularU1cellT2angle} \\
\phi_{T_3} (r_1,r_2,r_3;s) = & \: \chi_1 (r_1 - r_2) , n_{T_3} = 0, \label{regularU1cellT3angle} \\
\begin{split} \label{regularU1cellC6angle}
\phi_{\overline{C}_6} (r_1,r_2,r_3;s) = & - \chi_1 r_1 (r_2 - r_3) \\ & - [ 2 + (\delta_{s,2} - \delta_{s,3})] \chi_1 r_1 \\ & + \delta_{s,2} \chi_1 r_3 + \varphi_{\overline{C}_6} (s) ,
\end{split} \\
\begin{split} \label{regularU1cellSangle}
\phi_S (r_1,r_2,r_3;s) = & \: \chi_1 \left[ \frac{r_1 (r_1 + 1)}{2} - \frac{r_2 (r_2 + 1)}{2} - r_1 r_2 \right] \\
& + (2 + \delta_{s,1} - \delta_{s,2}) \chi_1 (r_3 + r_1) \\
& + (2 \delta_{s,1} - \delta_{s,2}) \chi_1 r_2  + \varphi_S (s) ,
\end{split} \\
\phi_\mathcal{T} = & \: 0, n_\mathcal{T} = 1 . \label{regularU1celltimeangle}
\end{align}
\end{subequations}
\endgroup
with
\begin{equation} \label{regularU1sublatticeangle}
\begin{aligned}[b]
& \varphi_{\overline{C}_6} (0,1,2) = 0, \varphi_{\overline{C}_6} (3) = \chi_1, \\
& \varphi_S (0,2) = 0, \varphi_S (1) = \chi_{\overline{C}_6 S}, \varphi_S (3) = \chi_1, \\
& \chi_1, \chi_{\overline{C}_6 S} \in \lbrace 0, \pi \rbrace .
\end{aligned}
\end{equation}
The two choices in each of $n_{\overline{C}_6}$, $n_S$, $\chi_1$, and $\chi_{\overline{C}_6 S}$ gives rise to $2^4=16$ $U(1)$ spin liquids. Notice that every $\phi_X$ is an integer multiple of $\pi$, which is invariant under a sign change. 

The point group generators of the breathing pyrochlore lattice are related to the regular one by $C_3 = \overline{C}_6^4$ and $\sigma = C_3 \Sigma C_3^{-1}$, where $\Sigma \equiv S \mathcal{I}$ is a reflection across the plane perpendicular to $[110]$ (c.f.~$\sigma$). For convenience of later calculations, we note that
\begingroup
\allowdisplaybreaks
\begin{subequations}
\begin{align}
\begin{split}
\mathcal{I} : (r_1,r_2,r_3;s) \longrightarrow (& \! -r_1 - \delta_{s,1} , -r_2 - \delta_{s,2} , \\ & \! -r_3 - \delta_{s,3};s) ,
\end{split} \\
\Sigma : (r_1,r_2,r_3;s) \longrightarrow (& r_1,r_2, -r_1 -r_2 -r_3;\Sigma (s)) ,
\end{align}
\end{subequations}
\endgroup
where $\Sigma (s)=3,1,2,0$ for $s=0,1,2,3$. We also note that $\overline{C}_6^3 = \mathcal{I}$~\cite{PhysRevB.100.075125,PhysRevB.104.054401}. 

From~\eqref{U1cellT1angle}-\eqref{U1celltimeangle}, Table~\ref{table:spinliquid}, and from~\eqref{regularU1cellT1angle}-\eqref{regularU1celltimeangle}, and~\eqref{regularU1sublatticeangle}, we see that $G_\mathcal{T}$ of both PSGs agree, while $G_{T_1}$, $G_{T_2}$, and $G_{T_3}$ agree upon the identification of $\theta_{T_2 T_1}$ with $\chi_1$. For the point group generators, let the gauge transformation parts of $(G_{\overline{C}_6} \overline{C}_6)^4$ be $(G_{\overline{C}_6} \overline{C}_6)^4 (G_S S) (G_{\overline{C}_6} \overline{C}_6)^{-1}$ by $G_{C_3}'$ and $G_\sigma'$ [which are expressed in terms of $\phi_{C_3, \sigma}'$ and $n_{C_3, \sigma}' $ according to~\eqref{U1gaugetransformform}] respectively. 

Let $i=(r_1,r_2,r_3;s)$. We start with
\begin{equation} \label{regularU1cellC6squareangle}
\begin{aligned}[b]
\phi_{\overline{C}_6^2} (i) &= \phi_{\overline{C}_6} (i) + \phi_{\overline{C}_6} (\overline{C}_6^{-1} (i)) + n_{\overline{C}_6} \pi \\
& = \chi_1 r_3 (r_1 - r_2) + \chi_1 (\delta_{s,1} + \delta_{s,3}) + n_{\overline{C}_6} \pi ,
\end{aligned}
\end{equation}
which further yields
\begin{equation}
\begin{aligned}[b]
\phi_{C_3}' (i) & = \phi_{\overline{C}_6^2} (i) + \phi_{\overline{C}_6^2} (\overline{C}_6^{-2} (i)) \\
& = - \chi_1 r_1 (r_2 - r_3) + \chi_1 (\delta_{s,1} + \delta_{s,2}) , n_{C_3}'=0 .
\end{aligned}
\end{equation}
The uniform additive factor $n_{\overline{C}_6} \pi$ in~\eqref{regularU1cellC6squareangle} can be further dropped without penalty. We proceed to calculate
\begin{equation}
\begin{aligned}[b]
\phi_\mathcal{I} (i) &= \phi_{\overline{C}_6} (i) + \phi_{\overline{C}_6^{2}} (\overline{C}_6^{-1} (i)) , \\
n_\mathcal{I} &= n_{\overline{C}_6} \in \lbrace 0,1 \rbrace ,
\end{aligned}
\end{equation}
which gives
\begingroup
\allowdisplaybreaks
\begin{subequations}
\begin{align}
& \phi_\mathcal{I} (r_1,r_2,r_3;0) = 0 , \\
& \phi_\mathcal{I} (r_1,r_2,r_3;1) = \chi_1 (r_2 - r_3 +1), \\
& \phi_\mathcal{I} (r_1,r_2,r_3;2) = \chi_1 (r_3 + 1) , \\
& \phi_\mathcal{I} (r_1,r_2,r_3;3) = \chi_1 .
\end{align}
\end{subequations}
\endgroup
Then, we calculate
\begin{equation}
\begin{aligned}[b]
\phi_\Sigma (i) = & \: \phi_S (i) + \phi_\mathcal{I} (S^{-1} (i)) \\
= & \: \chi_1 \left[ \frac{r_1 (r_1 + 1)}{2} - \frac{r_2 (r_2 + 1)}{2} - r_1 r_2 \right] \\ & + \chi_1 + \chi_{\overline{C}_6 S} \delta_{s,1} , \\
n_\Sigma = & \: (n_S + n_\mathcal{I}) \, \mathrm{mod} \, 2 \in \lbrace 0,1 \rbrace ,
\end{aligned}
\end{equation}
Finally, we calculate
\begin{equation}
\begin{aligned}[b]
\phi_\sigma' (i) = & \: \phi_{C_3}' (i) + \phi_\Sigma (C_3^{-1} (i)) - \phi_{C_3}' (\sigma^{-1} (i)) \\
= & - \chi_1 \left[ \frac{r_2 (r_2 -1)}{2} - \frac{r_3 (r_3 -1)}{2} + r_2 r_3 \right] \\
& + \chi_1 (\delta_{s,2} + \delta_{s,3}) + \chi_{\overline{C}_6 S} \delta_{s,2}, \\
n_\sigma' = & \: n_\Sigma \in \lbrace 0,1 \rbrace ,
\end{aligned}
\end{equation}
We further add $\chi_{\overline{C}_6 S} - \chi_1$ to $\phi_\sigma' (i)$ uniformly, and then perform a sublattice dependent gauge transformation $W_{0} = \exp (i \chi_{\overline{C}_6 S} \tau_3)$, $W_{1}=\exp (i \chi_1 \tau_3)$, $W_{2,3}=1$ so that
\begingroup
\allowdisplaybreaks
\begin{subequations}
\begin{align}
\phi_{C_3}' (r_1,r_2,r_3;s)  = & - \chi_1 r_1 (r_2 - r_3) , n_{C_3}'=0 ,  \label{regularU1C3angle} \\
\begin{split} \label{regularU1sigmaangle}
\phi_\sigma' (r_1,r_2,r_3;s) = & - \chi_1 \bigg[ \frac{r_2 (r_2 -1)}{2} - \frac{r_3 (r_3 -1)}{2} \\ & + r_2 r_3 \bigg] +\chi_{\overline{C}_6 S} \delta_{s,3}, n_\sigma' \in \lbrace 0,1 \rbrace , 
\end{split}
\end{align}
\end{subequations}
\endgroup
without affecting other gauges. 

We can now compare~\eqref{U1cellC3angle} and~\eqref{U1cellsigmaangle} with~\eqref{regularU1C3angle} and~\eqref{regularU1sigmaangle} respectively, and identify $\chi_{\overline{C}_6 S}$ with $p_{\sigma C_3}$ in Table~\ref{table:spinliquid}. We conclude that the 16 $U(1)$ spin liquids of the regular pyrochlore lattice~\cite{PhysRevB.104.054401} are continuously connected to the 8 $U(1)$ spin liquids of the breathing pyrochlore lattice with $\theta_{\sigma T_2}=0$, in the fashion of a two-to-one mapping defined by $n_\sigma = ( n_{\overline{C}_6} + n_S )$ mod 2. We see that the additional symmetries of the regular pyrochlore lattice, on the one hand forces $\theta_{\sigma T_2} = 0$, on the other hand give rise to two $\mathbb{Z}_2$ variables $n_{\overline{C}_6}$ and $n_S$ instead of the sole $n_\sigma$ in the breathing pyrochlore lattice. Therefore, while the number of $U(1)$ spin liquids are same in both lattices, they are not in a one-to-one correspondence.

\section{\label{appendix:analyticalsolution}Analytical solution of mean field theory}

For the $U(1)_\pi$ state, diagonalizing $d_\mathbf{k}$ yields the $16$ eigenvalues $\varepsilon_0 (\mathbf{k})$ \{8\}, $\varepsilon_{++} (\mathbf{k})$ \{2\}, $\varepsilon_{+-} (\mathbf{k})$ \{2\}, $\varepsilon_{-+} (\mathbf{k})$ \{2\}, $\varepsilon_{--} (\mathbf{k})$ \{2\}, where each curly bracket indicates the number of times that the corresponding eigenvalue appears,
\begingroup
\allowdisplaybreaks
\begin{subequations}
\begin{align}
\varepsilon_0 (\mathbf{k}) &= - \tilde{\chi}_1 - \tilde{\chi}_2 , \\
\begin{split}
\varepsilon_{+ \pm} (\mathbf{k}) &= \tilde{\chi}_1 + \tilde{\chi}_2 + \bigg[ 4 \tilde{\chi}_1^2 + 4 \tilde{\chi}_2^2 - 4 \tilde{\chi}_1 \tilde{\chi}_2 \\
& \qquad \qquad \quad \: \: \pm \lvert \tilde{\chi}_1 \tilde{\chi}_2 \rvert \sqrt{12 + 2 f (\mathbf{k})} \bigg]^{1/2} ,
\end{split} \\
\begin{split}
\varepsilon_{- \pm} (\mathbf{k}) &= \tilde{\chi}_1 + \tilde{\chi}_2 - \bigg[ 4 \tilde{\chi}_1^2 + 4 \tilde{\chi}_2^2 - 4 \tilde{\chi}_1 \tilde{\chi}_2 \\
& \qquad \qquad \quad \: \: \pm \lvert \tilde{\chi}_1 \tilde{\chi}_2 \rvert \sqrt{12 + 2 f (\mathbf{k})} \bigg]^{1/2} ,
\end{split}
\end{align}
\end{subequations}
\endgroup
and
\begin{equation}
\begin{aligned}[b]
f (\mathbf{k}) = & - \cos (2 k_1 - k_2) - \cos (k_2 - k_3) - \cos (k_3 - 2 k_1) \\
& + \cos (2 k_1) + \cos k_2 + \cos k_3
\, , \qquad k_i \in [0, 2 \pi) .
\end{aligned}
\end{equation}
Note that the function $f (\mathbf{k})$ defined here has nothing to do with that in Sec.~\ref{section:analyticalsolution}. We merely recycled the notation. $f (\mathbf{k})$ has maximum and minimum of $2$ and $-6$ respectively. 

The analysis proceeds very much along the same lines as in Sec.~\ref{section:analyticalsolution}, so we will only outline the steps and omit the details. Let us choose $\tilde{\chi_1}>0$ without loss of generality. We first assume that $\tilde{\chi}_2 > 0$. One can show that $\varepsilon_{+ \pm} (\mathbf{k}) , \varepsilon_{- \pm} (\mathbf{k}) > \varepsilon_0 (\mathbf{k})$ for all $\mathbf{k}$, so the dispersing bands are well separated from, and higher in energy than, the flat bands throughout the Brillouin zone. Then, filling the lower half of the energy eigenstates, i.e.,~all the flat bands, the total energy of a system with $N \times \mathcal{N}$ sites is given by
\begin{equation} \label{U1pifluxtotalenergy}
\begin{aligned}[b]
E_S & = \sum_\mathbf{k} \left[ 16 \varepsilon_0 (\mathbf{k}) + 24 \left( \frac{J_1}{4} \chi_1^2 + \frac{J_2}{4} \chi_2^2 \right) \right] \\
& = \sum_\mathbf{k} \left[ 16 \left( \frac{J_1}{4} \chi_1 + \frac{J_2}{4} \chi_2 \right) + 24 \left( \frac{J_1}{4} \chi_1^2 + \frac{J_2}{4} \chi_2^2 \right) \right] .
\end{aligned}
\end{equation}

Next, we assume that $\tilde{\chi}_2 < 0$, and without loss of generality $\lvert \chi_1 \rvert > \lvert \chi_2 \rvert$. In this case, the dispersing bands are well separated from the flat bands, with the $+ \pm$ ($- \pm$) bands lying above (below) the flat bands. One can show that, by filling half of the lower energy eigenstates, i.e.,~2 $-+$ bands, 2 $--$ bands, and 12 of the flat bands, the total energy $E_A$ thus obtained is strictly greater than $E_S$ in~\eqref{U1pifluxtotalenergy}. Therefore, we can exclude the case with $\tilde{\chi}_2 < 0$. 

Finally, minimizing \eqref{U1pifluxtotalenergy} with respect to $\chi_1$ and $\chi_2$ yields $\chi_{1,2} = - 1/3$ and the ground state energy per site $-(J_1+J_2)/24$.

\section{\label{appendix:effectivefieldtheory}Low energy effective field theory}

In this appendix, we provide details of the field theoretic treatment of the $U(1)_0$ state, from which we derive the low temperature heat capacity. Starting from \eqref{lagrangian} and \eqref{action}, we integrate out the spinons to obtain an effective action (of the gauge field),
\begin{equation}
\begin{aligned}[b]
Z & = \int D a \exp \left \lbrace 2 \ln \det \left[ \partial_\tau - i a_0 + \frac{1}{2m} ( - i \nabla - \mathbf{a})^2 \right] \right \rbrace \\
& \equiv \int D a \, e^{- S_\mathrm{eff} (a)} .
\end{aligned}
\end{equation}
Using the random phase approximation (RPA), we expand the logarithm up to one loop order~\cite{RevModPhys.78.17,PhysRevB.76.165104,PhysRevB.76.235124,nagaosatextbook}. RPA is formally justified by introducing $N$ species of fermions for large $N$, so that terms beyond one loop order and two external legs carry higher powers of $1/N$~\cite{PhysRevB.39.8988,PhysRevB.46.5621,POLCHINSKI1994617}. We merely assume here that such a large $N$ formulation can be extended to our case of $N=1$~\cite{PhysRevB.104.054401}. 

\begin{figure}
\subfloat[]{\label{figure:vertexa}
\includegraphics[scale=0.2]{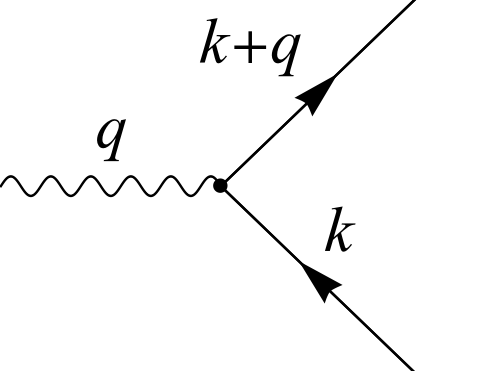}} \qquad \quad
\subfloat[]{\label{figure:vertexb}
\includegraphics[scale=0.2]{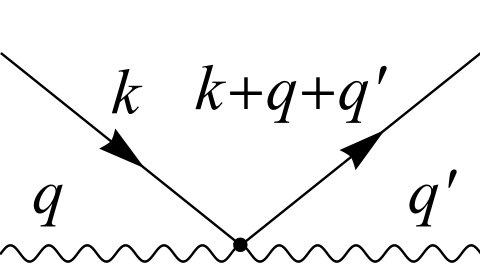}} \\
\subfloat[]{\label{figure:propagatora}
\includegraphics[scale=0.2]{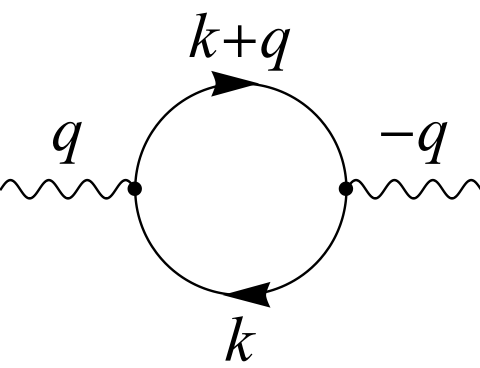}} \qquad \quad
\subfloat[]{\label{figure:propagatorb}
\includegraphics[scale=0.2]{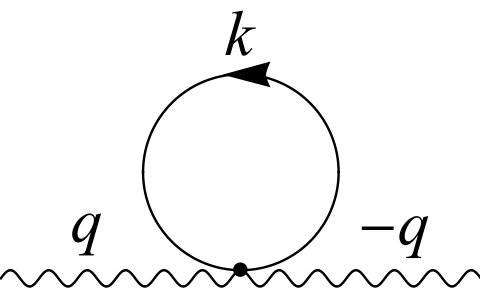}}
\caption{(a,b) The two interaction vertices in the low energy effective field theory of the $U(1)_0$ spin liquid, in which quadratically dispersing spinon excitations are coupled to a $U(1)$ gauge field. Straight (wavy) lines represent spinon (gauge) fields. (c,d) The two diagrams that contribute to, or, more precisely, generate the photon propagator within the random phase approximation. These are generated by (a) and (b) respectively. In each diagram,  the momentum of the photon is labeled such that it flows into the vertex.}
\end{figure}

RPA can be treated with the standard diagrammatic perturbation technique~\cite{nagaosatextbook}, as follows. Going back to~\eqref{lagrangian}, using the Coulomb gauge $\nabla \cdot \mathbf{a}=0$ and performing a Fourier transform,
\begingroup
\allowdisplaybreaks
\begin{subequations}
\begin{align}
\psi_\sigma (\mathbf{r}, \tau) &= \frac{1}{\sqrt{\beta V}} \sum_{\mathbf{k} n} \psi_\sigma (\mathbf{k},\omega_n) e^{i (- \omega_n \tau + \mathbf{k} \cdot \mathbf{r})} , \\
\mathbf{a} (\mathbf{r}, \tau) &= \frac{1}{\sqrt{\beta V}} \sum_{\mathbf{q} l} \mathbf{a} (\mathbf{q},\nu_l) e^{i (- \nu_l \tau + \mathbf{q} \cdot \mathbf{r})} ,
\end{align}
\end{subequations}
\endgroup
where $\beta=1/T$ is the inverse temperature, $V$ is the volume of the system, and $\omega_n = 2 \pi (n + 1) / \beta$ and $\nu_l = 2 \pi l / \beta$ are the fermionic and bosonic Matsubara frequencies respectively, the interaction Lagrangian in momentum space reads
\begin{equation}
\begin{aligned}[b]
L_\mathrm{int} = & \frac{1}{\sqrt{\beta V}} \sum_{kq} \bar{\psi}_\sigma (k+q) \frac{(\mathbf{k} - \mathbf{k}_\parallel) \cdot \mathbf{a} (q)}{m} \psi_\sigma (k) \\
& + \frac{1}{\beta V} \sum_{kqq'} \bar{\psi}_\sigma (k+q+q') \frac{\mathbf{a} (q) \cdot \mathbf{a} (q')}{m} \psi_\sigma (k) , 
\end{aligned}
\end{equation}
where $k = (\mathbf{k} , \omega_n)$ etc are the four-momenta, and $\mathbf{k}_\parallel$ is the component of $\mathbf{k}$ parallel to $\mathbf{q}$. The interaction vertices, as well as the two diagrams that contribute at one loop level, are depicted in Figs.~\ref{figure:vertexa}-\ref{figure:propagatorb}. We also assume that the temporal component $a_0$ of the gauge field is screened out by spinon density fluctuations~\cite{PhysRevB.46.5621,PhysRevB.50.17917,PhysRevB.76.165104} so that it can be neglected.

We first calculate the diagram in Fig.~\ref{figure:propagatora}, which we call $\Pi^a$. Let us choose a coordinate system in which $\mathbf{q}$ aligns in the $z$ direction, so that $a_z=0$ by the Coulomb gauge. For the transverse components of the gauge field, $\Pi_{ij}^a$ is nonzero only when $ij=xx$ and $yy$,
\begingroup
\allowdisplaybreaks
\begin{equation}
\begin{aligned}[b]
\Pi_{xx}^a (q) &= \frac{-2}{m^2 \beta V} \sum_{\mathbf{k} n} G (k+q) G (k) k_x^2 \\
&= \frac{- 2}{m^2 \beta} \sum_n \int_\mathbf{k} \frac{1}{i (\omega_n + \nu_l) - \varepsilon_{\mathbf{k}+\mathbf{q}}} \frac{1}{i \omega_n - \varepsilon_\mathbf{k}} k_x^2,
\end{aligned}
\end{equation}
\endgroup
where the minus sign comes from the fermion loop, the factor of 2 is due to the two spin species, $G (k)$ is the free fermion Green function, $\int_\mathbf{k}$ is the shorthand notation for $\int \mathrm{d}^3 \mathbf{k}/(2 \pi)^3$, and $\varepsilon_\mathbf{k} \equiv \lvert \mathbf{k} \rvert^2 / 2m$. The corresponding expression for $\Pi_{yy}^a (q)$ is obtained with the replacement $x \longrightarrow y$. Using the contour integral method~\cite{nagaosatextbook,mahantextbook,colemantextbook} to perform the sum over $\omega_n$,
\begin{equation} \label{photonpropagatoraraw}
\begin{aligned}[b]
\Pi_{xx}^a (q) &= -\frac{2}{m^2} \int_\mathbf{k} \frac{[ f (\varepsilon_{\mathbf{k} + \mathbf{q}}) - f (\varepsilon_\mathbf{k}) ] k_x^2} {- i \nu_l + (2 \mathbf{k} \cdot \mathbf{q} + \lvert \mathbf{q} \rvert^2) / 2m} , \\
& = -\frac{2}{m^2} \int_\mathbf{k} \frac{[ f (\varepsilon_{\mathbf{k} + \mathbf{q}/2}) - f (\varepsilon_{\mathbf{k} - \mathbf{q}/2}) ] k_x^2} {- i \nu_l + \mathbf{k} \cdot \mathbf{q} / m} ,
\end{aligned}
\end{equation}
where $f (\varepsilon) = 1 / [\exp (\beta \varepsilon) + 1]$ is the Fermi Dirac distribution, and we have shifted $\mathbf{k}$ by ${-\mathbf{q}/2}$ in the second line to obtain a more symmetric expression. It is very difficult if not impossible to evaluate the integral, so we study the small $\lvert \mathbf{q} \rvert$ limit. More precisely, we expand
\begin{equation} \label{taylorexpansion}
f (\varepsilon_{\mathbf{k} \pm \mathbf{q}/2}) = \sum_{n=0}^{\infty} \frac{1}{n!} \frac{\partial^n f (\varepsilon_\mathbf{k})}{\partial \varepsilon_\mathbf{k}^n} \left(\pm \frac{\mathbf{k} \cdot \mathbf{q}}{m} + \frac{\lvert \mathbf{q} \rvert^2}{2m} \right)^n
\end{equation}
with $\lvert \mathbf{q} \rvert/{\sqrt{mT}} \ll 1$ as the small parameter. The inclusion of $T$ in defining the small parameter is important because each derivative of $f (\varepsilon)$ with respect to $\varepsilon$ brings down a factor of $1/T$, so we require higher powers of $\lvert \mathbf{q} \rvert/\sqrt{mT}$ to be less important than lower powers to justify a truncation of the expansion \eqref{taylorexpansion} at finite $n$. (There should also be a constraint $\lvert \mathbf{k} \rvert \lesssim \sqrt{mT}$. Nevertheless, large momentum modes are thermally suppressed by the Fermi Dirac distribution function, so we will take the upper limits of the $\lvert \mathbf{k} \rvert$ integrals below to infinity.) Note that this is unlike the case of a spinon Fermi surface, where one can expand $\mathbf{q}$ against the Fermi wavevector $\mathbf{k}_\mathrm{F}$, i.e.,~taking $\lvert \mathbf{q} \rvert / k_\mathrm{F} \ll 1$ as the small parameter, without a direct comparison with $T$, and use a zero temperature approximation for $f (\varepsilon)$ which is effectively a step function~\cite{PhysRevB.76.165104,nagaosatextbook}. 

Let us first examine the case when $l \neq 0$. Expanding the numerator in~\eqref{photonpropagatoraraw} to the lowest nontrivial order in $\mathbf{q}$,
\begin{equation} \label{photonpropagatoradynamicsmallq}
\begin{aligned}[b]
\Pi_{xx}^a (q) & \approx -\frac{2}{m^2} \int_\mathbf{k} \frac{ \partial f (\varepsilon_\mathbf{k}) / \partial \varepsilon_\mathbf{k} \times \mathbf{k} \cdot \mathbf{q} / m \times k_x^2} {- i \nu_l + \mathbf{k} \cdot \mathbf{q} / m} \\
& = -\frac{2}{m^2} \int_\mathbf{k} \frac{\partial f (\varepsilon_\mathbf{k})}{\partial \varepsilon_\mathbf{k}} \frac{(\mathbf{k} \cdot \mathbf{q} / m)^2}{\nu_l^2 + (\mathbf{k} \cdot \mathbf{q} / m)^2} k_x^2 .
\end{aligned}
\end{equation}
Since $\nu_l \equiv 2 \pi l T$ and $\mathbf{k} \cdot \mathbf{q} / m T \ll 1$, we have $\mathbf{k} \cdot \mathbf{q} / m \ll \nu_l$ for $l \neq 0$, which allows us to approximate
\begingroup
\allowdisplaybreaks
\begin{equation} \label{photonpropagatoradynamicsmallqcontinue}
\begin{aligned}[b]
& \Pi_{xx}^a (\mathbf{q}, \nu_l \neq 0) \approx \frac{-2}{m^2} \int_\mathbf{k} \frac{\partial f (\varepsilon_\mathbf{k})}{\partial \varepsilon_\mathbf{k}} \left( \frac{\mathbf{k} \cdot \mathbf{q}}{m \nu_l} \right)^2 k_x^2 \\
& = \frac{-2/m^4}{(2 \pi)^3} \int_0^{2 \pi} \mathrm{d} \phi \, \cos^2 \phi \int_{-1}^{+1} \mathrm{d} (\cos \theta) \, \sin^2 \theta \cos^2 \theta \\
& \quad \times \int_0^{\infty} \mathrm{d} \lvert \mathbf{k} \rvert \, \lvert \mathbf{k} \rvert^6 \frac{\partial f (\varepsilon_\mathbf{k})}{\partial \varepsilon_\mathbf{k}} \frac{\lvert \mathbf{q} \rvert^2}{\nu_l^2} \\
& = \frac{64 \pi \sqrt{2/m}}{15 (2 \pi)^3} T^{5/2} \frac{\lvert \mathbf{q} \rvert^2}{\nu_l^2} \int_0^\infty \mathrm{d} x \, \frac{x^6 e^{x^2}}{(e^{x^2}+1)^2} \equiv c_3 T^{5/2} \frac{\lvert \mathbf{q} \rvert^2}{\nu_l^2} 
\, .
\end{aligned}
\end{equation}
\endgroup
The dimensionless $x$ integral evaluates to $1.44$ (to three significant figures). 

On the other hand, the diagram in Fig.~\ref{figure:propagatorb}, which we call $\Pi^b$, is much easier to evaluate than $\Pi^a$. As before, $\Pi_{ij}^b$ is nonzero only when $ij=xx$ and $yy$. Contracting the fermion lines forces $q=q'$, so
\begingroup
\allowdisplaybreaks
\begin{equation} \label{photonpropagatorbdynamic}
\begin{aligned}[b]
\Pi_{xx}^b (q) &= \frac{- 2}{m \beta V} \sum_{n \mathbf{k}} G (k) = - \frac{2}{m} \int_\mathbf{k} f (\mathbf{k}) \\
& = - \frac{8 \pi \sqrt{8m}}{(2 \pi)^3} T^{3/2} \int_0^\infty \mathrm{d} x \, \frac{x^2}{e^{x^2}+1} \equiv - c_2 T^{3/2} ,
\end{aligned}
\end{equation}
\endgroup
which is independent of $q$. The dimensionless $x$ integral evaluates to $0.339$ (to three significant figures). Adding up~\eqref{photonpropagatoradynamicsmallqcontinue} and~\eqref{photonpropagatorbdynamic} then yields~\eqref{photonpropagatordynamic}. 

Next, we examine the case when $l=0$. Expanding the numerator in~\eqref{photonpropagatoraraw} to the lowest nontrivial order in $\mathbf{q}$,
\begin{equation} \label{photonpropagatorastaticsmallq}
\begin{aligned}[b]
& \Pi_{xx}^a (\mathbf{q} , \nu_l=0) = -\frac{2}{m^2} \int_\mathbf{k} \frac{ [f (\varepsilon_{\mathbf{k}+\mathbf{q}/2}) - f (\varepsilon_{\mathbf{k}-\mathbf{q}/2})] k_x^2} {\mathbf{k} \cdot \mathbf{q} / m} \\
& \approx -\frac{2}{m^2} \int_\mathbf{k} \left[ \frac{\partial f (\varepsilon_\mathbf{k})}{\partial \varepsilon_\mathbf{k}} + \frac{\partial ^2 f (\varepsilon_\mathbf{k})}{\partial \varepsilon_\mathbf{k}^2} \frac{\lvert \mathbf{q} \rvert^2}{8m} + \frac{\partial ^3 f (\varepsilon_\mathbf{k})}{\partial \varepsilon_\mathbf{k}^3} \frac{( \mathbf{k} \cdot \mathbf{q} )^2}{24 m^2} \right] k_x^2 .
\end{aligned}
\end{equation}
To evaluate the terms in the square bracket, we note that
\begin{equation}
\varepsilon \frac{\partial f (\varepsilon)}{\partial \varepsilon} = - T \frac{\partial f (\varepsilon)}{\partial T} ,
\end{equation}
which allows us to replace the derivative with respect to energy by one with respect to temperature, and pull the temperature derivative out of the integral over $\mathbf{k}$. The first derivative term evaluates to
\begingroup
\allowdisplaybreaks
\begin{equation} \label{staticfirstderivative}
\begin{aligned}[b]
& - \frac{2}{m^2} \int_\mathbf{k} \frac{\partial f (\varepsilon_\mathbf{k})}{\partial \varepsilon_\mathbf{k}} \lvert \mathbf{k} \rvert^2 \sin^2 \theta \cos^2 \phi \\
& = -\frac{4}{m} \frac{4 \pi / 3}{(2 \pi)^3} \int_0^\infty \mathrm{d} \lvert \mathbf{k} \rvert \, \lvert \mathbf{k} \rvert^2 \varepsilon_\mathbf{k} \frac{\partial f (\varepsilon_\mathbf{k})}{\partial \varepsilon_\mathbf{k}} \\
& = \frac{16 \pi / 3}{(2 \pi)^3 m} T \frac{\partial}{\partial T} (2 m T)^{3/2} \int_0^\infty \mathrm{d} x \, \frac{x^2}{e^{x^2}+1} \\
& = \frac{8 \pi \sqrt{8m}}{(2 \pi)^3} T^{3/2} \int_0^\infty \mathrm{d} x \, \frac{x^2}{e^{x^2}+1} = c_2 T^{3/2} 
\, ,
\end{aligned}
\end{equation}
\endgroup
which is same as~\eqref{photonpropagatorbdynamic} apart from a minus sign. The second derivative term evaluates to
\begingroup
\allowdisplaybreaks
\begin{equation} \label{staticsecondderivative}
\begin{aligned}[b]
& - \frac{2}{m^2} \int_\mathbf{k} \frac{\partial^2 f (\varepsilon_\mathbf{k})}{\partial \varepsilon_\mathbf{k}^2}  \frac{\lvert \mathbf{q} \rvert^2}{8 m} \lvert \mathbf{k} \rvert^2 \sin^2 \theta \cos^2 \phi \\
&= - \frac{\lvert \mathbf{q} \rvert^2}{m} \frac{4 \pi / 3}{ (2 \pi)^3} \int_0^{\infty} \mathrm{d} \lvert \mathbf{k} \rvert \, \varepsilon_\mathbf{k}^2 \frac{\partial^2 f (\varepsilon_\mathbf{k})}{\partial \varepsilon_\mathbf{k}^2} \\
&= - \frac{\lvert \mathbf{q} \rvert^2}{m} \frac{4 \pi / 3}{ (2 \pi)^3} \left( 2 T \frac{\partial}{\partial T} + T^2 \frac{\partial^2}{\partial T^2} \right) \int_0^\infty \frac{\sqrt{2mT} \mathrm{d} x}{e^{x^2}+1} \\
& = -\frac{\sqrt{2/m}}{8 \pi^2} \sqrt{T} \lvert \mathbf{q} \rvert^2 \int_0^\infty \frac{\mathrm{d} x}{e^{x^2}+1} .
\end{aligned}
\end{equation}
\endgroup
The dimensionless $x$ integral evaluates to 0.536 (to three significant figures). Similarly, the third derivative term evaluates to
\begin{equation} \label{staticthirdderivative}
\begin{aligned}[b]
& -\frac{2}{m^2} \int_\mathbf{k} \frac{\partial^3 f (\varepsilon_\mathbf{k})}{\partial \varepsilon_\mathbf{k}^3} \frac{\lvert \mathbf{q} \rvert^2}{24 m^2} \lvert \mathbf{k} \rvert^4 \cos^2 \theta \sin^2 \theta \cos^2 \phi \\
& = \frac{\sqrt{2/m}}{24 \pi^2} \sqrt{T} \lvert \mathbf{q} \rvert^2 \int_0^\infty \, \frac{\mathrm{d} x}{e^{x^2}+1} .
\end{aligned}
\end{equation}

On the other hand, $\Pi^b (q)$ is still given by~\eqref{photonpropagatorbdynamic}, which cancels out~\eqref{staticfirstderivative}. We are left with~\eqref{staticsecondderivative} and~\eqref{staticthirdderivative}, and adding them up yields~\eqref{photonpropagatorstatic}. 

We can then write down the effective Lagrangian of the $U(1)$ gauge field~\eqref{effectivephotonlagrangian}, the corresponding partition function~\eqref{effectivepartitionfunction}, and the free energy~\eqref{freeenergy}. We now calculate the free energy contribution~\eqref{freeenergydynamic} from the $l \neq 0$ modes, as follows. Changing the summation over $\mathbf{q}$ to an integral, the free energy density reads
\begin{equation} \label{freeenergydensitydynamic}
\begin{aligned}[b]
f_\mathrm{dyn} & \equiv \frac{F_\mathrm{dyn}}{V} = T \sum_{l \neq 0} \int_\mathbf{q}  \ln \bigg( c_2 T^{3/2} - c_3 T^{5/2} \frac{\lvert \mathbf{q} \rvert^2}{\nu_l^2} \bigg) \\
& \approx T \sum_{l \neq 0} \int_\mathbf{q} \bigg[ \ln \left( c_2 T^{3/2} \right) - c_3' T \frac{\lvert \mathbf{q} \rvert^2}{\nu_l^2} \bigg] ,
\end{aligned}
\end{equation}
where $c_3' \equiv c_3/c_2$. Since we have assumed $\lvert \mathbf{q} \rvert \ll \sqrt{mT}$ from the beginning, the $\mathbf{q}$ integral should have an upper limit of the form $x_0 \sqrt{mT}$~\cite{PhysRevB.46.5621}, where $x_0 < 1$ is some small number, whose precise value is not important for the current analysis. Performing the integration over $\mathbf{q}$ yields
\begin{equation} \label{freeenergydensitydynamicraw}
f_\mathrm{dyn} = \frac{8 \pi T}{(2 \pi)^3} \sum_{l \in \mathbb{N}} \bigg[ \frac{(x_0 \sqrt{mT})^3}{3} \ln \left( c_2 T^{3/2} \right) - c_3' T \frac{(x_0 \sqrt{mT})^5}{5 \nu_l^2}  \bigg] .
\end{equation}
The first term is independent of $\nu_l$, so the summation over natural numbers yields a divergence, which can be regularized using the Riemann zeta function~\cite{lainetextbook},
\begin{equation} \label{zetafunction}
\zeta (s) = \sum_{n=1}^\infty n^{-s} = \frac{1}{\Gamma (s)} \int_0^\infty \mathrm{d} x \, \frac{x^{s-1}}{e^x-1} ,
\end{equation}
where
\begin{equation} \label{gammafunction}
\Gamma (s) = \int_0^\infty \mathrm{d} x \, x^{s-1} e^{-x}
\end{equation}
is the gamma function. In particular, we use $\zeta (0) = - 1/2$ and $\zeta(2) = \pi^2/6$ for the first and second terms in \eqref{freeenergydensitydynamicraw} respectively, and we obtain
\begin{equation} \label{freeenergydensitydynamicregularize}
f_\mathrm{dyn} = - \frac{8 \pi T}{(2 \pi)^3} \bigg[ \frac{( x_0 \sqrt{mT} )^3  \ln ( c_2 T^{3/2} )}{6} + \frac{c_3' (x_0 \sqrt{mT})^5}{120 T} \bigg] .
\end{equation}

We then calculate the free energy contribution~\eqref{freeenergystatic} from the $l=0$ mode. The free energy density reads
\begin{equation} \label{freeenergydensitystatic}
\begin{aligned}[b]
f_\mathrm{sta} & \equiv \frac{F_\mathrm{sta}}{V} = T \int_\mathbf{q} \ln \left( c_1 \sqrt{T} \lvert \mathbf{q} \rvert^2 \right) \\
&= \frac{4 \pi T}{(2 \pi)^3} \frac{(x_0 \sqrt{mT})^3}{3} \bigg[ \ln \left( c_1 x_0^2 m T^{3/2} \right) - \frac{2}{3} \bigg] .
\end{aligned}
\end{equation}

Adding up~\eqref{freeenergydensitydynamicregularize} and~\eqref{freeenergydensitystatic} yields
\begin{equation} \label{freeenergydensitytotal}
\begin{aligned}[b]
f &\equiv f_\mathrm{sta} + f_\mathrm{dyn} \\
&= \frac{(x_0 \sqrt{m})^3}{6 \pi^2}\bigg[ \ln \left( 0.0659 x_0^2 \right) - 0.0567 x_0^2 - \frac{2}{3} \bigg] T^{5/2} .
\end{aligned}
\end{equation}
With $x_0 \lesssim 1$, the square bracket in~\eqref{freeenergydensitytotal} evaluates to a negative number. The volumetric heat capacity is readily obtained as
\begin{equation}
\frac{C(T)}{V} = - T \frac{\partial^2 f}{\partial T^2} \sim T^{3/2} .
\end{equation}

\section{\label{appendix:dzyaloshinskiimoriya}Parton representation of Dzyaloshinskii-Moriya interaction}

We first introduce the triplet hopping and pairing channels, $\hat{\mathbf{E}}_{ij}$ and $\hat{\mathbf{D}}_{ij}$~\cite{PhysRevB.80.064410,PhysRevB.85.224428,PhysRevB.86.224417,s41598-019-47517-6},
\begin{subequations}
\begin{align}
\hat{E}_{ij}^\lambda &= \sum_{\alpha \beta} f_{i \alpha}^\dagger [\sigma^\lambda]_{\alpha \beta} f_{j \beta} , \\
\hat{D}_{ij}^\lambda &= \sum_{\alpha \beta} f_{i \alpha} [i \sigma^y \sigma^\lambda]_{\alpha \beta} f_{j \beta} .
\end{align}
\end{subequations}

Let $(\lambda, \mu, \nu)$ be a cyclic permutation of $(x,y,z)$. The Dzyaloshinskii-Moriya interaction in~\eqref{JDmodel} involves spin products in the combination
\begin{equation}
\pm \frac{\lvert D \rvert}{\sqrt{2}} \left( S_i^\mu S_j^\nu - S_i^\nu S_j^\mu \right) ,
\end{equation}
which, with the parton representation of spins \eqref{partonrepresentation}, can be expressed in terms of products of bond operators,
\begin{equation} \label{dzyaloshinskiimoriyapartonrepresentation}
\begin{aligned}
- \frac{\lvert D \rvert}{8 \sqrt{2}} \bigg[ & (\hat{\chi}_{ij} \pm i \hat{E}_{ij}^\lambda)^\dagger (\hat{\chi}_{ij} \pm i \hat{E}_{ij}^\lambda) + (\hat{\Delta}_{ij} \pm i \hat{D}_{ij}^\lambda)^\dagger (\hat{\Delta}_{ij} \pm i \hat{D}_{ij}^\lambda) \\
& + \hat{E}_{ij}^{\mu \dagger} \hat{E}_{ij}^\mu + \hat{E}_{ij}^{\nu \dagger} \hat{E}_{ij}^\nu + \hat{D}_{ij}^{\mu \dagger} \hat{D}_{ij}^\mu + \hat{D}_{ij}^{\nu \dagger} \hat{D}_{ij}^\nu \bigg] ,
\end{aligned}
\end{equation}
where we have used the identity $\hat{\chi}_{ij}^\dagger \hat{\chi}_{ij} + \hat{\Delta}_{ij}^\dagger \hat{\Delta}_{ij} = - \hat{\mathbf{E}}_{ij}^\dagger \cdot \hat{\mathbf{E}}_{ij} -  \hat{\mathbf{D}}_{ij}^\dagger \cdot \hat{\mathbf{D}}_{ij}$ (up to some constant) to ensure stability of the resulting mean field Hamiltonian~\cite{s41598-019-47517-6}. From \eqref{dzyaloshinskiimoriyapartonrepresentation}, a mean field decoupling (or, more formally, a Hubbard Stratonovich transformation) $\hat{O}^\dagger \hat{O} \approx \langle \hat{O}^\dagger \rangle \hat{O} + \hat{O}^\dagger \langle \hat{O} \rangle - \langle \hat{O}^\dagger \rangle \langle \hat{O} \rangle$ is now straightforward. For $U(1)$ spin liquids, we set the pairing terms $\Delta_{ij}$ and $\mathbf{D}_{ij}$ to be zero. 

\begin{table}
\caption{\label{table:tripletansatz}The relation between triplet hopping parameters $\mathbf{E}_{ij}$, which are purely imaginary if nonzero, in the $U(1)_0$ state. $s$ and $t$ are the sublattice indices of sites $i$ and $j$ respectively. The corresponding relation for the $U(1)_\pi$ state can be obtained from this table in conjunction with Fig.~\ref{figure:ansatzU1pi}.}
\begin{ruledtabular}
\begin{tabular}{cccc}
$s$ & $t$ & up tetrahedron & down tetrahedron \\ \hline
0 & 1 & $(0,+E_1^y,-E_1^y)$ & $(0,+E_2^y,-E_2^y)$ \\
0 & 2 & $(-E_1^y,0,+E_1^y)$ & $(-E_2^y,0,+E_2^y)$ \\
0 & 3 & $(+E_1^y,-E_1^y,0)$ & $(+E_2^y,-E_2^y,0)$ \\
1 & 2 & $(+E_1^y,+E_1^y,0)$ & $(+E_2^y,+E_2^y,0)$ \\
2 & 3 & $(0,+E_1^y,+E_1^y)$ & $(0,+E_2^y,+E_2^y)$ \\
3 & 1 & $(+E_1^y,0,+E_1^y)$ & $(+E_2^y,0,+E_2^y)$
\end{tabular}
\end{ruledtabular}
\end{table}

Similar to~\eqref{singlettrace}, the mean field Hamiltonian for the triplet channels can be written in terms of a trace~\cite{PhysRevB.95.054404},
\begin{equation} \label{triplettrace}
H^\mathrm{MF}_\mathrm{triplet} = \sum_{ij} \sum_{\lambda \in x,y,z} \mathrm{Tr} \left[ \sigma^\lambda \Psi_i u_{ij}^\lambda \Psi_j^\dagger \right] ,
\end{equation}
where the $2 \times 2$ matrix $u_{ij}^\lambda$ contains the coefficients of $\hat{E}^\lambda_{ij}$ and $\hat{D}^\lambda_{ij}$. For a space group element $X$, PSG constrains the triplet ansatze via
\begin{equation}
u_{X(i)X(j)}^\mu = \sum_\nu O_X^{\mu \nu} G_X (X(i)) u_{ij}^\nu G_X^\dagger (X(j)) ,
\end{equation}
where $O_X \in SO (3)$ encodes the $SU(2)$ spin rotation associated with $X$~\cite{PhysRevB.95.054404}. There are only two nontrivial $O_X$ in our problem,
\begin{equation}
O_{C_3} = \begin{pmatrix} 0 & 0 & 1 \\ 1 & 0 & 0 \\ 0 & 1 & 0 \end{pmatrix} , O_\sigma = \begin{pmatrix} -1 & 0 & 0 \\ 0 & 0 & 1 \\ 0 & 1 & 0 \end{pmatrix} .
\end{equation}
On the other hand, for the time reversal symmetry, 
\begin{equation}
u_{ij}^\lambda = - G_\mathcal{T} (i) u_{ij}^\lambda G_\mathcal{T}^\dagger (j) 
\, .
\end{equation}
An excellent account of the treatment of triplet ansatze can be found in Ref.~\onlinecite{PhysRevB.95.054404}, to which interested readers can refer for more information. See also Ref.~\onlinecite{s41598-019-47517-6}. Let $E_1^y$ ($E_2^y$) be the $y$ component of the triplet hopping parameter on the bond connecting sublattices $0$ and $1$ on an up (down) tetrahedron. Omitting the details of the calculation, we summarize the relation between the triplet ansatze $\mathbf{E}_{ij}=(E_{ij}^x, E_{ij}^y, E_{ij}^z)$ for the $U(1)_0$ state in Table~\ref{table:tripletansatz}, which can be extended to the $U(1)_\pi$ state by enlarging the unit cell and multiplying by $\pm 1$ as prescribed in Fig.~\ref{figure:ansatzU1pi}. In both states, a nonzero $E_{ij}^\lambda$ is purely imaginary due to time reversal symmetry. The relations between the singlet ansatze $\chi_{ij}$ remain the same as before.

\bibliography{reference220606}

\begin{thebibliography}{97}%
\makeatletter
\providecommand \@ifxundefined [1]{%
 \@ifx{#1\undefined}
}%
\providecommand \@ifnum [1]{%
 \ifnum #1\expandafter \@firstoftwo
 \else \expandafter \@secondoftwo
 \fi
}%
\providecommand \@ifx [1]{%
 \ifx #1\expandafter \@firstoftwo
 \else \expandafter \@secondoftwo
 \fi
}%
\providecommand \natexlab [1]{#1}%
\providecommand \enquote  [1]{``#1''}%
\providecommand \bibnamefont  [1]{#1}%
\providecommand \bibfnamefont [1]{#1}%
\providecommand \citenamefont [1]{#1}%
\providecommand \href@noop [0]{\@secondoftwo}%
\providecommand \href [0]{\begingroup \@sanitize@url \@href}%
\providecommand \@href[1]{\@@startlink{#1}\@@href}%
\providecommand \@@href[1]{\endgroup#1\@@endlink}%
\providecommand \@sanitize@url [0]{\catcode `\\12\catcode `\$12\catcode
  `\&12\catcode `\#12\catcode `\^12\catcode `\_12\catcode `\%12\relax}%
\providecommand \@@startlink[1]{}%
\providecommand \@@endlink[0]{}%
\providecommand \url  [0]{\begingroup\@sanitize@url \@url }%
\providecommand \@url [1]{\endgroup\@href {#1}{\urlprefix }}%
\providecommand \urlprefix  [0]{URL }%
\providecommand \Eprint [0]{\href }%
\providecommand \doibase [0]{https://doi.org/}%
\providecommand \selectlanguage [0]{\@gobble}%
\providecommand \bibinfo  [0]{\@secondoftwo}%
\providecommand \bibfield  [0]{\@secondoftwo}%
\providecommand \translation [1]{[#1]}%
\providecommand \BibitemOpen [0]{}%
\providecommand \bibitemStop [0]{}%
\providecommand \bibitemNoStop [0]{.\EOS\space}%
\providecommand \EOS [0]{\spacefactor3000\relax}%
\providecommand \BibitemShut  [1]{\csname bibitem#1\endcsname}%
\let\auto@bib@innerbib\@empty
\bibitem [{\citenamefont {Balents}(2010)}]{nature08917}%
  \BibitemOpen
  \bibfield  {author} {\bibinfo {author} {\bibfnamefont {L.}~\bibnamefont
  {Balents}},\ }\bibfield  {title} {\bibinfo {title} {Spin liquids in
  frustrated magnets},\ }\href {https://doi.org/10.1038/nature08917} {\bibfield
   {journal} {\bibinfo  {journal} {Nature}\ }\textbf {\bibinfo {volume}
  {464}},\ \bibinfo {pages} {199} (\bibinfo {year} {2010})}\BibitemShut
  {NoStop}%
\bibitem [{\citenamefont {Zhou}\ \emph {et~al.}(2017)\citenamefont {Zhou},
  \citenamefont {Kanoda},\ and\ \citenamefont {Ng}}]{RevModPhys.89.025003}%
  \BibitemOpen
  \bibfield  {author} {\bibinfo {author} {\bibfnamefont {Y.}~\bibnamefont
  {Zhou}}, \bibinfo {author} {\bibfnamefont {K.}~\bibnamefont {Kanoda}},\ and\
  \bibinfo {author} {\bibfnamefont {T.-K.}\ \bibnamefont {Ng}},\ }\bibfield
  {title} {\bibinfo {title} {Quantum spin liquid states},\ }\href
  {https://doi.org/10.1103/RevModPhys.89.025003} {\bibfield  {journal}
  {\bibinfo  {journal} {Rev. Mod. Phys.}\ }\textbf {\bibinfo {volume} {89}},\
  \bibinfo {pages} {025003} (\bibinfo {year} {2017})}\BibitemShut {NoStop}%
\bibitem [{\citenamefont {Knolle}\ and\ \citenamefont
  {Moessner}(2019)}]{annurev-conmatphys-031218-013401}%
  \BibitemOpen
  \bibfield  {author} {\bibinfo {author} {\bibfnamefont {J.}~\bibnamefont
  {Knolle}}\ and\ \bibinfo {author} {\bibfnamefont {R.}~\bibnamefont
  {Moessner}},\ }\bibfield  {title} {\bibinfo {title} {A field guide to spin
  liquids},\ }\href {https://doi.org/10.1146/annurev-conmatphys-031218-013401}
  {\bibfield  {journal} {\bibinfo  {journal} {Annual Review of Condensed Matter
  Physics}\ }\textbf {\bibinfo {volume} {10}},\ \bibinfo {pages} {451}
  (\bibinfo {year} {2019})}\BibitemShut {NoStop}%
\bibitem [{\citenamefont {Broholm}\ \emph {et~al.}(2020)\citenamefont
  {Broholm}, \citenamefont {Cava}, \citenamefont {Kivelson}, \citenamefont
  {Nocera}, \citenamefont {Norman},\ and\ \citenamefont
  {Senthil}}]{science.aay0668}%
  \BibitemOpen
  \bibfield  {author} {\bibinfo {author} {\bibfnamefont {C.}~\bibnamefont
  {Broholm}}, \bibinfo {author} {\bibfnamefont {R.~J.}\ \bibnamefont {Cava}},
  \bibinfo {author} {\bibfnamefont {S.~A.}\ \bibnamefont {Kivelson}}, \bibinfo
  {author} {\bibfnamefont {D.~G.}\ \bibnamefont {Nocera}}, \bibinfo {author}
  {\bibfnamefont {M.~R.}\ \bibnamefont {Norman}},\ and\ \bibinfo {author}
  {\bibfnamefont {T.}~\bibnamefont {Senthil}},\ }\bibfield  {title} {\bibinfo
  {title} {Quantum spin liquids},\ }\href
  {https://doi.org/10.1126/science.aay0668} {\bibfield  {journal} {\bibinfo
  {journal} {Science}\ }\textbf {\bibinfo {volume} {367}},\ \bibinfo {pages}
  {eaay0668} (\bibinfo {year} {2020})}\BibitemShut {NoStop}%
\bibitem [{\citenamefont {Ramirez}\ \emph {et~al.}(1999)\citenamefont
  {Ramirez}, \citenamefont {Hayashi}, \citenamefont {Cava}, \citenamefont
  {Siddharthan},\ and\ \citenamefont {Shastry}}]{nature20619}%
  \BibitemOpen
  \bibfield  {author} {\bibinfo {author} {\bibfnamefont {A.~P.}\ \bibnamefont
  {Ramirez}}, \bibinfo {author} {\bibfnamefont {A.}~\bibnamefont {Hayashi}},
  \bibinfo {author} {\bibfnamefont {R.~J.}\ \bibnamefont {Cava}}, \bibinfo
  {author} {\bibfnamefont {R.}~\bibnamefont {Siddharthan}},\ and\ \bibinfo
  {author} {\bibfnamefont {B.~S.}\ \bibnamefont {Shastry}},\ }\bibfield
  {title} {\bibinfo {title} {Zero-point entropy in ‘spin ice’},\ }\href
  {https://doi.org/10.1038/20619} {\bibfield  {journal} {\bibinfo  {journal}
  {Nature}\ }\textbf {\bibinfo {volume} {399}},\ \bibinfo {pages} {333}
  (\bibinfo {year} {1999})}\BibitemShut {NoStop}%
\bibitem [{\citenamefont {den Hertog}\ and\ \citenamefont
  {Gingras}(2000)}]{PhysRevLett.84.3430}%
  \BibitemOpen
  \bibfield  {author} {\bibinfo {author} {\bibfnamefont {B.~C.}\ \bibnamefont
  {den Hertog}}\ and\ \bibinfo {author} {\bibfnamefont {M.~J.~P.}\ \bibnamefont
  {Gingras}},\ }\bibfield  {title} {\bibinfo {title} {Dipolar interactions and
  origin of spin ice in {I}sing pyrochlore magnets},\ }\href
  {https://doi.org/10.1103/PhysRevLett.84.3430} {\bibfield  {journal} {\bibinfo
   {journal} {Phys. Rev. Lett.}\ }\textbf {\bibinfo {volume} {84}},\ \bibinfo
  {pages} {3430} (\bibinfo {year} {2000})}\BibitemShut {NoStop}%
\bibitem [{\citenamefont {Bramwell}\ and\ \citenamefont
  {Gingras}(2001)}]{science.1064761}%
  \BibitemOpen
  \bibfield  {author} {\bibinfo {author} {\bibfnamefont {S.~T.}\ \bibnamefont
  {Bramwell}}\ and\ \bibinfo {author} {\bibfnamefont {M.~J.~P.}\ \bibnamefont
  {Gingras}},\ }\bibfield  {title} {\bibinfo {title} {Spin ice state in
  frustrated magnetic pyrochlore materials},\ }\href
  {https://doi.org/10.1126/science.1064761} {\bibfield  {journal} {\bibinfo
  {journal} {Science}\ }\textbf {\bibinfo {volume} {294}},\ \bibinfo {pages}
  {1495} (\bibinfo {year} {2001})}\BibitemShut {NoStop}%
\bibitem [{\citenamefont {Castelnovo}\ \emph {et~al.}(2008)\citenamefont
  {Castelnovo}, \citenamefont {Moessner},\ and\ \citenamefont
  {Sondhi}}]{nature06433}%
  \BibitemOpen
  \bibfield  {author} {\bibinfo {author} {\bibfnamefont {C.}~\bibnamefont
  {Castelnovo}}, \bibinfo {author} {\bibfnamefont {R.}~\bibnamefont
  {Moessner}},\ and\ \bibinfo {author} {\bibfnamefont {S.~L.}\ \bibnamefont
  {Sondhi}},\ }\bibfield  {title} {\bibinfo {title} {Magnetic monopoles in spin
  ice},\ }\href {https://doi.org/10.1038/nature06433} {\bibfield  {journal}
  {\bibinfo  {journal} {Nature}\ }\textbf {\bibinfo {volume} {451}},\ \bibinfo
  {pages} {42} (\bibinfo {year} {2008})}\BibitemShut {NoStop}%
\bibitem [{\citenamefont {Morris}\ \emph {et~al.}(2009)\citenamefont {Morris},
  \citenamefont {Tennant}, \citenamefont {Grigera}, \citenamefont {Klemke},
  \citenamefont {Castelnovo}, \citenamefont {Moessner}, \citenamefont
  {Czternasty}, \citenamefont {Meissner}, \citenamefont {Rule}, \citenamefont
  {Hoffmann}, \citenamefont {Kiefer}, \citenamefont {Gerischer}, \citenamefont
  {Slobinsky},\ and\ \citenamefont {Perry}}]{science.1178868}%
  \BibitemOpen
  \bibfield  {author} {\bibinfo {author} {\bibfnamefont {D.~J.~P.}\
  \bibnamefont {Morris}}, \bibinfo {author} {\bibfnamefont {D.~A.}\
  \bibnamefont {Tennant}}, \bibinfo {author} {\bibfnamefont {S.~A.}\
  \bibnamefont {Grigera}}, \bibinfo {author} {\bibfnamefont {B.}~\bibnamefont
  {Klemke}}, \bibinfo {author} {\bibfnamefont {C.}~\bibnamefont {Castelnovo}},
  \bibinfo {author} {\bibfnamefont {R.}~\bibnamefont {Moessner}}, \bibinfo
  {author} {\bibfnamefont {C.}~\bibnamefont {Czternasty}}, \bibinfo {author}
  {\bibfnamefont {M.}~\bibnamefont {Meissner}}, \bibinfo {author}
  {\bibfnamefont {K.~C.}\ \bibnamefont {Rule}}, \bibinfo {author}
  {\bibfnamefont {J.-U.}\ \bibnamefont {Hoffmann}}, \bibinfo {author}
  {\bibfnamefont {K.}~\bibnamefont {Kiefer}}, \bibinfo {author} {\bibfnamefont
  {S.}~\bibnamefont {Gerischer}}, \bibinfo {author} {\bibfnamefont
  {D.}~\bibnamefont {Slobinsky}},\ and\ \bibinfo {author} {\bibfnamefont
  {R.~S.}\ \bibnamefont {Perry}},\ }\bibfield  {title} {\bibinfo {title} {Dirac
  strings and magnetic monopoles in the spin ice {Dy$_2$Ti$_2$O$_7$}},\ }\href
  {https://doi.org/10.1126/science.1178868} {\bibfield  {journal} {\bibinfo
  {journal} {Science}\ }\textbf {\bibinfo {volume} {326}},\ \bibinfo {pages}
  {411} (\bibinfo {year} {2009})}\BibitemShut {NoStop}%
\bibitem [{\citenamefont {Fennell}\ \emph {et~al.}(2009)\citenamefont
  {Fennell}, \citenamefont {Deen}, \citenamefont {Wildes}, \citenamefont
  {Schmalzl}, \citenamefont {Prabhakaran}, \citenamefont {Boothroyd},
  \citenamefont {Aldus}, \citenamefont {McMorrow},\ and\ \citenamefont
  {Bramwell}}]{science.1177582}%
  \BibitemOpen
  \bibfield  {author} {\bibinfo {author} {\bibfnamefont {T.}~\bibnamefont
  {Fennell}}, \bibinfo {author} {\bibfnamefont {P.~P.}\ \bibnamefont {Deen}},
  \bibinfo {author} {\bibfnamefont {A.~R.}\ \bibnamefont {Wildes}}, \bibinfo
  {author} {\bibfnamefont {K.}~\bibnamefont {Schmalzl}}, \bibinfo {author}
  {\bibfnamefont {D.}~\bibnamefont {Prabhakaran}}, \bibinfo {author}
  {\bibfnamefont {A.~T.}\ \bibnamefont {Boothroyd}}, \bibinfo {author}
  {\bibfnamefont {R.~J.}\ \bibnamefont {Aldus}}, \bibinfo {author}
  {\bibfnamefont {D.~F.}\ \bibnamefont {McMorrow}},\ and\ \bibinfo {author}
  {\bibfnamefont {S.~T.}\ \bibnamefont {Bramwell}},\ }\bibfield  {title}
  {\bibinfo {title} {Magnetic {C}oulomb phase in the spin ice
  {Ho$_2$Ti$_2$O$_7$}},\ }\href {https://doi.org/10.1126/science.1177582}
  {\bibfield  {journal} {\bibinfo  {journal} {Science}\ }\textbf {\bibinfo
  {volume} {326}},\ \bibinfo {pages} {415} (\bibinfo {year}
  {2009})}\BibitemShut {NoStop}%
\bibitem [{\citenamefont {Gingras}(2009)}]{science.1181510}%
  \BibitemOpen
  \bibfield  {author} {\bibinfo {author} {\bibfnamefont {M.~J.~P.}\
  \bibnamefont {Gingras}},\ }\bibfield  {title} {\bibinfo {title} {Observing
  monopoles in a magnetic analog of ice},\ }\href
  {https://doi.org/10.1126/science.1181510} {\bibfield  {journal} {\bibinfo
  {journal} {Science}\ }\textbf {\bibinfo {volume} {326}},\ \bibinfo {pages}
  {375} (\bibinfo {year} {2009})}\BibitemShut {NoStop}%
\bibitem [{\citenamefont {Castelnovo}\ \emph {et~al.}(2012)\citenamefont
  {Castelnovo}, \citenamefont {Moessner},\ and\ \citenamefont
  {Sondhi}}]{annurev-conmatphys-020911-125058}%
  \BibitemOpen
  \bibfield  {author} {\bibinfo {author} {\bibfnamefont {C.}~\bibnamefont
  {Castelnovo}}, \bibinfo {author} {\bibfnamefont {R.}~\bibnamefont
  {Moessner}},\ and\ \bibinfo {author} {\bibfnamefont {S.}~\bibnamefont
  {Sondhi}},\ }\bibfield  {title} {\bibinfo {title} {Spin ice,
  fractionalization, and topological order},\ }\href
  {https://doi.org/10.1146/annurev-conmatphys-020911-125058} {\bibfield
  {journal} {\bibinfo  {journal} {Annual Review of Condensed Matter Physics}\
  }\textbf {\bibinfo {volume} {3}},\ \bibinfo {pages} {35} (\bibinfo {year}
  {2012})}\BibitemShut {NoStop}%
\bibitem [{\citenamefont {Hermele}\ \emph {et~al.}(2004)\citenamefont
  {Hermele}, \citenamefont {Fisher},\ and\ \citenamefont
  {Balents}}]{PhysRevB.69.064404}%
  \BibitemOpen
  \bibfield  {author} {\bibinfo {author} {\bibfnamefont {M.}~\bibnamefont
  {Hermele}}, \bibinfo {author} {\bibfnamefont {M.~P.~A.}\ \bibnamefont
  {Fisher}},\ and\ \bibinfo {author} {\bibfnamefont {L.}~\bibnamefont
  {Balents}},\ }\bibfield  {title} {\bibinfo {title} {Pyrochlore photons: The
  {$U(1)$} spin liquid in a {$S=\frac{1}{2}$} three-dimensional frustrated
  magnet},\ }\href {https://doi.org/10.1103/PhysRevB.69.064404} {\bibfield
  {journal} {\bibinfo  {journal} {Phys. Rev. B}\ }\textbf {\bibinfo {volume}
  {69}},\ \bibinfo {pages} {064404} (\bibinfo {year} {2004})}\BibitemShut
  {NoStop}%
\bibitem [{\citenamefont {Banerjee}\ \emph {et~al.}(2008)\citenamefont
  {Banerjee}, \citenamefont {Isakov}, \citenamefont {Damle},\ and\
  \citenamefont {Kim}}]{PhysRevLett.100.047208}%
  \BibitemOpen
  \bibfield  {author} {\bibinfo {author} {\bibfnamefont {A.}~\bibnamefont
  {Banerjee}}, \bibinfo {author} {\bibfnamefont {S.~V.}\ \bibnamefont
  {Isakov}}, \bibinfo {author} {\bibfnamefont {K.}~\bibnamefont {Damle}},\ and\
  \bibinfo {author} {\bibfnamefont {Y.~B.}\ \bibnamefont {Kim}},\ }\bibfield
  {title} {\bibinfo {title} {Unusual liquid state of hard-core bosons on the
  pyrochlore lattice},\ }\href {https://doi.org/10.1103/PhysRevLett.100.047208}
  {\bibfield  {journal} {\bibinfo  {journal} {Phys. Rev. Lett.}\ }\textbf
  {\bibinfo {volume} {100}},\ \bibinfo {pages} {047208} (\bibinfo {year}
  {2008})}\BibitemShut {NoStop}%
\bibitem [{\citenamefont {Savary}\ and\ \citenamefont
  {Balents}(2012)}]{PhysRevLett.108.037202}%
  \BibitemOpen
  \bibfield  {author} {\bibinfo {author} {\bibfnamefont {L.}~\bibnamefont
  {Savary}}\ and\ \bibinfo {author} {\bibfnamefont {L.}~\bibnamefont
  {Balents}},\ }\bibfield  {title} {\bibinfo {title} {Coulombic quantum liquids
  in spin-$1/2$ pyrochlores},\ }\href
  {https://doi.org/10.1103/PhysRevLett.108.037202} {\bibfield  {journal}
  {\bibinfo  {journal} {Phys. Rev. Lett.}\ }\textbf {\bibinfo {volume} {108}},\
  \bibinfo {pages} {037202} (\bibinfo {year} {2012})}\BibitemShut {NoStop}%
\bibitem [{\citenamefont {Lee}\ \emph {et~al.}(2012)\citenamefont {Lee},
  \citenamefont {Onoda},\ and\ \citenamefont {Balents}}]{PhysRevB.86.104412}%
  \BibitemOpen
  \bibfield  {author} {\bibinfo {author} {\bibfnamefont {S.}~\bibnamefont
  {Lee}}, \bibinfo {author} {\bibfnamefont {S.}~\bibnamefont {Onoda}},\ and\
  \bibinfo {author} {\bibfnamefont {L.}~\bibnamefont {Balents}},\ }\bibfield
  {title} {\bibinfo {title} {Generic quantum spin ice},\ }\href
  {https://doi.org/10.1103/PhysRevB.86.104412} {\bibfield  {journal} {\bibinfo
  {journal} {Phys. Rev. B}\ }\textbf {\bibinfo {volume} {86}},\ \bibinfo
  {pages} {104412} (\bibinfo {year} {2012})}\BibitemShut {NoStop}%
\bibitem [{\citenamefont {Benton}\ \emph {et~al.}(2018)\citenamefont {Benton},
  \citenamefont {Jaubert}, \citenamefont {Singh}, \citenamefont {Oitmaa},\ and\
  \citenamefont {Shannon}}]{PhysRevLett.121.067201}%
  \BibitemOpen
  \bibfield  {author} {\bibinfo {author} {\bibfnamefont {O.}~\bibnamefont
  {Benton}}, \bibinfo {author} {\bibfnamefont {L.~D.~C.}\ \bibnamefont
  {Jaubert}}, \bibinfo {author} {\bibfnamefont {R.~R.~P.}\ \bibnamefont
  {Singh}}, \bibinfo {author} {\bibfnamefont {J.}~\bibnamefont {Oitmaa}},\ and\
  \bibinfo {author} {\bibfnamefont {N.}~\bibnamefont {Shannon}},\ }\bibfield
  {title} {\bibinfo {title} {Quantum spin ice with frustrated transverse
  exchange: From a $\ensuremath{\pi}$-flux phase to a nematic quantum spin
  liquid},\ }\href {https://doi.org/10.1103/PhysRevLett.121.067201} {\bibfield
  {journal} {\bibinfo  {journal} {Phys. Rev. Lett.}\ }\textbf {\bibinfo
  {volume} {121}},\ \bibinfo {pages} {067201} (\bibinfo {year}
  {2018})}\BibitemShut {NoStop}%
\bibitem [{\citenamefont {Gardner}\ \emph {et~al.}(2010)\citenamefont
  {Gardner}, \citenamefont {Gingras},\ and\ \citenamefont
  {Greedan}}]{RevModPhys.82.53}%
  \BibitemOpen
  \bibfield  {author} {\bibinfo {author} {\bibfnamefont {J.~S.}\ \bibnamefont
  {Gardner}}, \bibinfo {author} {\bibfnamefont {M.~J.~P.}\ \bibnamefont
  {Gingras}},\ and\ \bibinfo {author} {\bibfnamefont {J.~E.}\ \bibnamefont
  {Greedan}},\ }\bibfield  {title} {\bibinfo {title} {Magnetic pyrochlore
  oxides},\ }\href {https://doi.org/10.1103/RevModPhys.82.53} {\bibfield
  {journal} {\bibinfo  {journal} {Rev. Mod. Phys.}\ }\textbf {\bibinfo {volume}
  {82}},\ \bibinfo {pages} {53} (\bibinfo {year} {2010})}\BibitemShut {NoStop}%
\bibitem [{\citenamefont {Ross}\ \emph {et~al.}(2011)\citenamefont {Ross},
  \citenamefont {Savary}, \citenamefont {Gaulin},\ and\ \citenamefont
  {Balents}}]{PhysRevX.1.021002}%
  \BibitemOpen
  \bibfield  {author} {\bibinfo {author} {\bibfnamefont {K.~A.}\ \bibnamefont
  {Ross}}, \bibinfo {author} {\bibfnamefont {L.}~\bibnamefont {Savary}},
  \bibinfo {author} {\bibfnamefont {B.~D.}\ \bibnamefont {Gaulin}},\ and\
  \bibinfo {author} {\bibfnamefont {L.}~\bibnamefont {Balents}},\ }\bibfield
  {title} {\bibinfo {title} {Quantum excitations in quantum spin ice},\ }\href
  {https://doi.org/10.1103/PhysRevX.1.021002} {\bibfield  {journal} {\bibinfo
  {journal} {Phys. Rev. X}\ }\textbf {\bibinfo {volume} {1}},\ \bibinfo {pages}
  {021002} (\bibinfo {year} {2011})}\BibitemShut {NoStop}%
\bibitem [{\citenamefont {Gingras}\ and\ \citenamefont
  {McClarty}(2014)}]{Gingras_2014}%
  \BibitemOpen
  \bibfield  {author} {\bibinfo {author} {\bibfnamefont {M.~J.~P.}\
  \bibnamefont {Gingras}}\ and\ \bibinfo {author} {\bibfnamefont {P.~A.}\
  \bibnamefont {McClarty}},\ }\bibfield  {title} {\bibinfo {title} {Quantum
  spin ice: a search for gapless quantum spin liquids in pyrochlore magnets},\
  }\href {https://doi.org/10.1088/0034-4885/77/5/056501} {\bibfield  {journal}
  {\bibinfo  {journal} {Reports on Progress in Physics}\ }\textbf {\bibinfo
  {volume} {77}},\ \bibinfo {pages} {056501} (\bibinfo {year}
  {2014})}\BibitemShut {NoStop}%
\bibitem [{\citenamefont {Yan}\ \emph {et~al.}(2017)\citenamefont {Yan},
  \citenamefont {Benton}, \citenamefont {Jaubert},\ and\ \citenamefont
  {Shannon}}]{PhysRevB.95.094422}%
  \BibitemOpen
  \bibfield  {author} {\bibinfo {author} {\bibfnamefont {H.}~\bibnamefont
  {Yan}}, \bibinfo {author} {\bibfnamefont {O.}~\bibnamefont {Benton}},
  \bibinfo {author} {\bibfnamefont {L.}~\bibnamefont {Jaubert}},\ and\ \bibinfo
  {author} {\bibfnamefont {N.}~\bibnamefont {Shannon}},\ }\bibfield  {title}
  {\bibinfo {title} {Theory of multiple-phase competition in pyrochlore magnets
  with anisotropic exchange with application to
  {${\mathrm{Yb}}_{2}{\mathrm{Ti}}_{2}{\mathrm{O}}_{7},
  {\mathrm{Er}}_{2}{\mathrm{Ti}}_{2}{\mathrm{O}}_{7}$, and
  ${\mathrm{Er}}_{2}{\mathrm{Sn}}_{2}{\mathrm{O}}_{7}$}},\ }\href
  {https://doi.org/10.1103/PhysRevB.95.094422} {\bibfield  {journal} {\bibinfo
  {journal} {Phys. Rev. B}\ }\textbf {\bibinfo {volume} {95}},\ \bibinfo
  {pages} {094422} (\bibinfo {year} {2017})}\BibitemShut {NoStop}%
\bibitem [{\citenamefont {Rau}\ and\ \citenamefont
  {Gingras}(2019)}]{annurev-conmatphys-022317-110520}%
  \BibitemOpen
  \bibfield  {author} {\bibinfo {author} {\bibfnamefont {J.~G.}\ \bibnamefont
  {Rau}}\ and\ \bibinfo {author} {\bibfnamefont {M.~J.}\ \bibnamefont
  {Gingras}},\ }\bibfield  {title} {\bibinfo {title} {Frustrated quantum
  rare-earth pyrochlores},\ }\href
  {https://doi.org/10.1146/annurev-conmatphys-022317-110520} {\bibfield
  {journal} {\bibinfo  {journal} {Annual Review of Condensed Matter Physics}\
  }\textbf {\bibinfo {volume} {10}},\ \bibinfo {pages} {357} (\bibinfo {year}
  {2019})}\BibitemShut {NoStop}%
\bibitem [{\citenamefont {Benton}\ and\ \citenamefont
  {Shannon}(2015)}]{JPSJ.84.104710}%
  \BibitemOpen
  \bibfield  {author} {\bibinfo {author} {\bibfnamefont {O.}~\bibnamefont
  {Benton}}\ and\ \bibinfo {author} {\bibfnamefont {N.}~\bibnamefont
  {Shannon}},\ }\bibfield  {title} {\bibinfo {title} {Ground state selection
  and spin-liquid behaviour in the classical {H}eisenberg model on the
  breathing pyrochlore lattice},\ }\href
  {https://doi.org/10.7566/JPSJ.84.104710} {\bibfield  {journal} {\bibinfo
  {journal} {Journal of the Physical Society of Japan}\ }\textbf {\bibinfo
  {volume} {84}},\ \bibinfo {pages} {104710} (\bibinfo {year}
  {2015})}\BibitemShut {NoStop}%
\bibitem [{\citenamefont {Savary}\ \emph {et~al.}(2016)\citenamefont {Savary},
  \citenamefont {Wang}, \citenamefont {Kee}, \citenamefont {Kim}, \citenamefont
  {Yu},\ and\ \citenamefont {Chen}}]{PhysRevB.94.075146}%
  \BibitemOpen
  \bibfield  {author} {\bibinfo {author} {\bibfnamefont {L.}~\bibnamefont
  {Savary}}, \bibinfo {author} {\bibfnamefont {X.}~\bibnamefont {Wang}},
  \bibinfo {author} {\bibfnamefont {H.-Y.}\ \bibnamefont {Kee}}, \bibinfo
  {author} {\bibfnamefont {Y.~B.}\ \bibnamefont {Kim}}, \bibinfo {author}
  {\bibfnamefont {Y.}~\bibnamefont {Yu}},\ and\ \bibinfo {author}
  {\bibfnamefont {G.}~\bibnamefont {Chen}},\ }\bibfield  {title} {\bibinfo
  {title} {Quantum spin ice on the breathing pyrochlore lattice},\ }\href
  {https://doi.org/10.1103/PhysRevB.94.075146} {\bibfield  {journal} {\bibinfo
  {journal} {Phys. Rev. B}\ }\textbf {\bibinfo {volume} {94}},\ \bibinfo
  {pages} {075146} (\bibinfo {year} {2016})}\BibitemShut {NoStop}%
\bibitem [{\citenamefont {Tsunetsugu}(2017)}]{ptx023}%
  \BibitemOpen
  \bibfield  {author} {\bibinfo {author} {\bibfnamefont {H.}~\bibnamefont
  {Tsunetsugu}},\ }\bibfield  {title} {\bibinfo {title} {Theory of
  antiferromagnetic {H}eisenberg spins on a breathing pyrochlore lattice},\
  }\href {https://doi.org/10.1093/ptep/ptx023} {\bibfield  {journal} {\bibinfo
  {journal} {Progress of Theoretical and Experimental Physics}\ }\textbf
  {\bibinfo {volume} {2017}},\ \bibinfo {pages} {033I01} (\bibinfo {year}
  {2017})}\BibitemShut {NoStop}%
\bibitem [{\citenamefont {Essafi}\ \emph {et~al.}(2017)\citenamefont {Essafi},
  \citenamefont {Jaubert},\ and\ \citenamefont {Udagawa}}]{Essafi_2017}%
  \BibitemOpen
  \bibfield  {author} {\bibinfo {author} {\bibfnamefont {K.}~\bibnamefont
  {Essafi}}, \bibinfo {author} {\bibfnamefont {L.~D.~C.}\ \bibnamefont
  {Jaubert}},\ and\ \bibinfo {author} {\bibfnamefont {M.}~\bibnamefont
  {Udagawa}},\ }\bibfield  {title} {\bibinfo {title} {Flat bands and {D}irac
  cones in breathing lattices},\ }\href
  {https://doi.org/10.1088/1361-648x/aa782f} {\bibfield  {journal} {\bibinfo
  {journal} {Journal of Physics: Condensed Matter}\ }\textbf {\bibinfo {volume}
  {29}},\ \bibinfo {pages} {315802} (\bibinfo {year} {2017})}\BibitemShut
  {NoStop}%
\bibitem [{\citenamefont {Ezawa}(2018)}]{PhysRevLett.120.026801}%
  \BibitemOpen
  \bibfield  {author} {\bibinfo {author} {\bibfnamefont {M.}~\bibnamefont
  {Ezawa}},\ }\bibfield  {title} {\bibinfo {title} {Higher-order topological
  insulators and semimetals on the breathing kagome and pyrochlore lattices},\
  }\href {https://doi.org/10.1103/PhysRevLett.120.026801} {\bibfield  {journal}
  {\bibinfo  {journal} {Phys. Rev. Lett.}\ }\textbf {\bibinfo {volume} {120}},\
  \bibinfo {pages} {026801} (\bibinfo {year} {2018})}\BibitemShut {NoStop}%
\bibitem [{\citenamefont {Aoyama}\ and\ \citenamefont
  {Kawamura}(2019)}]{PhysRevB.99.144406}%
  \BibitemOpen
  \bibfield  {author} {\bibinfo {author} {\bibfnamefont {K.}~\bibnamefont
  {Aoyama}}\ and\ \bibinfo {author} {\bibfnamefont {H.}~\bibnamefont
  {Kawamura}},\ }\bibfield  {title} {\bibinfo {title} {Spin ordering induced by
  lattice distortions in classical {H}eisenberg antiferromagnets on the
  breathing pyrochlore lattice},\ }\href
  {https://doi.org/10.1103/PhysRevB.99.144406} {\bibfield  {journal} {\bibinfo
  {journal} {Phys. Rev. B}\ }\textbf {\bibinfo {volume} {99}},\ \bibinfo
  {pages} {144406} (\bibinfo {year} {2019})}\BibitemShut {NoStop}%
\bibitem [{\citenamefont {Aoyama}\ and\ \citenamefont
  {Kawamura}()}]{2206.03707}%
  \BibitemOpen
  \bibfield  {author} {\bibinfo {author} {\bibfnamefont {K.}~\bibnamefont
  {Aoyama}}\ and\ \bibinfo {author} {\bibfnamefont {H.}~\bibnamefont
  {Kawamura}},\ }\bibfield  {title} {\bibinfo {title} {Hedgehog lattice and
  field-induced chirality in breathing-pyrochlore {H}eisenberg
  antiferromagnets},\ }\href {https://arxiv.org/abs/2206.03707} {\ }\Eprint
  {https://arxiv.org/abs/2206.03707} {arXiv:2206.03707} \BibitemShut {NoStop}%
\bibitem [{\citenamefont {Okamoto}\ \emph {et~al.}(2013)\citenamefont
  {Okamoto}, \citenamefont {Nilsen}, \citenamefont {Attfield},\ and\
  \citenamefont {Hiroi}}]{PhysRevLett.110.097203}%
  \BibitemOpen
  \bibfield  {author} {\bibinfo {author} {\bibfnamefont {Y.}~\bibnamefont
  {Okamoto}}, \bibinfo {author} {\bibfnamefont {G.~J.}\ \bibnamefont {Nilsen}},
  \bibinfo {author} {\bibfnamefont {J.~P.}\ \bibnamefont {Attfield}},\ and\
  \bibinfo {author} {\bibfnamefont {Z.}~\bibnamefont {Hiroi}},\ }\bibfield
  {title} {\bibinfo {title} {Breathing pyrochlore lattice realized in
  {$A$}-site ordered spinel oxides {${\mathrm{LiGaCr}}_{4}{\mathrm{O}}_{8}$ and
  ${\mathrm{LiInCr}}_{4}{\mathrm{O}}_{8}$}},\ }\href
  {https://doi.org/10.1103/PhysRevLett.110.097203} {\bibfield  {journal}
  {\bibinfo  {journal} {Phys. Rev. Lett.}\ }\textbf {\bibinfo {volume} {110}},\
  \bibinfo {pages} {097203} (\bibinfo {year} {2013})}\BibitemShut {NoStop}%
\bibitem [{\citenamefont {Tanaka}\ \emph {et~al.}(2014)\citenamefont {Tanaka},
  \citenamefont {Yoshida}, \citenamefont {Takigawa}, \citenamefont {Okamoto},\
  and\ \citenamefont {Hiroi}}]{PhysRevLett.113.227204}%
  \BibitemOpen
  \bibfield  {author} {\bibinfo {author} {\bibfnamefont {Y.}~\bibnamefont
  {Tanaka}}, \bibinfo {author} {\bibfnamefont {M.}~\bibnamefont {Yoshida}},
  \bibinfo {author} {\bibfnamefont {M.}~\bibnamefont {Takigawa}}, \bibinfo
  {author} {\bibfnamefont {Y.}~\bibnamefont {Okamoto}},\ and\ \bibinfo {author}
  {\bibfnamefont {Z.}~\bibnamefont {Hiroi}},\ }\bibfield  {title} {\bibinfo
  {title} {Novel phase transitions in the breathing pyrochlore lattice:
  {$^{7}\mathrm{Li}\text{\ensuremath{-}}\mathrm{NMR}$ on
  ${\mathrm{LiInCr}}_{4}{\mathrm{O}}_{8}$ and
  ${\mathrm{LiGaCr}}_{4}{\mathrm{O}}_{8}$}},\ }\href
  {https://doi.org/10.1103/PhysRevLett.113.227204} {\bibfield  {journal}
  {\bibinfo  {journal} {Phys. Rev. Lett.}\ }\textbf {\bibinfo {volume} {113}},\
  \bibinfo {pages} {227204} (\bibinfo {year} {2014})}\BibitemShut {NoStop}%
\bibitem [{\citenamefont {Okamoto}\ \emph {et~al.}(2015)\citenamefont
  {Okamoto}, \citenamefont {Nilsen}, \citenamefont {Nakazono},\ and\
  \citenamefont {Hiroi}}]{JPSJ.84.043707}%
  \BibitemOpen
  \bibfield  {author} {\bibinfo {author} {\bibfnamefont {Y.}~\bibnamefont
  {Okamoto}}, \bibinfo {author} {\bibfnamefont {G.~J.}\ \bibnamefont {Nilsen}},
  \bibinfo {author} {\bibfnamefont {T.}~\bibnamefont {Nakazono}},\ and\
  \bibinfo {author} {\bibfnamefont {Z.}~\bibnamefont {Hiroi}},\ }\bibfield
  {title} {\bibinfo {title} {Magnetic phase diagram of the breathing pyrochlore
  antiferromagnet {LiGa$_{1-x}$In$_x$Cr$_4$O$_8$}},\ }\href
  {https://doi.org/10.7566/JPSJ.84.043707} {\bibfield  {journal} {\bibinfo
  {journal} {Journal of the Physical Society of Japan}\ }\textbf {\bibinfo
  {volume} {84}},\ \bibinfo {pages} {043707} (\bibinfo {year}
  {2015})}\BibitemShut {NoStop}%
\bibitem [{\citenamefont {Nilsen}\ \emph {et~al.}(2015)\citenamefont {Nilsen},
  \citenamefont {Okamoto}, \citenamefont {Masuda}, \citenamefont
  {Rodriguez-Carvajal}, \citenamefont {Mutka}, \citenamefont {Hansen},\ and\
  \citenamefont {Hiroi}}]{PhysRevB.91.174435}%
  \BibitemOpen
  \bibfield  {author} {\bibinfo {author} {\bibfnamefont {G.~J.}\ \bibnamefont
  {Nilsen}}, \bibinfo {author} {\bibfnamefont {Y.}~\bibnamefont {Okamoto}},
  \bibinfo {author} {\bibfnamefont {T.}~\bibnamefont {Masuda}}, \bibinfo
  {author} {\bibfnamefont {J.}~\bibnamefont {Rodriguez-Carvajal}}, \bibinfo
  {author} {\bibfnamefont {H.}~\bibnamefont {Mutka}}, \bibinfo {author}
  {\bibfnamefont {T.}~\bibnamefont {Hansen}},\ and\ \bibinfo {author}
  {\bibfnamefont {Z.}~\bibnamefont {Hiroi}},\ }\bibfield  {title} {\bibinfo
  {title} {Complex magnetostructural order in the frustrated spinel
  {${\mathrm{LiInCr}}_{4}{\mathrm{O}}_{8}$}},\ }\href
  {https://doi.org/10.1103/PhysRevB.91.174435} {\bibfield  {journal} {\bibinfo
  {journal} {Phys. Rev. B}\ }\textbf {\bibinfo {volume} {91}},\ \bibinfo
  {pages} {174435} (\bibinfo {year} {2015})}\BibitemShut {NoStop}%
\bibitem [{\citenamefont {Lee}\ \emph {et~al.}(2016)\citenamefont {Lee},
  \citenamefont {Do}, \citenamefont {Lee}, \citenamefont {Choi}, \citenamefont
  {Lee}, \citenamefont {Choi}, \citenamefont {Reyes}, \citenamefont {Kuhns},
  \citenamefont {Ozarowski},\ and\ \citenamefont {Choi}}]{PhysRevB.93.174402}%
  \BibitemOpen
  \bibfield  {author} {\bibinfo {author} {\bibfnamefont {S.}~\bibnamefont
  {Lee}}, \bibinfo {author} {\bibfnamefont {S.-H.}\ \bibnamefont {Do}},
  \bibinfo {author} {\bibfnamefont {W.-J.}\ \bibnamefont {Lee}}, \bibinfo
  {author} {\bibfnamefont {Y.~S.}\ \bibnamefont {Choi}}, \bibinfo {author}
  {\bibfnamefont {M.}~\bibnamefont {Lee}}, \bibinfo {author} {\bibfnamefont
  {E.~S.}\ \bibnamefont {Choi}}, \bibinfo {author} {\bibfnamefont {A.~P.}\
  \bibnamefont {Reyes}}, \bibinfo {author} {\bibfnamefont {P.~L.}\ \bibnamefont
  {Kuhns}}, \bibinfo {author} {\bibfnamefont {A.}~\bibnamefont {Ozarowski}},\
  and\ \bibinfo {author} {\bibfnamefont {K.-Y.}\ \bibnamefont {Choi}},\
  }\bibfield  {title} {\bibinfo {title} {Multistage symmetry breaking in the
  breathing pyrochlore lattice
  {${\mathrm{Li}(\mathrm{Ga},\mathrm{In})\mathrm{Cr}}_{4}{\mathrm{O}}_{8}$}},\
  }\href {https://doi.org/10.1103/PhysRevB.93.174402} {\bibfield  {journal}
  {\bibinfo  {journal} {Phys. Rev. B}\ }\textbf {\bibinfo {volume} {93}},\
  \bibinfo {pages} {174402} (\bibinfo {year} {2016})}\BibitemShut {NoStop}%
\bibitem [{\citenamefont {Okamoto}\ \emph {et~al.}(2017)\citenamefont
  {Okamoto}, \citenamefont {Nakamura}, \citenamefont {Miyake}, \citenamefont
  {Takeyama}, \citenamefont {Tokunaga}, \citenamefont {Matsuo}, \citenamefont
  {Kindo},\ and\ \citenamefont {Hiroi}}]{PhysRevB.95.134438}%
  \BibitemOpen
  \bibfield  {author} {\bibinfo {author} {\bibfnamefont {Y.}~\bibnamefont
  {Okamoto}}, \bibinfo {author} {\bibfnamefont {D.}~\bibnamefont {Nakamura}},
  \bibinfo {author} {\bibfnamefont {A.}~\bibnamefont {Miyake}}, \bibinfo
  {author} {\bibfnamefont {S.}~\bibnamefont {Takeyama}}, \bibinfo {author}
  {\bibfnamefont {M.}~\bibnamefont {Tokunaga}}, \bibinfo {author}
  {\bibfnamefont {A.}~\bibnamefont {Matsuo}}, \bibinfo {author} {\bibfnamefont
  {K.}~\bibnamefont {Kindo}},\ and\ \bibinfo {author} {\bibfnamefont
  {Z.}~\bibnamefont {Hiroi}},\ }\bibfield  {title} {\bibinfo {title} {Magnetic
  transitions under ultrahigh magnetic fields of up to 130 {T} in the breathing
  pyrochlore antiferromagnet {${\mathrm{LiInCr}}_{4}{\mathrm{O}}_{8}$}},\
  }\href {https://doi.org/10.1103/PhysRevB.95.134438} {\bibfield  {journal}
  {\bibinfo  {journal} {Phys. Rev. B}\ }\textbf {\bibinfo {volume} {95}},\
  \bibinfo {pages} {134438} (\bibinfo {year} {2017})}\BibitemShut {NoStop}%
\bibitem [{\citenamefont {Okamoto}\ \emph {et~al.}(2018)\citenamefont
  {Okamoto}, \citenamefont {Mori}, \citenamefont {Katayama}, \citenamefont
  {Miyake}, \citenamefont {Tokunaga}, \citenamefont {Matsuo}, \citenamefont
  {Kindo},\ and\ \citenamefont {Takenaka}}]{JPSJ.87.034709}%
  \BibitemOpen
  \bibfield  {author} {\bibinfo {author} {\bibfnamefont {Y.}~\bibnamefont
  {Okamoto}}, \bibinfo {author} {\bibfnamefont {M.}~\bibnamefont {Mori}},
  \bibinfo {author} {\bibfnamefont {N.}~\bibnamefont {Katayama}}, \bibinfo
  {author} {\bibfnamefont {A.}~\bibnamefont {Miyake}}, \bibinfo {author}
  {\bibfnamefont {M.}~\bibnamefont {Tokunaga}}, \bibinfo {author}
  {\bibfnamefont {A.}~\bibnamefont {Matsuo}}, \bibinfo {author} {\bibfnamefont
  {K.}~\bibnamefont {Kindo}},\ and\ \bibinfo {author} {\bibfnamefont
  {K.}~\bibnamefont {Takenaka}},\ }\bibfield  {title} {\bibinfo {title}
  {Magnetic and structural properties of {A}-site ordered chromium spinel
  sulfides: Alternating antiferromagnetic and ferromagnetic interactions in the
  breathing pyrochlore lattice},\ }\href
  {https://doi.org/10.7566/JPSJ.87.034709} {\bibfield  {journal} {\bibinfo
  {journal} {Journal of the Physical Society of Japan}\ }\textbf {\bibinfo
  {volume} {87}},\ \bibinfo {pages} {034709} (\bibinfo {year}
  {2018})}\BibitemShut {NoStop}%
\bibitem [{\citenamefont {Kimura}\ \emph {et~al.}(2014)\citenamefont {Kimura},
  \citenamefont {Nakatsuji},\ and\ \citenamefont
  {Kimura}}]{PhysRevB.90.060414}%
  \BibitemOpen
  \bibfield  {author} {\bibinfo {author} {\bibfnamefont {K.}~\bibnamefont
  {Kimura}}, \bibinfo {author} {\bibfnamefont {S.}~\bibnamefont {Nakatsuji}},\
  and\ \bibinfo {author} {\bibfnamefont {T.}~\bibnamefont {Kimura}},\
  }\bibfield  {title} {\bibinfo {title} {Experimental realization of a quantum
  breathing pyrochlore antiferromagnet},\ }\href
  {https://doi.org/10.1103/PhysRevB.90.060414} {\bibfield  {journal} {\bibinfo
  {journal} {Phys. Rev. B}\ }\textbf {\bibinfo {volume} {90}},\ \bibinfo
  {pages} {060414} (\bibinfo {year} {2014})}\BibitemShut {NoStop}%
\bibitem [{\citenamefont {Haku}\ \emph
  {et~al.}(2016{\natexlab{a}})\citenamefont {Haku}, \citenamefont {Soda},
  \citenamefont {Sera}, \citenamefont {Kimura}, \citenamefont {Itoh},
  \citenamefont {Yokoo},\ and\ \citenamefont {Masuda}}]{JPSJ.85.034721}%
  \BibitemOpen
  \bibfield  {author} {\bibinfo {author} {\bibfnamefont {T.}~\bibnamefont
  {Haku}}, \bibinfo {author} {\bibfnamefont {M.}~\bibnamefont {Soda}}, \bibinfo
  {author} {\bibfnamefont {M.}~\bibnamefont {Sera}}, \bibinfo {author}
  {\bibfnamefont {K.}~\bibnamefont {Kimura}}, \bibinfo {author} {\bibfnamefont
  {S.}~\bibnamefont {Itoh}}, \bibinfo {author} {\bibfnamefont {T.}~\bibnamefont
  {Yokoo}},\ and\ \bibinfo {author} {\bibfnamefont {T.}~\bibnamefont
  {Masuda}},\ }\bibfield  {title} {\bibinfo {title} {Crystal field excitations
  in the breathing pyrochlore antiferromagnet {Ba$_3$Yb$_2$Zn$_5$O$_{11}$}},\
  }\href {https://doi.org/10.7566/JPSJ.85.034721} {\bibfield  {journal}
  {\bibinfo  {journal} {Journal of the Physical Society of Japan}\ }\textbf
  {\bibinfo {volume} {85}},\ \bibinfo {pages} {034721} (\bibinfo {year}
  {2016}{\natexlab{a}})}\BibitemShut {NoStop}%
\bibitem [{\citenamefont {Rau}\ \emph {et~al.}(2016)\citenamefont {Rau},
  \citenamefont {Wu}, \citenamefont {May}, \citenamefont {Poudel},
  \citenamefont {Winn}, \citenamefont {Garlea}, \citenamefont {Huq},
  \citenamefont {Whitfield}, \citenamefont {Taylor}, \citenamefont {Lumsden},
  \citenamefont {Gingras},\ and\ \citenamefont
  {Christianson}}]{PhysRevLett.116.257204}%
  \BibitemOpen
  \bibfield  {author} {\bibinfo {author} {\bibfnamefont {J.~G.}\ \bibnamefont
  {Rau}}, \bibinfo {author} {\bibfnamefont {L.~S.}\ \bibnamefont {Wu}},
  \bibinfo {author} {\bibfnamefont {A.~F.}\ \bibnamefont {May}}, \bibinfo
  {author} {\bibfnamefont {L.}~\bibnamefont {Poudel}}, \bibinfo {author}
  {\bibfnamefont {B.}~\bibnamefont {Winn}}, \bibinfo {author} {\bibfnamefont
  {V.~O.}\ \bibnamefont {Garlea}}, \bibinfo {author} {\bibfnamefont
  {A.}~\bibnamefont {Huq}}, \bibinfo {author} {\bibfnamefont {P.}~\bibnamefont
  {Whitfield}}, \bibinfo {author} {\bibfnamefont {A.~E.}\ \bibnamefont
  {Taylor}}, \bibinfo {author} {\bibfnamefont {M.~D.}\ \bibnamefont {Lumsden}},
  \bibinfo {author} {\bibfnamefont {M.~J.~P.}\ \bibnamefont {Gingras}},\ and\
  \bibinfo {author} {\bibfnamefont {A.~D.}\ \bibnamefont {Christianson}},\
  }\bibfield  {title} {\bibinfo {title} {Anisotropic exchange within decoupled
  tetrahedra in the quantum breathing pyrochlore
  {${\mathrm{Ba}}_{3}{\mathrm{Yb}}_{2}{\mathrm{Zn}}_{5}{\mathrm{O}}_{11}$}},\
  }\href {https://doi.org/10.1103/PhysRevLett.116.257204} {\bibfield  {journal}
  {\bibinfo  {journal} {Phys. Rev. Lett.}\ }\textbf {\bibinfo {volume} {116}},\
  \bibinfo {pages} {257204} (\bibinfo {year} {2016})}\BibitemShut {NoStop}%
\bibitem [{\citenamefont {Haku}\ \emph
  {et~al.}(2016{\natexlab{b}})\citenamefont {Haku}, \citenamefont {Kimura},
  \citenamefont {Matsumoto}, \citenamefont {Soda}, \citenamefont {Sera},
  \citenamefont {Yu}, \citenamefont {Mole}, \citenamefont {Takeuchi},
  \citenamefont {Nakatsuji}, \citenamefont {Kono}, \citenamefont {Sakakibara},
  \citenamefont {Chang},\ and\ \citenamefont {Masuda}}]{PhysRevB.93.220407}%
  \BibitemOpen
  \bibfield  {author} {\bibinfo {author} {\bibfnamefont {T.}~\bibnamefont
  {Haku}}, \bibinfo {author} {\bibfnamefont {K.}~\bibnamefont {Kimura}},
  \bibinfo {author} {\bibfnamefont {Y.}~\bibnamefont {Matsumoto}}, \bibinfo
  {author} {\bibfnamefont {M.}~\bibnamefont {Soda}}, \bibinfo {author}
  {\bibfnamefont {M.}~\bibnamefont {Sera}}, \bibinfo {author} {\bibfnamefont
  {D.}~\bibnamefont {Yu}}, \bibinfo {author} {\bibfnamefont {R.~A.}\
  \bibnamefont {Mole}}, \bibinfo {author} {\bibfnamefont {T.}~\bibnamefont
  {Takeuchi}}, \bibinfo {author} {\bibfnamefont {S.}~\bibnamefont {Nakatsuji}},
  \bibinfo {author} {\bibfnamefont {Y.}~\bibnamefont {Kono}}, \bibinfo {author}
  {\bibfnamefont {T.}~\bibnamefont {Sakakibara}}, \bibinfo {author}
  {\bibfnamefont {L.-J.}\ \bibnamefont {Chang}},\ and\ \bibinfo {author}
  {\bibfnamefont {T.}~\bibnamefont {Masuda}},\ }\bibfield  {title} {\bibinfo
  {title} {Low-energy excitations and ground-state selection in the quantum
  breathing pyrochlore antiferromagnet
  {${\mathrm{Ba}}_{3}{\mathrm{Yb}}_{2}{\mathrm{Zn}}_{5}{\mathrm{O}}_{11}$}},\
  }\href {https://doi.org/10.1103/PhysRevB.93.220407} {\bibfield  {journal}
  {\bibinfo  {journal} {Phys. Rev. B}\ }\textbf {\bibinfo {volume} {93}},\
  \bibinfo {pages} {220407} (\bibinfo {year} {2016}{\natexlab{b}})}\BibitemShut
  {NoStop}%
\bibitem [{\citenamefont {Rau}\ \emph {et~al.}(2018)\citenamefont {Rau},
  \citenamefont {Wu}, \citenamefont {May}, \citenamefont {Taylor},
  \citenamefont {Liu}, \citenamefont {Higgins}, \citenamefont {Butch},
  \citenamefont {Ross}, \citenamefont {Nair}, \citenamefont {Lumsden},
  \citenamefont {Gingras},\ and\ \citenamefont {Christianson}}]{Rau_2018}%
  \BibitemOpen
  \bibfield  {author} {\bibinfo {author} {\bibfnamefont {J.~G.}\ \bibnamefont
  {Rau}}, \bibinfo {author} {\bibfnamefont {L.~S.}\ \bibnamefont {Wu}},
  \bibinfo {author} {\bibfnamefont {A.~F.}\ \bibnamefont {May}}, \bibinfo
  {author} {\bibfnamefont {A.~E.}\ \bibnamefont {Taylor}}, \bibinfo {author}
  {\bibfnamefont {I.-L.}\ \bibnamefont {Liu}}, \bibinfo {author} {\bibfnamefont
  {J.}~\bibnamefont {Higgins}}, \bibinfo {author} {\bibfnamefont {N.~P.}\
  \bibnamefont {Butch}}, \bibinfo {author} {\bibfnamefont {K.~A.}\ \bibnamefont
  {Ross}}, \bibinfo {author} {\bibfnamefont {H.~S.}\ \bibnamefont {Nair}},
  \bibinfo {author} {\bibfnamefont {M.~D.}\ \bibnamefont {Lumsden}}, \bibinfo
  {author} {\bibfnamefont {M.~J.~P.}\ \bibnamefont {Gingras}},\ and\ \bibinfo
  {author} {\bibfnamefont {A.~D.}\ \bibnamefont {Christianson}},\ }\bibfield
  {title} {\bibinfo {title} {Behavior of the breathing pyrochlore lattice
  {Ba$_3$Yb$_2$Zn$_5$O$_{11}$} in applied magnetic field},\ }\href
  {https://doi.org/10.1088/1361-648x/aae45a} {\bibfield  {journal} {\bibinfo
  {journal} {Journal of Physics: Condensed Matter}\ }\textbf {\bibinfo {volume}
  {30}},\ \bibinfo {pages} {455801} (\bibinfo {year} {2018})}\BibitemShut
  {NoStop}%
\bibitem [{\citenamefont {Dissanayake}\ \emph {et~al.}()\citenamefont
  {Dissanayake}, \citenamefont {Shi}, \citenamefont {Rau}, \citenamefont {Bag},
  \citenamefont {Steinhardt}, \citenamefont {Butch}, \citenamefont {Frontzek},
  \citenamefont {Podlesnyak}, \citenamefont {Graf}, \citenamefont
  {Marjerrison}, \citenamefont {Liu}, \citenamefont {Gingras},\ and\
  \citenamefont {Haravifard}}]{2111.06293}%
  \BibitemOpen
  \bibfield  {author} {\bibinfo {author} {\bibfnamefont {S.}~\bibnamefont
  {Dissanayake}}, \bibinfo {author} {\bibfnamefont {Z.}~\bibnamefont {Shi}},
  \bibinfo {author} {\bibfnamefont {J.~G.}\ \bibnamefont {Rau}}, \bibinfo
  {author} {\bibfnamefont {R.}~\bibnamefont {Bag}}, \bibinfo {author}
  {\bibfnamefont {W.}~\bibnamefont {Steinhardt}}, \bibinfo {author}
  {\bibfnamefont {N.~P.}\ \bibnamefont {Butch}}, \bibinfo {author}
  {\bibfnamefont {M.}~\bibnamefont {Frontzek}}, \bibinfo {author}
  {\bibfnamefont {A.}~\bibnamefont {Podlesnyak}}, \bibinfo {author}
  {\bibfnamefont {D.}~\bibnamefont {Graf}}, \bibinfo {author} {\bibfnamefont
  {C.}~\bibnamefont {Marjerrison}}, \bibinfo {author} {\bibfnamefont
  {J.}~\bibnamefont {Liu}}, \bibinfo {author} {\bibfnamefont {M.~J.~P.}\
  \bibnamefont {Gingras}},\ and\ \bibinfo {author} {\bibfnamefont
  {S.}~\bibnamefont {Haravifard}},\ }\bibfield  {title} {\bibinfo {title}
  {Towards understanding the magnetic properties of the breathing pyrochlore
  compound {Ba$_3$Yb$_2$Zn$_5$O$_{11}$}: A single crystal study},\ }\href
  {https://arxiv.org/abs/2111.06293} {\ }\Eprint
  {https://arxiv.org/abs/2111.06293} {arXiv:2111.06293} \BibitemShut {NoStop}%
\bibitem [{\citenamefont {Yan}\ \emph {et~al.}(2020)\citenamefont {Yan},
  \citenamefont {Benton}, \citenamefont {Jaubert},\ and\ \citenamefont
  {Shannon}}]{PhysRevLett.124.127203}%
  \BibitemOpen
  \bibfield  {author} {\bibinfo {author} {\bibfnamefont {H.}~\bibnamefont
  {Yan}}, \bibinfo {author} {\bibfnamefont {O.}~\bibnamefont {Benton}},
  \bibinfo {author} {\bibfnamefont {L.~D.~C.}\ \bibnamefont {Jaubert}},\ and\
  \bibinfo {author} {\bibfnamefont {N.}~\bibnamefont {Shannon}},\ }\bibfield
  {title} {\bibinfo {title} {Rank--2 {$U(1)$} spin liquid on the breathing
  pyrochlore lattice},\ }\href {https://doi.org/10.1103/PhysRevLett.124.127203}
  {\bibfield  {journal} {\bibinfo  {journal} {Phys. Rev. Lett.}\ }\textbf
  {\bibinfo {volume} {124}},\ \bibinfo {pages} {127203} (\bibinfo {year}
  {2020})}\BibitemShut {NoStop}%
\bibitem [{\citenamefont {Zhang}\ \emph {et~al.}(2022)\citenamefont {Zhang},
  \citenamefont {Buessen},\ and\ \citenamefont {Kim}}]{PhysRevB.105.L060408}%
  \BibitemOpen
  \bibfield  {author} {\bibinfo {author} {\bibfnamefont {E.~Z.}\ \bibnamefont
  {Zhang}}, \bibinfo {author} {\bibfnamefont {F.~L.}\ \bibnamefont {Buessen}},\
  and\ \bibinfo {author} {\bibfnamefont {Y.~B.}\ \bibnamefont {Kim}},\
  }\bibfield  {title} {\bibinfo {title} {Dynamical signatures of rank-2
  {$U(1)$} spin liquids},\ }\href
  {https://doi.org/10.1103/PhysRevB.105.L060408} {\bibfield  {journal}
  {\bibinfo  {journal} {Phys. Rev. B}\ }\textbf {\bibinfo {volume} {105}},\
  \bibinfo {pages} {L060408} (\bibinfo {year} {2022})}\BibitemShut {NoStop}%
\bibitem [{\citenamefont {Han}\ \emph {et~al.}(2022)\citenamefont {Han},
  \citenamefont {Patri},\ and\ \citenamefont {Kim}}]{PhysRevB.105.235120}%
  \BibitemOpen
  \bibfield  {author} {\bibinfo {author} {\bibfnamefont {S.}~\bibnamefont
  {Han}}, \bibinfo {author} {\bibfnamefont {A.~S.}\ \bibnamefont {Patri}},\
  and\ \bibinfo {author} {\bibfnamefont {Y.~B.}\ \bibnamefont {Kim}},\
  }\bibfield  {title} {\bibinfo {title} {Realization of fractonic quantum
  phases in the breathing pyrochlore lattice},\ }\href
  {https://doi.org/10.1103/PhysRevB.105.235120} {\bibfield  {journal} {\bibinfo
   {journal} {Phys. Rev. B}\ }\textbf {\bibinfo {volume} {105}},\ \bibinfo
  {pages} {235120} (\bibinfo {year} {2022})}\BibitemShut {NoStop}%
\bibitem [{\citenamefont {Pace}\ \emph {et~al.}()\citenamefont {Pace},
  \citenamefont {Castelnovo},\ and\ \citenamefont {Laumann}}]{2109.06890}%
  \BibitemOpen
  \bibfield  {author} {\bibinfo {author} {\bibfnamefont {S.~D.}\ \bibnamefont
  {Pace}}, \bibinfo {author} {\bibfnamefont {C.}~\bibnamefont {Castelnovo}},\
  and\ \bibinfo {author} {\bibfnamefont {C.~R.}\ \bibnamefont {Laumann}},\
  }\bibfield  {title} {\bibinfo {title} {Dynamical axions in {$U(1)$} quantum
  spin liquids},\ }\href {https://arxiv.org/abs/2109.06890} {\ }\Eprint
  {https://arxiv.org/abs/2109.06890} {arXiv:2109.06890} \BibitemShut {NoStop}%
\bibitem [{\citenamefont {Liu}\ \emph {et~al.}(2019)\citenamefont {Liu},
  \citenamefont {Hal\'asz},\ and\ \citenamefont
  {Balents}}]{PhysRevB.100.075125}%
  \BibitemOpen
  \bibfield  {author} {\bibinfo {author} {\bibfnamefont {C.}~\bibnamefont
  {Liu}}, \bibinfo {author} {\bibfnamefont {G.~B.}\ \bibnamefont {Hal\'asz}},\
  and\ \bibinfo {author} {\bibfnamefont {L.}~\bibnamefont {Balents}},\
  }\bibfield  {title} {\bibinfo {title} {Competing orders in pyrochlore magnets
  from a {$\mathbb{Z}_2$} spin liquid perspective},\ }\href
  {https://doi.org/10.1103/PhysRevB.100.075125} {\bibfield  {journal} {\bibinfo
   {journal} {Phys. Rev. B}\ }\textbf {\bibinfo {volume} {100}},\ \bibinfo
  {pages} {075125} (\bibinfo {year} {2019})}\BibitemShut {NoStop}%
\bibitem [{\citenamefont {Liu}\ \emph {et~al.}(2021)\citenamefont {Liu},
  \citenamefont {Hal\'asz},\ and\ \citenamefont
  {Balents}}]{PhysRevB.104.054401}%
  \BibitemOpen
  \bibfield  {author} {\bibinfo {author} {\bibfnamefont {C.}~\bibnamefont
  {Liu}}, \bibinfo {author} {\bibfnamefont {G.~B.}\ \bibnamefont {Hal\'asz}},\
  and\ \bibinfo {author} {\bibfnamefont {L.}~\bibnamefont {Balents}},\
  }\bibfield  {title} {\bibinfo {title} {Symmetric {$U(1)$} and
  {$\mathbb{Z}_2$} spin liquids on the pyrochlore lattice},\ }\href
  {https://doi.org/10.1103/PhysRevB.104.054401} {\bibfield  {journal} {\bibinfo
   {journal} {Phys. Rev. B}\ }\textbf {\bibinfo {volume} {104}},\ \bibinfo
  {pages} {054401} (\bibinfo {year} {2021})}\BibitemShut {NoStop}%
\bibitem [{\citenamefont {Desrochers}\ \emph {et~al.}(2022)\citenamefont
  {Desrochers}, \citenamefont {Chern},\ and\ \citenamefont
  {Kim}}]{PhysRevB.105.035149}%
  \BibitemOpen
  \bibfield  {author} {\bibinfo {author} {\bibfnamefont {F.}~\bibnamefont
  {Desrochers}}, \bibinfo {author} {\bibfnamefont {L.~E.}\ \bibnamefont
  {Chern}},\ and\ \bibinfo {author} {\bibfnamefont {Y.~B.}\ \bibnamefont
  {Kim}},\ }\bibfield  {title} {\bibinfo {title} {Competing {$U(1)$} and
  {$\mathbb{Z}_2$} dipolar-octupolar quantum spin liquids on the pyrochlore
  lattice: Application to
  {${\mathrm{Ce}}_{2}{\mathrm{Zr}}_{2}{\mathrm{O}}_{7}$}},\ }\href
  {https://doi.org/10.1103/PhysRevB.105.035149} {\bibfield  {journal} {\bibinfo
   {journal} {Phys. Rev. B}\ }\textbf {\bibinfo {volume} {105}},\ \bibinfo
  {pages} {035149} (\bibinfo {year} {2022})}\BibitemShut {NoStop}%
\bibitem [{\citenamefont {Schneider}\ \emph {et~al.}(2022)\citenamefont
  {Schneider}, \citenamefont {Halimeh},\ and\ \citenamefont
  {Punk}}]{PhysRevB.105.125122}%
  \BibitemOpen
  \bibfield  {author} {\bibinfo {author} {\bibfnamefont {B.}~\bibnamefont
  {Schneider}}, \bibinfo {author} {\bibfnamefont {J.~C.}\ \bibnamefont
  {Halimeh}},\ and\ \bibinfo {author} {\bibfnamefont {M.}~\bibnamefont
  {Punk}},\ }\bibfield  {title} {\bibinfo {title} {Projective symmetry group
  classification of chiral {$\mathbb{Z}_{2}$} spin liquids on the pyrochlore
  lattice: Application to the spin-$\frac{1}{2}$ {XXZ} {H}eisenberg model},\
  }\href {https://doi.org/10.1103/PhysRevB.105.125122} {\bibfield  {journal}
  {\bibinfo  {journal} {Phys. Rev. B}\ }\textbf {\bibinfo {volume} {105}},\
  \bibinfo {pages} {125122} (\bibinfo {year} {2022})}\BibitemShut {NoStop}%
\bibitem [{\citenamefont {Wen}(2002)}]{PhysRevB.65.165113}%
  \BibitemOpen
  \bibfield  {author} {\bibinfo {author} {\bibfnamefont {X.-G.}\ \bibnamefont
  {Wen}},\ }\bibfield  {title} {\bibinfo {title} {Quantum orders and symmetric
  spin liquids},\ }\href {https://doi.org/10.1103/PhysRevB.65.165113}
  {\bibfield  {journal} {\bibinfo  {journal} {Phys. Rev. B}\ }\textbf {\bibinfo
  {volume} {65}},\ \bibinfo {pages} {165113} (\bibinfo {year}
  {2002})}\BibitemShut {NoStop}%
\bibitem [{\citenamefont {Lu}\ \emph {et~al.}(2011)\citenamefont {Lu},
  \citenamefont {Ran},\ and\ \citenamefont {Lee}}]{PhysRevB.83.224413}%
  \BibitemOpen
  \bibfield  {author} {\bibinfo {author} {\bibfnamefont {Y.-M.}\ \bibnamefont
  {Lu}}, \bibinfo {author} {\bibfnamefont {Y.}~\bibnamefont {Ran}},\ and\
  \bibinfo {author} {\bibfnamefont {P.~A.}\ \bibnamefont {Lee}},\ }\bibfield
  {title} {\bibinfo {title} {{$\mathbb{Z}_2$} spin liquids in the
  {$S=\frac{1}{2}$} {H}eisenberg model on the kagome lattice: A projective
  symmetry-group study of {S}chwinger fermion mean-field states},\ }\href
  {https://doi.org/10.1103/PhysRevB.83.224413} {\bibfield  {journal} {\bibinfo
  {journal} {Phys. Rev. B}\ }\textbf {\bibinfo {volume} {83}},\ \bibinfo
  {pages} {224413} (\bibinfo {year} {2011})}\BibitemShut {NoStop}%
\bibitem [{\citenamefont {Lu}\ and\ \citenamefont
  {Ran}(2011)}]{PhysRevB.84.024420}%
  \BibitemOpen
  \bibfield  {author} {\bibinfo {author} {\bibfnamefont {Y.-M.}\ \bibnamefont
  {Lu}}\ and\ \bibinfo {author} {\bibfnamefont {Y.}~\bibnamefont {Ran}},\
  }\bibfield  {title} {\bibinfo {title} {{$\mathbb{Z}_2$} spin liquid and
  chiral antiferromagnetic phase in the {H}ubbard model on a honeycomb
  lattice},\ }\href {https://doi.org/10.1103/PhysRevB.84.024420} {\bibfield
  {journal} {\bibinfo  {journal} {Phys. Rev. B}\ }\textbf {\bibinfo {volume}
  {84}},\ \bibinfo {pages} {024420} (\bibinfo {year} {2011})}\BibitemShut
  {NoStop}%
\bibitem [{\citenamefont {Huang}\ \emph {et~al.}(2017)\citenamefont {Huang},
  \citenamefont {Kim},\ and\ \citenamefont {Lu}}]{PhysRevB.95.054404}%
  \BibitemOpen
  \bibfield  {author} {\bibinfo {author} {\bibfnamefont {B.}~\bibnamefont
  {Huang}}, \bibinfo {author} {\bibfnamefont {Y.~B.}\ \bibnamefont {Kim}},\
  and\ \bibinfo {author} {\bibfnamefont {Y.-M.}\ \bibnamefont {Lu}},\
  }\bibfield  {title} {\bibinfo {title} {Interplay of nonsymmorphic symmetry
  and spin-orbit coupling in hyperkagome spin liquids: Applications to
  {${\mathrm{Na}}_{4}{\mathrm{Ir}}_{3}{\mathrm{O}}_{8}$}},\ }\href
  {https://doi.org/10.1103/PhysRevB.95.054404} {\bibfield  {journal} {\bibinfo
  {journal} {Phys. Rev. B}\ }\textbf {\bibinfo {volume} {95}},\ \bibinfo
  {pages} {054404} (\bibinfo {year} {2017})}\BibitemShut {NoStop}%
\bibitem [{\citenamefont {Schaffer}\ \emph {et~al.}(2017)\citenamefont
  {Schaffer}, \citenamefont {Huh}, \citenamefont {Hwang},\ and\ \citenamefont
  {Kim}}]{PhysRevB.95.054410}%
  \BibitemOpen
  \bibfield  {author} {\bibinfo {author} {\bibfnamefont {R.}~\bibnamefont
  {Schaffer}}, \bibinfo {author} {\bibfnamefont {Y.}~\bibnamefont {Huh}},
  \bibinfo {author} {\bibfnamefont {K.}~\bibnamefont {Hwang}},\ and\ \bibinfo
  {author} {\bibfnamefont {Y.~B.}\ \bibnamefont {Kim}},\ }\bibfield  {title}
  {\bibinfo {title} {Quantum spin liquid in a breathing kagome lattice},\
  }\href {https://doi.org/10.1103/PhysRevB.95.054410} {\bibfield  {journal}
  {\bibinfo  {journal} {Phys. Rev. B}\ }\textbf {\bibinfo {volume} {95}},\
  \bibinfo {pages} {054410} (\bibinfo {year} {2017})}\BibitemShut {NoStop}%
\bibitem [{\citenamefont {Huang}\ \emph {et~al.}(2018)\citenamefont {Huang},
  \citenamefont {Choi}, \citenamefont {Kim},\ and\ \citenamefont
  {Lu}}]{PhysRevB.97.195141}%
  \BibitemOpen
  \bibfield  {author} {\bibinfo {author} {\bibfnamefont {B.}~\bibnamefont
  {Huang}}, \bibinfo {author} {\bibfnamefont {W.}~\bibnamefont {Choi}},
  \bibinfo {author} {\bibfnamefont {Y.~B.}\ \bibnamefont {Kim}},\ and\ \bibinfo
  {author} {\bibfnamefont {Y.-M.}\ \bibnamefont {Lu}},\ }\bibfield  {title}
  {\bibinfo {title} {Classification and properties of quantum spin liquids on
  the hyperhoneycomb lattice},\ }\href
  {https://doi.org/10.1103/PhysRevB.97.195141} {\bibfield  {journal} {\bibinfo
  {journal} {Phys. Rev. B}\ }\textbf {\bibinfo {volume} {97}},\ \bibinfo
  {pages} {195141} (\bibinfo {year} {2018})}\BibitemShut {NoStop}%
\bibitem [{\citenamefont {Chern}\ and\ \citenamefont
  {Kim}(2021)}]{PhysRevB.104.094413}%
  \BibitemOpen
  \bibfield  {author} {\bibinfo {author} {\bibfnamefont {L.~E.}\ \bibnamefont
  {Chern}}\ and\ \bibinfo {author} {\bibfnamefont {Y.~B.}\ \bibnamefont
  {Kim}},\ }\bibfield  {title} {\bibinfo {title} {Theoretical study of quantum
  spin liquids in {$S=\frac{1}{2}$} hyper-hyperkagome magnets: Classification,
  heat capacity, and dynamical spin structure factor},\ }\href
  {https://doi.org/10.1103/PhysRevB.104.094413} {\bibfield  {journal} {\bibinfo
   {journal} {Phys. Rev. B}\ }\textbf {\bibinfo {volume} {104}},\ \bibinfo
  {pages} {094413} (\bibinfo {year} {2021})}\BibitemShut {NoStop}%
\bibitem [{\citenamefont {Baskaran}\ \emph {et~al.}(1987)\citenamefont
  {Baskaran}, \citenamefont {Zou},\ and\ \citenamefont
  {Anderson}}]{BASKARAN1987973}%
  \BibitemOpen
  \bibfield  {author} {\bibinfo {author} {\bibfnamefont {G.}~\bibnamefont
  {Baskaran}}, \bibinfo {author} {\bibfnamefont {Z.}~\bibnamefont {Zou}},\ and\
  \bibinfo {author} {\bibfnamefont {P.}~\bibnamefont {Anderson}},\ }\bibfield
  {title} {\bibinfo {title} {The resonating valence bond state and high-{$T_c$}
  superconductivity -- a mean field theory},\ }\href
  {https://doi.org/https://doi.org/10.1016/0038-1098(87)90642-9} {\bibfield
  {journal} {\bibinfo  {journal} {Solid State Communications}\ }\textbf
  {\bibinfo {volume} {63}},\ \bibinfo {pages} {973} (\bibinfo {year}
  {1987})}\BibitemShut {NoStop}%
\bibitem [{\citenamefont {Baskaran}\ and\ \citenamefont
  {Anderson}(1988)}]{PhysRevB.37.580}%
  \BibitemOpen
  \bibfield  {author} {\bibinfo {author} {\bibfnamefont {G.}~\bibnamefont
  {Baskaran}}\ and\ \bibinfo {author} {\bibfnamefont {P.~W.}\ \bibnamefont
  {Anderson}},\ }\bibfield  {title} {\bibinfo {title} {Gauge theory of
  high-temperature superconductors and strongly correlated {F}ermi systems},\
  }\href {https://doi.org/10.1103/PhysRevB.37.580} {\bibfield  {journal}
  {\bibinfo  {journal} {Phys. Rev. B}\ }\textbf {\bibinfo {volume} {37}},\
  \bibinfo {pages} {580} (\bibinfo {year} {1988})}\BibitemShut {NoStop}%
\bibitem [{\citenamefont {Affleck}\ \emph {et~al.}(1988)\citenamefont
  {Affleck}, \citenamefont {Zou}, \citenamefont {Hsu},\ and\ \citenamefont
  {Anderson}}]{PhysRevB.38.745}%
  \BibitemOpen
  \bibfield  {author} {\bibinfo {author} {\bibfnamefont {I.}~\bibnamefont
  {Affleck}}, \bibinfo {author} {\bibfnamefont {Z.}~\bibnamefont {Zou}},
  \bibinfo {author} {\bibfnamefont {T.}~\bibnamefont {Hsu}},\ and\ \bibinfo
  {author} {\bibfnamefont {P.~W.}\ \bibnamefont {Anderson}},\ }\bibfield
  {title} {\bibinfo {title} {{$SU(2)$} gauge symmetry of the large-{$U$} limit
  of the {H}ubbard model},\ }\href {https://doi.org/10.1103/PhysRevB.38.745}
  {\bibfield  {journal} {\bibinfo  {journal} {Phys. Rev. B}\ }\textbf {\bibinfo
  {volume} {38}},\ \bibinfo {pages} {745} (\bibinfo {year} {1988})}\BibitemShut
  {NoStop}%
\bibitem [{\citenamefont {Wen}(1991)}]{PhysRevB.44.2664}%
  \BibitemOpen
  \bibfield  {author} {\bibinfo {author} {\bibfnamefont {X.~G.}\ \bibnamefont
  {Wen}},\ }\bibfield  {title} {\bibinfo {title} {Mean-field theory of
  spin-liquid states with finite energy gap and topological orders},\ }\href
  {https://doi.org/10.1103/PhysRevB.44.2664} {\bibfield  {journal} {\bibinfo
  {journal} {Phys. Rev. B}\ }\textbf {\bibinfo {volume} {44}},\ \bibinfo
  {pages} {2664} (\bibinfo {year} {1991})}\BibitemShut {NoStop}%
\bibitem [{spi({\natexlab{a}})}]{spinonnote}%
  \BibitemOpen
  \href@noop {} {} ({\natexlab{a}}),\ \bibinfo {note} {throughout this paper,
  the term \textit{spinon} denotes the $f$-fermion in the parton construction
  \eqref{partonrepresentation}, which is not to be confused with the magnetic
  monopole -- a defect of the 2-in-2-out configuration -- in a spin
  ice.}\BibitemShut {Stop}%
\bibitem [{\citenamefont {Ioffe}\ and\ \citenamefont
  {Larkin}(1989)}]{PhysRevB.39.8988}%
  \BibitemOpen
  \bibfield  {author} {\bibinfo {author} {\bibfnamefont {L.~B.}\ \bibnamefont
  {Ioffe}}\ and\ \bibinfo {author} {\bibfnamefont {A.~I.}\ \bibnamefont
  {Larkin}},\ }\bibfield  {title} {\bibinfo {title} {Gapless fermions and gauge
  fields in dielectrics},\ }\href {https://doi.org/10.1103/PhysRevB.39.8988}
  {\bibfield  {journal} {\bibinfo  {journal} {Phys. Rev. B}\ }\textbf {\bibinfo
  {volume} {39}},\ \bibinfo {pages} {8988} (\bibinfo {year}
  {1989})}\BibitemShut {NoStop}%
\bibitem [{\citenamefont {Lee}(1989)}]{PhysRevLett.63.680}%
  \BibitemOpen
  \bibfield  {author} {\bibinfo {author} {\bibfnamefont {P.~A.}\ \bibnamefont
  {Lee}},\ }\bibfield  {title} {\bibinfo {title} {Gauge field, {Aharonov-Bohm}
  flux, and high-{$T_c$} superconductivity},\ }\href
  {https://doi.org/10.1103/PhysRevLett.63.680} {\bibfield  {journal} {\bibinfo
  {journal} {Phys. Rev. Lett.}\ }\textbf {\bibinfo {volume} {63}},\ \bibinfo
  {pages} {680} (\bibinfo {year} {1989})}\BibitemShut {NoStop}%
\bibitem [{\citenamefont {Lee}\ and\ \citenamefont
  {Nagaosa}(1992)}]{PhysRevB.46.5621}%
  \BibitemOpen
  \bibfield  {author} {\bibinfo {author} {\bibfnamefont {P.~A.}\ \bibnamefont
  {Lee}}\ and\ \bibinfo {author} {\bibfnamefont {N.}~\bibnamefont {Nagaosa}},\
  }\bibfield  {title} {\bibinfo {title} {Gauge theory of the normal state of
  high-${T_c}$ superconductors},\ }\href
  {https://doi.org/10.1103/PhysRevB.46.5621} {\bibfield  {journal} {\bibinfo
  {journal} {Phys. Rev. B}\ }\textbf {\bibinfo {volume} {46}},\ \bibinfo
  {pages} {5621} (\bibinfo {year} {1992})}\BibitemShut {NoStop}%
\bibitem [{\citenamefont {Polchinski}(1994)}]{POLCHINSKI1994617}%
  \BibitemOpen
  \bibfield  {author} {\bibinfo {author} {\bibfnamefont {J.}~\bibnamefont
  {Polchinski}},\ }\bibfield  {title} {\bibinfo {title} {Low-energy dynamics of
  the spinon-gauge system},\ }\href
  {https://doi.org/10.1016/0550-3213(94)90449-9} {\bibfield  {journal}
  {\bibinfo  {journal} {Nuclear Physics B}\ }\textbf {\bibinfo {volume}
  {422}},\ \bibinfo {pages} {617} (\bibinfo {year} {1994})}\BibitemShut
  {NoStop}%
\bibitem [{\citenamefont {Kim}\ \emph {et~al.}(1994)\citenamefont {Kim},
  \citenamefont {Furusaki}, \citenamefont {Wen},\ and\ \citenamefont
  {Lee}}]{PhysRevB.50.17917}%
  \BibitemOpen
  \bibfield  {author} {\bibinfo {author} {\bibfnamefont {Y.~B.}\ \bibnamefont
  {Kim}}, \bibinfo {author} {\bibfnamefont {A.}~\bibnamefont {Furusaki}},
  \bibinfo {author} {\bibfnamefont {X.-G.}\ \bibnamefont {Wen}},\ and\ \bibinfo
  {author} {\bibfnamefont {P.~A.}\ \bibnamefont {Lee}},\ }\bibfield  {title}
  {\bibinfo {title} {Gauge-invariant response functions of fermions coupled to
  a gauge field},\ }\href {https://doi.org/10.1103/PhysRevB.50.17917}
  {\bibfield  {journal} {\bibinfo  {journal} {Phys. Rev. B}\ }\textbf {\bibinfo
  {volume} {50}},\ \bibinfo {pages} {17917} (\bibinfo {year}
  {1994})}\BibitemShut {NoStop}%
\bibitem [{\citenamefont {Kim}\ \emph {et~al.}(1995)\citenamefont {Kim},
  \citenamefont {Lee},\ and\ \citenamefont {Wen}}]{PhysRevB.52.17275}%
  \BibitemOpen
  \bibfield  {author} {\bibinfo {author} {\bibfnamefont {Y.~B.}\ \bibnamefont
  {Kim}}, \bibinfo {author} {\bibfnamefont {P.~A.}\ \bibnamefont {Lee}},\ and\
  \bibinfo {author} {\bibfnamefont {X.-G.}\ \bibnamefont {Wen}},\ }\bibfield
  {title} {\bibinfo {title} {Quantum {B}oltzmann equation of composite fermions
  interacting with a gauge field},\ }\href
  {https://doi.org/10.1103/PhysRevB.52.17275} {\bibfield  {journal} {\bibinfo
  {journal} {Phys. Rev. B}\ }\textbf {\bibinfo {volume} {52}},\ \bibinfo
  {pages} {17275} (\bibinfo {year} {1995})}\BibitemShut {NoStop}%
\bibitem [{\citenamefont {Kim}\ and\ \citenamefont
  {Lee}(1996)}]{PhysRevB.54.2715}%
  \BibitemOpen
  \bibfield  {author} {\bibinfo {author} {\bibfnamefont {Y.~B.}\ \bibnamefont
  {Kim}}\ and\ \bibinfo {author} {\bibfnamefont {P.~A.}\ \bibnamefont {Lee}},\
  }\bibfield  {title} {\bibinfo {title} {Specific heat and validity of the
  quasiparticle approximation in the half-filled {L}andau level},\ }\href
  {https://doi.org/10.1103/PhysRevB.54.2715} {\bibfield  {journal} {\bibinfo
  {journal} {Phys. Rev. B}\ }\textbf {\bibinfo {volume} {54}},\ \bibinfo
  {pages} {2715} (\bibinfo {year} {1996})}\BibitemShut {NoStop}%
\bibitem [{\citenamefont {Senthil}\ \emph {et~al.}(2004)\citenamefont
  {Senthil}, \citenamefont {Vojta},\ and\ \citenamefont
  {Sachdev}}]{PhysRevB.69.035111}%
  \BibitemOpen
  \bibfield  {author} {\bibinfo {author} {\bibfnamefont {T.}~\bibnamefont
  {Senthil}}, \bibinfo {author} {\bibfnamefont {M.}~\bibnamefont {Vojta}},\
  and\ \bibinfo {author} {\bibfnamefont {S.}~\bibnamefont {Sachdev}},\
  }\bibfield  {title} {\bibinfo {title} {Weak magnetism and non-{F}ermi liquids
  near heavy-fermion critical points},\ }\href
  {https://doi.org/10.1103/PhysRevB.69.035111} {\bibfield  {journal} {\bibinfo
  {journal} {Phys. Rev. B}\ }\textbf {\bibinfo {volume} {69}},\ \bibinfo
  {pages} {035111} (\bibinfo {year} {2004})}\BibitemShut {NoStop}%
\bibitem [{\citenamefont {Motrunich}(2005)}]{PhysRevB.72.045105}%
  \BibitemOpen
  \bibfield  {author} {\bibinfo {author} {\bibfnamefont {O.~I.}\ \bibnamefont
  {Motrunich}},\ }\bibfield  {title} {\bibinfo {title} {Variational study of
  triangular lattice spin-$1/2$ model with ring exchanges and spin liquid state
  in
  {$\ensuremath{\kappa}\text{\ensuremath{-}}{(\mathrm{ET})}_{2}{\mathrm{Cu}}_{2}{(\mathrm{CN})}_{3}$}},\
  }\href {https://doi.org/10.1103/PhysRevB.72.045105} {\bibfield  {journal}
  {\bibinfo  {journal} {Phys. Rev. B}\ }\textbf {\bibinfo {volume} {72}},\
  \bibinfo {pages} {045105} (\bibinfo {year} {2005})}\BibitemShut {NoStop}%
\bibitem [{\citenamefont {Lee}\ and\ \citenamefont
  {Lee}(2005)}]{PhysRevLett.95.036403}%
  \BibitemOpen
  \bibfield  {author} {\bibinfo {author} {\bibfnamefont {S.-S.}\ \bibnamefont
  {Lee}}\ and\ \bibinfo {author} {\bibfnamefont {P.~A.}\ \bibnamefont {Lee}},\
  }\bibfield  {title} {\bibinfo {title} {{$U(1)$} gauge theory of the {H}ubbard
  model: Spin liquid states and possible application to
  {$\ensuremath{\kappa}\mathrm{\text{\ensuremath{-}}}(\mathrm{BEDT}\mathrm{\text{\ensuremath{-}}}\mathrm{TTF}{)}_{2}{\mathrm{Cu}}_{2}(\mathrm{CN}{)}_{3}$}},\
  }\href {https://doi.org/10.1103/PhysRevLett.95.036403} {\bibfield  {journal}
  {\bibinfo  {journal} {Phys. Rev. Lett.}\ }\textbf {\bibinfo {volume} {95}},\
  \bibinfo {pages} {036403} (\bibinfo {year} {2005})}\BibitemShut {NoStop}%
\bibitem [{\citenamefont {Lee}\ \emph {et~al.}(2006)\citenamefont {Lee},
  \citenamefont {Nagaosa},\ and\ \citenamefont {Wen}}]{RevModPhys.78.17}%
  \BibitemOpen
  \bibfield  {author} {\bibinfo {author} {\bibfnamefont {P.~A.}\ \bibnamefont
  {Lee}}, \bibinfo {author} {\bibfnamefont {N.}~\bibnamefont {Nagaosa}},\ and\
  \bibinfo {author} {\bibfnamefont {X.-G.}\ \bibnamefont {Wen}},\ }\bibfield
  {title} {\bibinfo {title} {Doping a {M}ott insulator: Physics of
  high-temperature superconductivity},\ }\href
  {https://doi.org/10.1103/RevModPhys.78.17} {\bibfield  {journal} {\bibinfo
  {journal} {Rev. Mod. Phys.}\ }\textbf {\bibinfo {volume} {78}},\ \bibinfo
  {pages} {17} (\bibinfo {year} {2006})}\BibitemShut {NoStop}%
\bibitem [{\citenamefont {Nave}\ \emph {et~al.}(2007)\citenamefont {Nave},
  \citenamefont {Lee},\ and\ \citenamefont {Lee}}]{PhysRevB.76.165104}%
  \BibitemOpen
  \bibfield  {author} {\bibinfo {author} {\bibfnamefont {C.~P.}\ \bibnamefont
  {Nave}}, \bibinfo {author} {\bibfnamefont {S.-S.}\ \bibnamefont {Lee}},\ and\
  \bibinfo {author} {\bibfnamefont {P.~A.}\ \bibnamefont {Lee}},\ }\bibfield
  {title} {\bibinfo {title} {Susceptibility of a spinon {F}ermi surface coupled
  to a {$U(1)$} gauge field},\ }\href
  {https://doi.org/10.1103/PhysRevB.76.165104} {\bibfield  {journal} {\bibinfo
  {journal} {Phys. Rev. B}\ }\textbf {\bibinfo {volume} {76}},\ \bibinfo
  {pages} {165104} (\bibinfo {year} {2007})}\BibitemShut {NoStop}%
\bibitem [{\citenamefont {Nave}\ and\ \citenamefont
  {Lee}(2007)}]{PhysRevB.76.235124}%
  \BibitemOpen
  \bibfield  {author} {\bibinfo {author} {\bibfnamefont {C.~P.}\ \bibnamefont
  {Nave}}\ and\ \bibinfo {author} {\bibfnamefont {P.~A.}\ \bibnamefont {Lee}},\
  }\bibfield  {title} {\bibinfo {title} {Transport properties of a spinon
  {F}ermi surface coupled to a {$U(1)$} gauge field},\ }\href
  {https://doi.org/10.1103/PhysRevB.76.235124} {\bibfield  {journal} {\bibinfo
  {journal} {Phys. Rev. B}\ }\textbf {\bibinfo {volume} {76}},\ \bibinfo
  {pages} {235124} (\bibinfo {year} {2007})}\BibitemShut {NoStop}%
\bibitem [{\citenamefont {Tinkham}(2003)}]{tinkhamtextbook}%
  \BibitemOpen
  \bibfield  {author} {\bibinfo {author} {\bibfnamefont {M.}~\bibnamefont
  {Tinkham}},\ }\href@noop {} {\emph {\bibinfo {title} {Group Theory and
  Quantum Mechanics}}}\ (\bibinfo  {publisher} {Dover Publications},\ \bibinfo
  {address} {Mineola, New York},\ \bibinfo {year} {2003})\BibitemShut {NoStop}%
\bibitem [{\citenamefont {Dresselhaus}\ \emph {et~al.}(2008)\citenamefont
  {Dresselhaus}, \citenamefont {Dresselhaus},\ and\ \citenamefont
  {Jorio}}]{dresselhaustextbook}%
  \BibitemOpen
  \bibfield  {author} {\bibinfo {author} {\bibfnamefont {M.~S.}\ \bibnamefont
  {Dresselhaus}}, \bibinfo {author} {\bibfnamefont {G.}~\bibnamefont
  {Dresselhaus}},\ and\ \bibinfo {author} {\bibfnamefont {A.}~\bibnamefont
  {Jorio}},\ }\href {https://doi.org/10.1007/978-3-540-32899-5} {\emph
  {\bibinfo {title} {Group Theory: Application to the Physics of Condensed
  Matter}}}\ (\bibinfo  {publisher} {Springer},\ \bibinfo {address} {Berlin,
  Heidelberg},\ \bibinfo {year} {2008})\BibitemShut {NoStop}%
\bibitem [{\citenamefont {Shindou}\ and\ \citenamefont
  {Momoi}(2009)}]{PhysRevB.80.064410}%
  \BibitemOpen
  \bibfield  {author} {\bibinfo {author} {\bibfnamefont {R.}~\bibnamefont
  {Shindou}}\ and\ \bibinfo {author} {\bibfnamefont {T.}~\bibnamefont
  {Momoi}},\ }\bibfield  {title} {\bibinfo {title} {{$SU(2)$} slave-boson
  formulation of spin nematic states in {$S=\frac{1}{2}$} frustrated
  ferromagnets},\ }\href {https://doi.org/10.1103/PhysRevB.80.064410}
  {\bibfield  {journal} {\bibinfo  {journal} {Phys. Rev. B}\ }\textbf {\bibinfo
  {volume} {80}},\ \bibinfo {pages} {064410} (\bibinfo {year}
  {2009})}\BibitemShut {NoStop}%
\bibitem [{\citenamefont {Bhattacharjee}\ \emph {et~al.}(2012)\citenamefont
  {Bhattacharjee}, \citenamefont {Kim}, \citenamefont {Lee},\ and\
  \citenamefont {Lee}}]{PhysRevB.85.224428}%
  \BibitemOpen
  \bibfield  {author} {\bibinfo {author} {\bibfnamefont {S.}~\bibnamefont
  {Bhattacharjee}}, \bibinfo {author} {\bibfnamefont {Y.~B.}\ \bibnamefont
  {Kim}}, \bibinfo {author} {\bibfnamefont {S.-S.}\ \bibnamefont {Lee}},\ and\
  \bibinfo {author} {\bibfnamefont {D.-H.}\ \bibnamefont {Lee}},\ }\bibfield
  {title} {\bibinfo {title} {Fractionalized topological insulators from
  frustrated spin models in three dimensions},\ }\href
  {https://doi.org/10.1103/PhysRevB.85.224428} {\bibfield  {journal} {\bibinfo
  {journal} {Phys. Rev. B}\ }\textbf {\bibinfo {volume} {85}},\ \bibinfo
  {pages} {224428} (\bibinfo {year} {2012})}\BibitemShut {NoStop}%
\bibitem [{\citenamefont {Schaffer}\ \emph {et~al.}(2012)\citenamefont
  {Schaffer}, \citenamefont {Bhattacharjee},\ and\ \citenamefont
  {Kim}}]{PhysRevB.86.224417}%
  \BibitemOpen
  \bibfield  {author} {\bibinfo {author} {\bibfnamefont {R.}~\bibnamefont
  {Schaffer}}, \bibinfo {author} {\bibfnamefont {S.}~\bibnamefont
  {Bhattacharjee}},\ and\ \bibinfo {author} {\bibfnamefont {Y.~B.}\
  \bibnamefont {Kim}},\ }\bibfield  {title} {\bibinfo {title} {Quantum phase
  transition in {H}eisenberg-{K}itaev model},\ }\href
  {https://doi.org/10.1103/PhysRevB.86.224417} {\bibfield  {journal} {\bibinfo
  {journal} {Phys. Rev. B}\ }\textbf {\bibinfo {volume} {86}},\ \bibinfo
  {pages} {224417} (\bibinfo {year} {2012})}\BibitemShut {NoStop}%
\bibitem [{\citenamefont {Chern}\ and\ \citenamefont
  {Kim}(2019)}]{s41598-019-47517-6}%
  \BibitemOpen
  \bibfield  {author} {\bibinfo {author} {\bibfnamefont {L.~E.}\ \bibnamefont
  {Chern}}\ and\ \bibinfo {author} {\bibfnamefont {Y.~B.}\ \bibnamefont
  {Kim}},\ }\bibfield  {title} {\bibinfo {title} {Magnetic order with
  fractionalized excitations in pyrochlore magnets with strong spin-orbit
  coupling},\ }\href {https://doi.org/10.1038/s41598-019-47517-6} {\bibfield
  {journal} {\bibinfo  {journal} {Scientific Reports}\ }\textbf {\bibinfo
  {volume} {9}},\ \bibinfo {pages} {10974} (\bibinfo {year}
  {2019})}\BibitemShut {NoStop}%
\bibitem [{spi({\natexlab{b}})}]{spinmodelnote}%
  \BibitemOpen
  \href@noop {} {} ({\natexlab{b}}),\ \bibinfo {note} {there is also a bond
  dependent Ising interaction two orders of magnitude smaller than the
  Heisenberg interaction, which we neglect in this work.}\BibitemShut {Stop}%
\bibitem [{\citenamefont {Polyakov}(1977)}]{POLYAKOV1977429}%
  \BibitemOpen
  \bibfield  {author} {\bibinfo {author} {\bibfnamefont {A.}~\bibnamefont
  {Polyakov}},\ }\bibfield  {title} {\bibinfo {title} {Quark confinement and
  topology of gauge theories},\ }\href
  {https://doi.org/10.1016/0550-3213(77)90086-4} {\bibfield  {journal}
  {\bibinfo  {journal} {Nuclear Physics B}\ }\textbf {\bibinfo {volume}
  {120}},\ \bibinfo {pages} {429} (\bibinfo {year} {1977})}\BibitemShut
  {NoStop}%
\bibitem [{\citenamefont {Setyawan}\ and\ \citenamefont
  {Curtarolo}(2010)}]{SETYAWAN2010299}%
  \BibitemOpen
  \bibfield  {author} {\bibinfo {author} {\bibfnamefont {W.}~\bibnamefont
  {Setyawan}}\ and\ \bibinfo {author} {\bibfnamefont {S.}~\bibnamefont
  {Curtarolo}},\ }\bibfield  {title} {\bibinfo {title} {High-throughput
  electronic band structure calculations: Challenges and tools},\ }\href
  {https://doi.org/10.1016/j.commatsci.2010.05.010} {\bibfield  {journal}
  {\bibinfo  {journal} {Computational Materials Science}\ }\textbf {\bibinfo
  {volume} {49}},\ \bibinfo {pages} {299} (\bibinfo {year} {2010})}\BibitemShut
  {NoStop}%
\bibitem [{\citenamefont {Burnell}\ \emph {et~al.}(2009)\citenamefont
  {Burnell}, \citenamefont {Chakravarty},\ and\ \citenamefont
  {Sondhi}}]{PhysRevB.79.144432}%
  \BibitemOpen
  \bibfield  {author} {\bibinfo {author} {\bibfnamefont {F.~J.}\ \bibnamefont
  {Burnell}}, \bibinfo {author} {\bibfnamefont {S.}~\bibnamefont
  {Chakravarty}},\ and\ \bibinfo {author} {\bibfnamefont {S.~L.}\ \bibnamefont
  {Sondhi}},\ }\bibfield  {title} {\bibinfo {title} {Monopole flux state on the
  pyrochlore lattice},\ }\href {https://doi.org/10.1103/PhysRevB.79.144432}
  {\bibfield  {journal} {\bibinfo  {journal} {Phys. Rev. B}\ }\textbf {\bibinfo
  {volume} {79}},\ \bibinfo {pages} {144432} (\bibinfo {year}
  {2009})}\BibitemShut {NoStop}%
\bibitem [{\citenamefont {Holstein}\ \emph {et~al.}(1973)\citenamefont
  {Holstein}, \citenamefont {Norton},\ and\ \citenamefont
  {Pincus}}]{PhysRevB.8.2649}%
  \BibitemOpen
  \bibfield  {author} {\bibinfo {author} {\bibfnamefont {T.}~\bibnamefont
  {Holstein}}, \bibinfo {author} {\bibfnamefont {R.~E.}\ \bibnamefont
  {Norton}},\ and\ \bibinfo {author} {\bibfnamefont {P.}~\bibnamefont
  {Pincus}},\ }\bibfield  {title} {\bibinfo {title} {{de Haas-van Alphen}
  effect and the specific heat of an electron gas},\ }\href
  {https://doi.org/10.1103/PhysRevB.8.2649} {\bibfield  {journal} {\bibinfo
  {journal} {Phys. Rev. B}\ }\textbf {\bibinfo {volume} {8}},\ \bibinfo {pages}
  {2649} (\bibinfo {year} {1973})}\BibitemShut {NoStop}%
\bibitem [{\citenamefont {Reizer}(1989)}]{PhysRevB.40.11571}%
  \BibitemOpen
  \bibfield  {author} {\bibinfo {author} {\bibfnamefont {M.~Y.}\ \bibnamefont
  {Reizer}},\ }\bibfield  {title} {\bibinfo {title} {Relativistic effects in
  the electron density of states, specific heat, and the electron spectrum of
  normal metals},\ }\href {https://doi.org/10.1103/PhysRevB.40.11571}
  {\bibfield  {journal} {\bibinfo  {journal} {Phys. Rev. B}\ }\textbf {\bibinfo
  {volume} {40}},\ \bibinfo {pages} {11571} (\bibinfo {year}
  {1989})}\BibitemShut {NoStop}%
\bibitem [{\citenamefont {Nagaosa}(1999)}]{nagaosatextbook}%
  \BibitemOpen
  \bibfield  {author} {\bibinfo {author} {\bibfnamefont {N.}~\bibnamefont
  {Nagaosa}},\ }\href {https://doi.org/10.1007/978-3-662-03774-4} {\emph
  {\bibinfo {title} {Quantum Field Theory in Condensed Matter Physics}}}\
  (\bibinfo  {publisher} {Springer},\ \bibinfo {address} {Berlin, Heidelberg},\
  \bibinfo {year} {1999})\BibitemShut {NoStop}%
\bibitem [{\citenamefont {Tsunetsugu}(2001{\natexlab{a}})}]{JPSJ.70.640}%
  \BibitemOpen
  \bibfield  {author} {\bibinfo {author} {\bibfnamefont {H.}~\bibnamefont
  {Tsunetsugu}},\ }\bibfield  {title} {\bibinfo {title} {Antiferromagnetic
  quantum spins on the pyrochlore lattice},\ }\href
  {https://doi.org/10.1143/JPSJ.70.640} {\bibfield  {journal} {\bibinfo
  {journal} {Journal of the Physical Society of Japan}\ }\textbf {\bibinfo
  {volume} {70}},\ \bibinfo {pages} {640} (\bibinfo {year}
  {2001}{\natexlab{a}})}\BibitemShut {NoStop}%
\bibitem [{\citenamefont
  {Tsunetsugu}(2001{\natexlab{b}})}]{PhysRevB.65.024415}%
  \BibitemOpen
  \bibfield  {author} {\bibinfo {author} {\bibfnamefont {H.}~\bibnamefont
  {Tsunetsugu}},\ }\bibfield  {title} {\bibinfo {title} {Spin-singlet order in
  a pyrochlore antiferromagnet},\ }\href
  {https://doi.org/10.1103/PhysRevB.65.024415} {\bibfield  {journal} {\bibinfo
  {journal} {Phys. Rev. B}\ }\textbf {\bibinfo {volume} {65}},\ \bibinfo
  {pages} {024415} (\bibinfo {year} {2001}{\natexlab{b}})}\BibitemShut
  {NoStop}%
\bibitem [{\citenamefont {Canals}\ and\ \citenamefont
  {Lacroix}(1998)}]{PhysRevLett.80.2933}%
  \BibitemOpen
  \bibfield  {author} {\bibinfo {author} {\bibfnamefont {B.}~\bibnamefont
  {Canals}}\ and\ \bibinfo {author} {\bibfnamefont {C.}~\bibnamefont
  {Lacroix}},\ }\bibfield  {title} {\bibinfo {title} {Pyrochlore
  antiferromagnet: A three-dimensional quantum spin liquid},\ }\href
  {https://doi.org/10.1103/PhysRevLett.80.2933} {\bibfield  {journal} {\bibinfo
   {journal} {Phys. Rev. Lett.}\ }\textbf {\bibinfo {volume} {80}},\ \bibinfo
  {pages} {2933} (\bibinfo {year} {1998})}\BibitemShut {NoStop}%
\bibitem [{\citenamefont {Canals}\ and\ \citenamefont
  {Lacroix}(2000)}]{PhysRevB.61.1149}%
  \BibitemOpen
  \bibfield  {author} {\bibinfo {author} {\bibfnamefont {B.}~\bibnamefont
  {Canals}}\ and\ \bibinfo {author} {\bibfnamefont {C.}~\bibnamefont
  {Lacroix}},\ }\bibfield  {title} {\bibinfo {title} {Quantum spin liquid: The
  {H}eisenberg antiferromagnet on the three-dimensional pyrochlore lattice},\
  }\href {https://doi.org/10.1103/PhysRevB.61.1149} {\bibfield  {journal}
  {\bibinfo  {journal} {Phys. Rev. B}\ }\textbf {\bibinfo {volume} {61}},\
  \bibinfo {pages} {1149} (\bibinfo {year} {2000})}\BibitemShut {NoStop}%
\bibitem [{\citenamefont {Henley}(2010)}]{annurev-conmatphys-070909-104138}%
  \BibitemOpen
  \bibfield  {author} {\bibinfo {author} {\bibfnamefont {C.~L.}\ \bibnamefont
  {Henley}},\ }\bibfield  {title} {\bibinfo {title} {The ``{C}oulomb phase'' in
  frustrated systems},\ }\href
  {https://doi.org/10.1146/annurev-conmatphys-070909-104138} {\bibfield
  {journal} {\bibinfo  {journal} {Annual Review of Condensed Matter Physics}\
  }\textbf {\bibinfo {volume} {1}},\ \bibinfo {pages} {179} (\bibinfo {year}
  {2010})}\BibitemShut {NoStop}%
\bibitem [{\citenamefont {Hagym\'asi}\ \emph {et~al.}(2021)\citenamefont
  {Hagym\'asi}, \citenamefont {Sch\"afer}, \citenamefont {Moessner},\ and\
  \citenamefont {Luitz}}]{PhysRevLett.126.117204}%
  \BibitemOpen
  \bibfield  {author} {\bibinfo {author} {\bibfnamefont {I.}~\bibnamefont
  {Hagym\'asi}}, \bibinfo {author} {\bibfnamefont {R.}~\bibnamefont
  {Sch\"afer}}, \bibinfo {author} {\bibfnamefont {R.}~\bibnamefont
  {Moessner}},\ and\ \bibinfo {author} {\bibfnamefont {D.~J.}\ \bibnamefont
  {Luitz}},\ }\bibfield  {title} {\bibinfo {title} {Possible inversion symmetry
  breaking in the {$S=1/2$} pyrochlore {H}eisenberg magnet},\ }\href
  {https://doi.org/10.1103/PhysRevLett.126.117204} {\bibfield  {journal}
  {\bibinfo  {journal} {Phys. Rev. Lett.}\ }\textbf {\bibinfo {volume} {126}},\
  \bibinfo {pages} {117204} (\bibinfo {year} {2021})}\BibitemShut {NoStop}%
\bibitem [{\citenamefont {Mahan}(2000)}]{mahantextbook}%
  \BibitemOpen
  \bibfield  {author} {\bibinfo {author} {\bibfnamefont {G.~D.}\ \bibnamefont
  {Mahan}},\ }\href@noop {} {\emph {\bibinfo {title} {Many-Particle
  Physics}}},\ \bibinfo {edition} {3rd}\ ed.\ (\bibinfo  {publisher} {Kluwer
  Academic/Plenum Publishers},\ \bibinfo {address} {New York, New York},\
  \bibinfo {year} {2000})\BibitemShut {NoStop}%
\bibitem [{\citenamefont {Coleman}(2015)}]{colemantextbook}%
  \BibitemOpen
  \bibfield  {author} {\bibinfo {author} {\bibfnamefont {P.}~\bibnamefont
  {Coleman}},\ }\href {https://doi.org/10.1017/CBO9781139020916} {\emph
  {\bibinfo {title} {Introduction to Many-Body Physics}}}\ (\bibinfo
  {publisher} {Cambridge University Press},\ \bibinfo {year}
  {2015})\BibitemShut {NoStop}%
\bibitem [{\citenamefont {Laine}\ and\ \citenamefont
  {Vuorinen}(2016)}]{lainetextbook}%
  \BibitemOpen
  \bibfield  {author} {\bibinfo {author} {\bibfnamefont {M.}~\bibnamefont
  {Laine}}\ and\ \bibinfo {author} {\bibfnamefont {A.}~\bibnamefont
  {Vuorinen}},\ }\href {https://doi.org/10.1007/978-3-319-31933-9} {\emph
  {\bibinfo {title} {Basics of Thermal Field Theory: A Tutorial on Perturbative
  Computations}}}\ (\bibinfo  {publisher} {Springer International Publishing},\
  \bibinfo {address} {Switzerland},\ \bibinfo {year} {2016})\BibitemShut
  {NoStop}%
\end{thebibliography}%

\end{document}